\documentclass[reprint,aip,jmp,showpacs,onecolumn,nofootinbib]{revtex4-2}

\usepackage{amsthm,amssymb,amsmath}				
\usepackage{enumerate}							
\usepackage{graphicx}							
\usepackage[colorlinks = true,linkcolor = blue, 
            urlcolor  = blue, citecolor = blue, 
            anchorcolor = blue]{hyperref}       
\usepackage{geometry}							
\usepackage{hyperref}							    
\usepackage{booktabs}
\usepackage{xcolor}									
\usepackage{verbatim}								
\usepackage[normalem]{ulem}							
\usepackage{fancybox}								
\usepackage{bm}										
\usepackage{lipsum}									

\begin{document}

\title{
Exact solution and coherent states of an 
asymmetric oscillator with position-dependent mass
}

\author{Bruno G.\ da Costa}
\email{bruno.costa@ifsertao-pe.edu.br}
\affiliation{Instituto Federal de Educa\c{c}\~ao, Ci\^encia e Tecnologia do Sert\~ao Pernambucano,
             Rua Maria Luiza de Ara\'ujo Gomes Cabral s/n, 56316-686 Petrolina, Pernambuco, Brazil}
\author{Ignacio S.\ Gomez}
\email{ignacio.gomez@uesb.edu.br}
\affiliation{Departamento de Ci\^encias Exatas e Naturais, 
             Universidade Estadual do Sudoeste da Bahia, 
             Rodovia BR 415, km 03 s/n, 45700-000 Itapetinga, Bahia, Brazil}
\author{Biswanath Rath}
\email{biswanathrath10@gmail.com}
\affiliation{Department of Physics, Maharaja Sriram Chandra Bhanja Deo University, Baripada 757003, India}
\date{\today}

\begin{abstract}
We revisit the problem of the deformed oscillator 
with position-dependent mass 
[da Costa {\it et al.}, J. Math. Phys. {\bf 62}, 092101 (2021)]
in the classical and quantum formalisms,
by introducing the effect of the mass function
in both kinetic and potential energies.
The resulting Hamiltonian is mapped 
into a Morse oscillator
by means of a point canonical transformation from 
the usual phase space $(x, p)$ to a deformed one 
$(x_\gamma, \Pi_\gamma)$.
Similar to the Morse potential, 
the deformed oscillator presents bound trajectories 
in phase space corresponding to an anharmonic oscillatory motion 
in classical formalism and, therefore, bound states 
with a discrete spectrum in quantum formalism.
On the other hand, open trajectories in phase space 
are associated with scattering states and continuous energy spectrum.
Employing the factorization method,
we investigate the properties of
the coherent states, such as 
the time evolution and their uncertainties. 
A fast localization, 
classical and quantum, is reported for 
the coherent states due to the asymmetrical 
position-dependent mass.
An oscillation of the time evolution of 
the  uncertainty relationship is also observed, 
whose amplitude increases as the deformation increases.   
\end{abstract}

\maketitle 

\section{Introduction}

It is attributed to Schr\"odinger
for introducing the idea of coherent states 
in a seminal work of 1926 about the simple harmonic oscillator. 
In his work, a superposition of quantum states is 
constructed in order to reproduce the dynamics of 
its classical analog.\cite{Schrodinger-1926} 
Later, in the beginning of the 1960s,
Glauber, Klauder, and Sudarshan were pioneers 
in applying the coherent states in quantum optics.
\cite{Glauber-1963,Klauder-1963a,Klauder-1963b,Sudarshan-1963}
The term ``coherent states"
\cite{Glauber-1963,Klauder-1963a,Klauder-1963b,Sudarshan-1963,
Barut-Girardello-1971,Perelomov-1972,Gazeau-2009}
was employed for the first time by Glauber in his investigations 
about electromagnetic radiation.
Following the definition introduced by Glauber,
the coherent states are the eigenstates 
of the annihilation operator of the quantum harmonic oscillator. 
It can be shown that the coherent states for the harmonic oscillator 
are Gaussian wave-packet, which satisfy 
the minimization of the uncertainty principle.\cite{Gazeau-2009}

From the practical viewpoint, it is worth 
mentioning that the coherent states are more simple 
to be prepared in laboratory than the other states. 
Lasers in special conditions produce light beams of states 
sufficiently near the coherent states.\cite{Gerry-et-al-2005}
Other standard example is the superposition 
of two coherent states evolving
in different regions of the phase space, 
which gives place to the so-called ``cat state'',
also referred to as Schr\"odinger's cat.
\cite{Monroe-et-al-1996,Laghaout-2013}
From the techniques to produce the coherent states,
it is possible to generate a wide class of states: 
(i) superposition of two or more coherent states, \cite{Lutterbach-Davidovich-1997}
(ii) phase states,\cite{Aragao-et-al-2004}
(iii) addition or subtraction of photons,\cite{Zavatta-et-al-2007}
(iv) squeezed states,\cite{Walls-1983} etc.

Another important issue in quantum mechanics is 
the concept of position-dependent effective mass.
Along the last few decades, it has attracted the interest of several researchers 
due to its wide applicability: 
semiconductors,
\cite{Bastard-1975,
      vonroos_1983,
      BenDaniel-Duke-1966,
      Gora-Williams-1969,
      Zhu-Kroemer-1983,
      Li-Kuhn-1993,
      Morrow-Brownstein-1984,
      Mustafa-Mazharimousavi-2007}
nonlinear optics,\cite{Li-Guo-Jiang-Hu}
quantum liquids,\cite{Saavedra_1994}
many body theory,\cite{Bencheikh-2004}
molecular physics,\cite{Yu-Dong-Sun-2004,Christiansen-Cunha-2014}
quantum information entropy,\cite{Yanez-Navarro}
relativistic quantum mechanics,\cite{Aydogdu-Arda-Sever-2012,Merad-etal_2019}
nuclear physics,\cite{Alimohammadi-Hassanabadi-Zare-2017}
magnetic monopoles,\cite{Schmidt-2018,Jesus-2019}
nonlinear oscillations,\cite{Mathews-Lakshmanan-1974,Mathews-Lakshmanan-1975,Tiwari-2013,
Bagchi-2015,Carinena-2015,Ruby-2015,Schulze-Halberg-Roy-2016,Karthiga-et-al-2017}
semiconfined harmonic oscillator,\cite{Quesne-2022,Jafarov-2020,Jafarov-2021,Jafarov-2022}
factorization methods and supersymmetry,
\cite{Plastino-etal-1999,Bravo-PRD-2016,Amir-Iqbal-2016,Karthiga-2018,Mustafa-2020}
coherent states,\cite{Ruby-Senthilvelan-2010,Amir-Iqbal-2015,Amir-Iqbal-2016-CS,Tchoffo-2019}
etc.
The mathematical description of quantum systems 
with position-dependent mass (PDM)
is based on the non-commutativity between 
the mass and the linear momentum operators, 
which leads to the ordering problem 
for the kinetic energy operator.\cite{vonroos_1983,Mustafa-Mazharimousavi-2007}
By means of methods such as canonical point transformation,\cite{Mustafa-Mazharimousavi-2007,Bravo-PRD-2016} 
supersymmetric quantum mechanics\cite{Plastino-etal-1999} or numerical integration,
solutions of wave equations for different mass functions have been 
obtained for different potentials of interest.
Some theoretical studies have been developed with 
the aim of introducing the effect of a PDM by means 
of deformed algebraic structures.\cite{CostaFilho-Almeida-Farias-AndradeJr-2011,
      CostaFilho-Alencar-Skagerstam-AndradeJr-2013,Aguiar-2020,
      Costa-Borges-2014,
      Costa-Borges-2018,
      Costa-Silva-Gomez-2021,
      Costa-Gomez-Portesi-2020,
      Jamshir-Lari-Hassanabadi-2021}
More specifically, the effect of a PDM can be described
by a Schr\"odinger equation where the usual derivative 
is replaced by a deformed derivative operator, and 
the Hamiltonian operator is expressed in terms 
of a deformed linear momentum operator.\cite{CostaFilho-Almeida-Farias-AndradeJr-2011,
											 Costa-Borges-2018,
											 Costa-Gomez-Portesi-2020}
For instance, Costa Filho et al.\cite{CostaFilho-Almeida-Farias-AndradeJr-2011}
introduced a displacement operator
that leads to non-additive spatial translations
$\hat{\mathcal{T}}_\gamma (\varepsilon ) |x\rangle = |x + (1 + \gamma x)\varepsilon \rangle$,
in which $\gamma$ is a deformation parameter 
with inverse length dimension.
The generator of the deformed translations 
corresponds to a position-dependent linear momentum 
$\hat{p}_\gamma = -i\hbar(1+\gamma x)\frac{\textrm{d}}{\textrm{d}x}$,
and therefore a particle with PDM.
Within this approach, 
a harmonic oscillator with PDM subjected to 
a quadratic potential 
has been proposed.\cite{Costa-Borges-2018,
                        CostaFilho-Alencar-Skagerstam-AndradeJr-2013,Aguiar-2020,
                        Costa-Silva-Gomez-2021}
The deformed harmonic oscillator presents an anharmonic spectrum, 
which it can describe diatomic molecules. 
From factorization methods, the coherent states 
of the deformed oscillator have been obtained recently in 
Ref.~\onlinecite{Costa-Silva-Gomez-2021}.

This work is a continuation 
of Ref.~\onlinecite{Costa-Silva-Gomez-2021}.
However, we address here the problem of 
the deformed oscillator subject to an asymmetric potential
(which also includes the effect of PDM),
as well as properties of its coherent states.
This paper is organized as follows: 
In Sec. \ref{sec:classical-deformed-formalism},
we study the deformed classical oscillator provided 
with a PDM potential term. 
Section \ref{sec:quantum-deformed-formalism}
is devoted to the quantum treatment, 
where the eigenfunctions, the energies,
and the expected values are calculated. 
Next, in Sec. \ref{CS-deformed-oscillator},
we calculate the coherent states by means of 
the factorization method. 
The time evolution of the states
and the uncertainties are analyzed.
Finally, in Sec. \ref{sec:final-remarks},
the conclusions are outlined.

\section{\label{sec:classical-deformed-formalism}
		 Deformed classical oscillator with position-dependent mass}

Let us initially address the problem of an one-dimensional 
classical system with PDM characterized by the Lagrangian
\begin{equation}
\label{eq:Lagragian_PDM}
\mathcal{L}(x, \dot{x}) = \frac{1}{2}m(x) \left(\dot{x}^2 - \omega_0^2 x^2\right),
\end{equation}
where the factor $m(x)$ is the function mass
and $V(x) = \frac{1}{2}m(x)\omega_0^2 x^2$ corresponds
to the ``{\it quadratic}'' potential.
From Legendre transformation, the Hamiltonian associated 
with Eq.~(\ref{eq:Lagragian_PDM}) takes the form
\begin{equation}
\label{eq:hamiltonian_x_p}
 \mathcal{H}(x , p) = \dot{x}p-\mathcal{L} 
                    = \frac{p^2}{2m(x)} +  \frac{1}{2}m(x) \omega_0^2 x^2,
\end{equation}
with linear momentum $p=\partial \mathcal{L}/\partial \dot{x} = m(x) \dot{x}$.
The Euler--Lagrange equation leads to the equation of motion 
for oscillator with PDM,
\begin{equation}
\label{eq:newton-PDM}
m(x) \left( \ddot{x} + \omega_0^2 x \right)
+ \frac{1}{2} m'(x) \left( \dot{x}^2 + \omega_0^2 x^2 \right) = 0, 
\qquad [m'(x) = \textrm{d}m(x)/\textrm{d}x],
\end{equation}
which corresponds to a Li\'enard-type nonlinear oscillator
$\ddot{x} + r(x)\dot{x}^2 + s(x)=0$ with
$r(x) = \frac{1}{2} \frac{m'(x)}{m(x)}$
and
$s(x) = \omega_0^2 \left( x + \frac{1}{2} \frac{m'(x)}{m(x)} x^2 \right)$.\cite{Tiwari-2013}
For $m(x)$ not dependent on the position,
we recover the equation of motion for the standard oscillator.
Even for systems with PDM, the quantity 
$E = \frac{1}{2} m(x) (\dot{x}^2 + \omega_0^2 x^2)$ 
(total energy) is an integral of motion.

A lot of studies have been devoted to this nonlinear oscillator, 
its quantum version, as well as some of its generalizations
(see, for instance,
Refs.~\onlinecite{Mathews-Lakshmanan-1974,
                  Mathews-Lakshmanan-1975,
                  Tiwari-2013,
                  Bagchi-2015}).
According to Bravo and Plyushchay,\cite{Bravo-PRD-2016}
in a more general approach,
it is possible to transform the problem of the kinetic term with PDM 
in one dimension [Eq.~(\ref{eq:hamiltonian_x_p})] 
into the problem of a particle with constant mass in a curved space.

In the case where the mass function takes the form $m(x) = m_0/(1 + \kappa^2 x^2)$,
the Lagrangian (\ref{eq:Lagragian_PDM}) becomes the well-known 
Mathews--Lakshmanan (ML) oscillator,\cite{Mathews-Lakshmanan-1974,Mathews-Lakshmanan-1975}
\begin{equation}
\label{eq:Lagragian_ML}
\mathcal{L}_{\textrm{ML}} (x, \dot{x}) = 
\frac{m_0}{2} \left( \frac{\dot{x}^2 - \omega_0^2 x^2}{1+\kappa^2 x^2} \right).
\end{equation}
The parameter $\kappa$ (with units of inverse length) controls 
the deformation in relation to the standard oscillator.
The ML oscillator was introduced as an one-dimensional analog
of a Lagrangian density for the scalar field.
The equation of motion for system (\ref{eq:Lagragian_ML}) is
\begin{equation}
\ddot{x} -\frac{\kappa^2 x \dot{x}^2}{1+\kappa^2 x^2} 
+ \frac{\omega_0^2 x}{1+\kappa^2 x^2} = 0,
\end{equation}
and that despite its non-linear structure, 
it admits solutions in the form of a simple harmonic oscillator 
for classical bound states, 
but with amplitude depending on the frequency of oscillation.
Recently, the ML oscillator 
has been revisited in context of a $\kappa$-deformed algebraic structure
that emerges from the so called Kappa statistics.\cite{Costa-Gomez-Portesi-2020}
It is important to mention that the
ML oscillator in the standard space 
(or $\kappa$-deformed oscillator)
is equivalent to the P\"{o}sch--Teller potential 
problem in a $\kappa$-deformed space.

Within the formalism of 
the displacement operator previously introduced by Costa Filho et al.,
the problem of a harmonic oscillator with PDM
has been investigated in Refs.
\onlinecite{CostaFilho-Almeida-Farias-AndradeJr-2011,
         CostaFilho-Alencar-Skagerstam-AndradeJr-2013,Aguiar-2020,
         Costa-Borges-2018,
         Costa-Silva-Gomez-2021}.
The mass function in this approach has the form
\begin{equation}
\label{eq:m(x)}
 m(x) = \frac{m_0}{(1+\gamma x)^2},
 \qquad (x > -1/\gamma{~\textrm{and}~}\gamma >0),
\end{equation}
so the Lagrangian becomes
\begin{equation}
\label{eq:Lagragian_gamma}
\mathcal{L} (x, \dot{x}) = 
        \frac{m_0}{2} 
        \left[ \frac{\dot{x}^2 - \omega_0^2 x^2}{(1+\gamma x)^2} \right],
\end{equation}
and the corresponding Hamiltonian is
\begin{equation}
\label{eq:H(x,p)}
 \mathcal{H}(x , p) = \frac{(1+\gamma x)^2 p^2}{2m_0} 
					  + \frac{m_0\omega_0^2 x^2}{2(1+\gamma x)^2}.
\end{equation}
The potential term 
$V(x)=\frac{m_0 \omega_0^2 x^2}{2(1+\gamma x)^2}$
is semiconfined since
$\lim_{x\rightarrow -\frac{1}{\gamma}}V(x) = +\infty$
and
$\lim_{x\rightarrow +\infty}V(x) = W_\gamma$ 
with well depth
$W_\gamma = m_0\omega_0^2/2\gamma^2$
depending on the deformation parameter $\gamma$.
Other oscillators with PDM subject to semiconfined potentials have been reported in
Refs.~\onlinecite{Quesne-2022,Jafarov-2020,Jafarov-2021,Jafarov-2022}.

\begin{figure}[!htb]
\begin{minipage}[h]{\linewidth}
\includegraphics[width=0.40\linewidth]{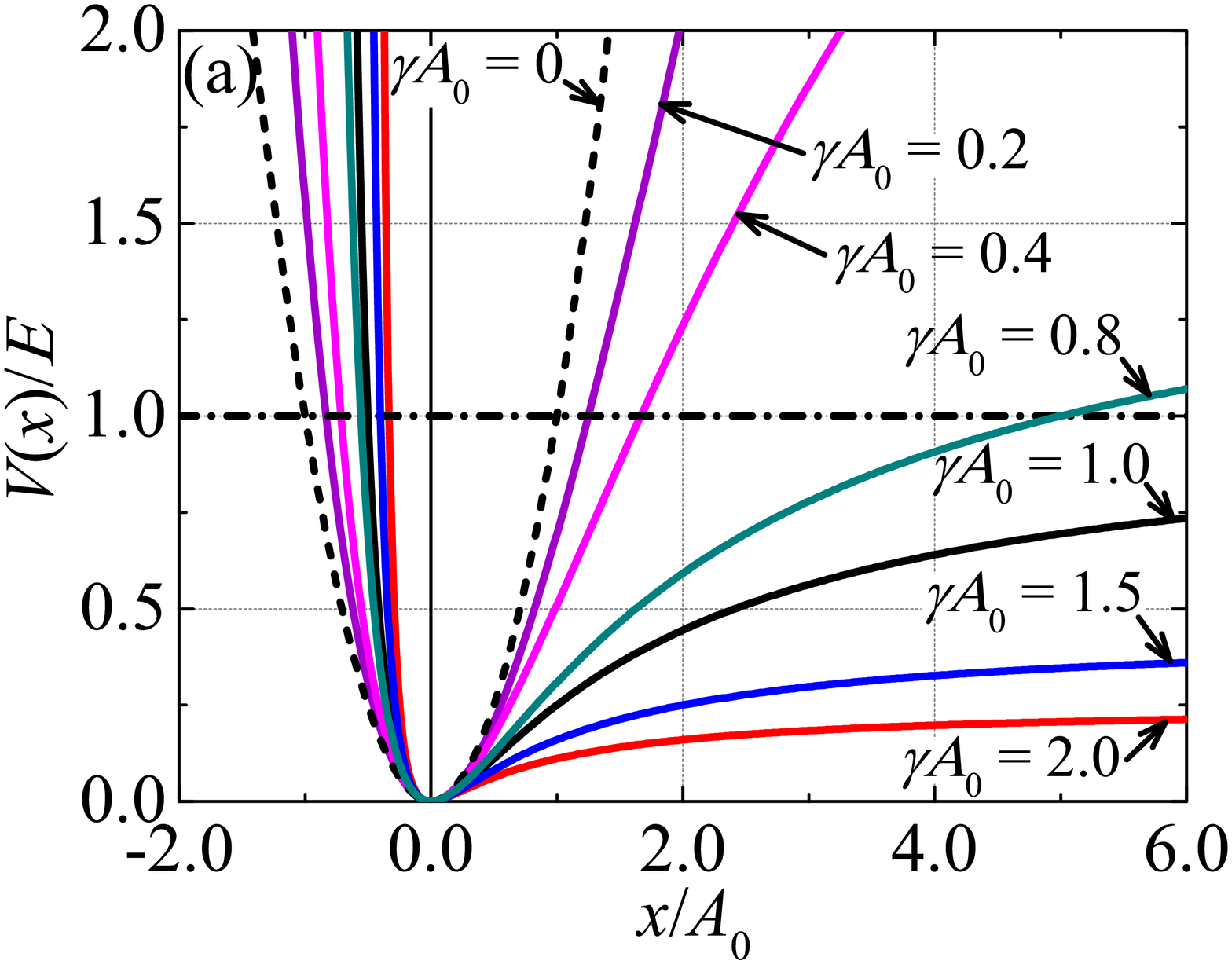}
\includegraphics[width=0.40\linewidth]{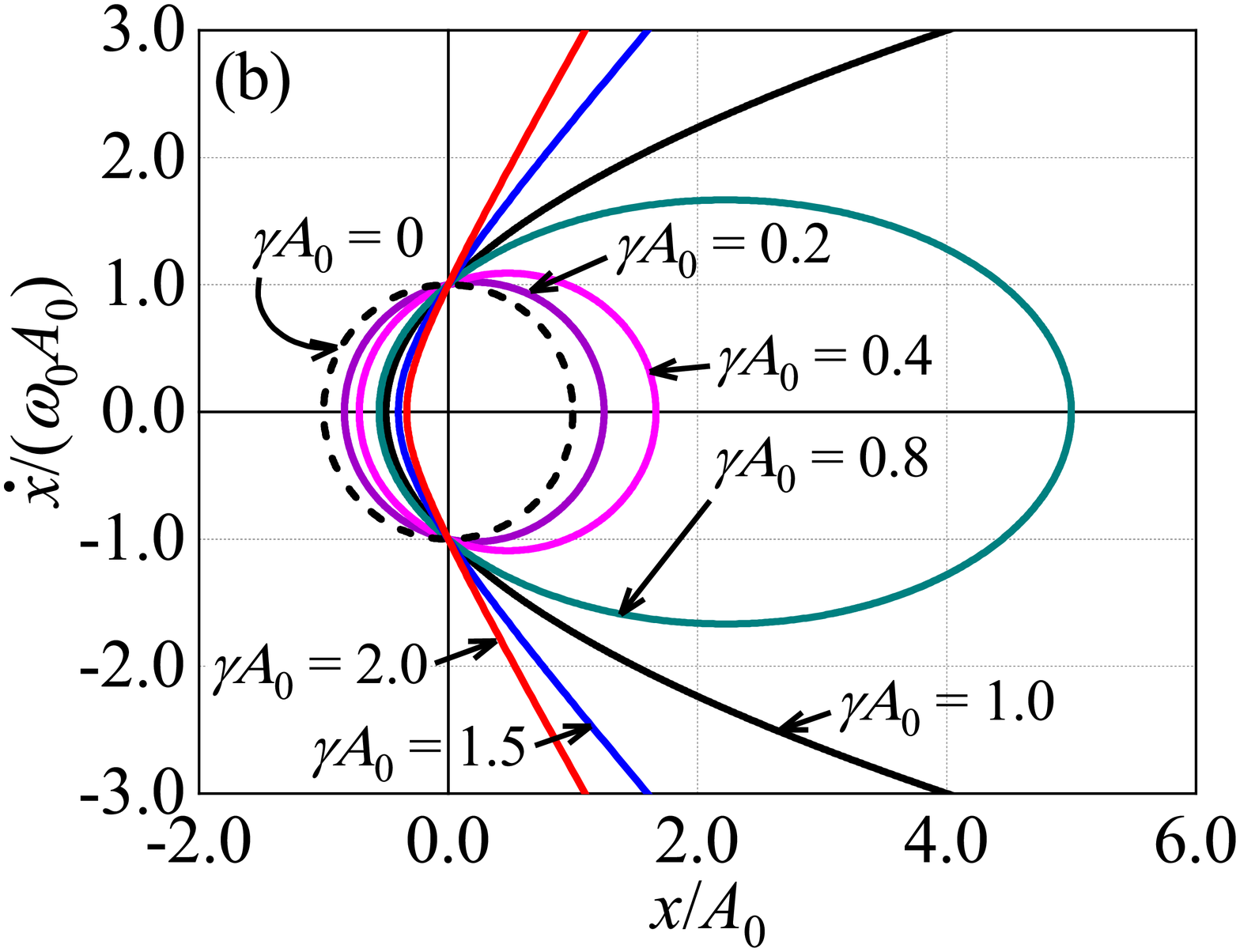}
\end{minipage}
\caption{\label{fig:1}
(a) Potentials $V(x)=\frac{1}{2}m(x) \omega_0^2 x^2$ 
for the oscillator with PDM given by Eq.~(\ref{eq:m(x)}) and
(b)
phase portraits $(x, \dot{x})$ for deformation parameters:
$\gamma A_0$ = 0 (usual case --- circular trajectory);
0.2, 0.4, and 0.8 (elliptical trajectories);
1.0 (parabolic trajectory); and
1.5 and 2.0 (hyperbolic trajectories) [Eq.~(\ref{eq:conic-sections})].
For $\gamma A_0=0$, we recover the usual oscillator trajectory
$\frac{x^2}{A_0^2} + \frac{\dot{x}^2}{A_0^2 \omega_0^2} = 1$.
For $\gamma A_0 = 1$, the trajectory is the parabola  
$\frac{\dot{x}^2}{\omega_0^2 A_0^2} = \frac{2x}{A_0} + 1$.}
\end{figure}

The motion equation is
\begin{equation}
\mathcal{D}_\gamma^2 x(t) = -\frac{\omega_0^2 x}{(1+\gamma x)^3},
\end{equation}
with $\mathcal{D}_\gamma x(t) = 
\frac{1}{1+\gamma x} \frac{\textrm{d}x}{\textrm{d}t}$
being a deformed derivative operator,
or more explicitly,
\begin{equation}
\label{eq:motion-equation-m(x)}
 \ddot{x} - \frac{\gamma \dot{x}^2}{1+\gamma x}  + \frac{\omega_0^2 x}{1+\gamma x} = 0.
\end{equation}
Considering  the energy of the oscillator
$\mathcal{H} = E = \frac{1}{2}m_0\omega_0^2 A_0^2 > 0$, 
we can write $\gamma A_0 = \sqrt{E/W_\gamma}$,
that is, the deformation parameter is related 
to the ratio between the energy and the depth of the potential well.
The phase space portrait $(x,\dot{x})$ presents 
classical bound trajectories for $0 \leq \gamma A_0 < 1$
and half-infinite trajectories for $\gamma A_0 \geq 1$  
(for simplicity, we restrict the analysis to cases with $\gamma > 0$).
The equations of the path are conic sections:
ellipse ($0 \leq \gamma A_0 < 1$ --- circle corresponds to usual case $\gamma A_0 = 0$), 
parabola ($\gamma A_0 = 1$), and hyperbola ($\gamma A_0 > 1$) and are expressed by
\begin{subequations}
\label{eq:conic-sections}
\begin{align}
\label{eq:ellipse}
\frac{(x-\gamma A_0 A_\gamma)^2}{A_\gamma^2} + \frac{\dot{x}^2}{\Omega_\gamma^2 A_\gamma^2} = 1
& \qquad (0 \leq \gamma A_0 < 1),
\\
\label{eq:parabola}
\frac{\dot{x}^2}{\omega_0^2 A_0^2} = \frac{2x}{A_0} + 1
& \qquad (\gamma A_0 = 1),
\\
\label{eq:hyperbole}
\frac{(x-\gamma A_0 A_\gamma)^2}{A_\gamma^2} - \frac{\dot{x}^2}{\Lambda_\gamma^2 A_\gamma^2} = 1
& \qquad (\gamma A_0 > 1),
\end{align}
\end{subequations}
with $\Omega_\gamma = \omega_0 \sqrt{1-\gamma^2 A_0^2}$,
$\Lambda_\gamma = i\Omega_\gamma$,
and $A_\gamma = A_0/(1-\gamma^2 A_0^2)$.
Figure~\ref{fig:1}
shows (a) the  asymmetric potential $V(x)$ and 
(b) the corresponding phase space portrait $(x,\dot{x})$ 
for different deformation parameters $\gamma A_0$.
The solution of Eq.~(\ref{eq:motion-equation-m(x)}) is
\begin{equation}
\label{eq:x_classic(t)}
x(t) = 
\left\{
\begin{array}{ll}
A_\gamma (\cos [\Omega_\gamma (t-t_0)] + \gamma A_0), & 0 \leq \gamma A_0 < 1, \\[4pt]
A_\gamma (-\cosh [\Lambda_\gamma (t-t_0)] + \gamma A_0), & \gamma A_0 \geq 1.
\end{array}
\right.
\end{equation}
Since $A_0^2 = 2E/m_0 \omega_0^2$, so   
$\Omega_\gamma$ (and $\Lambda_\gamma$)
results dependent 
on the energy of the system for $\gamma \neq 0$.
For $\gamma A_0 = 0$, we recover the standard oscillator $x(t) = A_0 \cos [\omega_0 (t-t_0)]$,
and for $\gamma A_0 = 1$, we obtain a motion under constant force
$x(t) = \frac{A_0}{2}[\omega_0^2 (t-t_0)^2 - 1]$.
Equation (\ref{eq:x_classic(t)}) can rewritten in the form
\begin{equation}
\label{eq:x_classic(t)-phase}
x(t) = \frac{A_0\cos \theta_\gamma (t)}{1-\gamma A_0 \cos \theta_\gamma (t)},
\end{equation}
with the deformed phase
\begin{equation}
\label{eq:deformed-phase}
\theta_\gamma (t) = \left\{
\begin{array}{ll}
	\displaystyle  2\,\textrm{tan}^{-1} \left\{ 
	\sqrt {\frac{1 - \gamma A_0}{1 + \gamma A_0}}\textrm{tan} \left[ 
	\frac{1}{2} \Omega_\gamma (t-t_0) \right] \right\}, & 0 \leq \gamma A_0 < 1, \\[4pt]
	\pi 
	\displaystyle - 2\,\textrm{tan}^{-1} \left\{ 
	\sqrt {\frac{\gamma A_0 +1}{\gamma A_0-1}}\textrm{tanh} \left[ 
	\frac{1}{2} \Lambda_\gamma (t-t_0) \right] \right\}, & \gamma A_0 \geq 1.
\end{array}
\right.
\end{equation}
For closed orbits ($0 \leq \gamma A_0 < 1$),
the PDM effect produces oscillations with
period $\tau_\gamma = \frac{2\pi}{\omega_0 \sqrt{1-\gamma^2 A_0^2}}$
around the equilibrium position 
$x_{\textrm{eq.}}=\gamma A_0^2/(1-\gamma^2 A_0^2)$.
In this case,
the position is confined between $x_{\textrm{min}} = -A_0/(1+\gamma A_0)$
and 
$x_{{\textrm{max}}} = A_0/(1-\gamma A_0)$.
On the other hand, for open orbits ($\gamma A_0 \geq 1$), 
the motion loses its oscillatory character 
and the particle moves in an
infinite half-line within the interval $x_{\textrm{min}} < x < \infty$.

Furthermore, the linear momentum evolves over time according to
\begin{equation}
\label{eq:p_classic(t)}
p(t) = - m_0 \omega_0  A_0\sin \theta_\gamma (t) [1-\gamma A_0 \cos \theta_\gamma (t)].
\end{equation}

In general for systems with 
position-dependent kinetic energy terms, 
Bravo and Plyushchay showed that the classical Hamiltonians 
can be conveniently turned into systems with constant mass
through the point canonical transformation
\cite{Bravo-PRD-2016}
\begin{subequations}
\label{eq:classical-dynamical-variables-Bravo}
\begin{align}
\eta & = \int^x \frac{\textrm{d}y}{\sqrt{g(y)}},\\
\Pi & = g(x) p,
\end{align}
\end{subequations}
with $g(x) = \sqrt{m_0/m(x)}$, and
$\eta$ being a deformed space and
$\Pi$ named linear pseudomomentum,
which satisfy the Poisson bracket 
$\{ \eta, \Pi \}_{x,p} = 1.$
In particular, for the mass function (\ref{eq:m(x)}),
from Eq.~(\ref{eq:classical-dynamical-variables-Bravo}),
we obtain\cite{CostaFilho-Alencar-Skagerstam-AndradeJr-2013,Costa-Borges-2018}
\begin{subequations}
\label{eq:classical-dynamical-variables}
\begin{align}
\eta & = \frac{\ln (1 + \gamma x)}{\gamma} \equiv x_\gamma, \\
\Pi & = (1+\gamma x)p \equiv \Pi_\gamma,
\end{align}
\end{subequations}
and Hamiltonian (\ref{eq:H(x,p)}) 
of a particle with PDM 
in the usual phase space $(x,p)$ is mapped into the
Hamiltonian of a Morse oscillator 
in the deformed phase space
$(x_\gamma, \Pi_\gamma)$,
\begin{equation}
\label{eq:hamiltonian_x_gamma_p_gamma}
\mathcal{K}(x_\gamma, \Pi_\gamma) = 
\frac{1}{2m_0}\Pi_\gamma^2 + W_\gamma (e^{-\gamma x_\gamma} - 1)^2,
\end{equation}
with the binding energy $W_\gamma = m_0 \omega_0^2/2\gamma^2$
and the parameter of anamorticity $\gamma$.
Therefore, the time evolution of the position $x_\gamma(t)$ and
the linear momentum $\Pi_\gamma (t)$ is
\begin{subequations}
\begin{align}
\label{eq:xgamma_classic(t)}
x_\gamma (t) &= -\frac{\ln (1 - \gamma A_0 \cos \theta_\gamma (t))}{\gamma},\\
\label{eq:Pi_classic(t)}
\Pi_\gamma (t) & = -m_0 \omega_0 A_0 \sin \theta_\gamma (t).
\end{align}
\end{subequations}
Since $\gamma^2 A_0^2 = E/W_\gamma$,
the particle presents closed path in the phase space $(x_\gamma, p_\gamma)$ 
for $E < W_\gamma$.

\begin{figure}[htb]
\begin{minipage}[h]{0.32\linewidth}
\includegraphics[width=\linewidth]{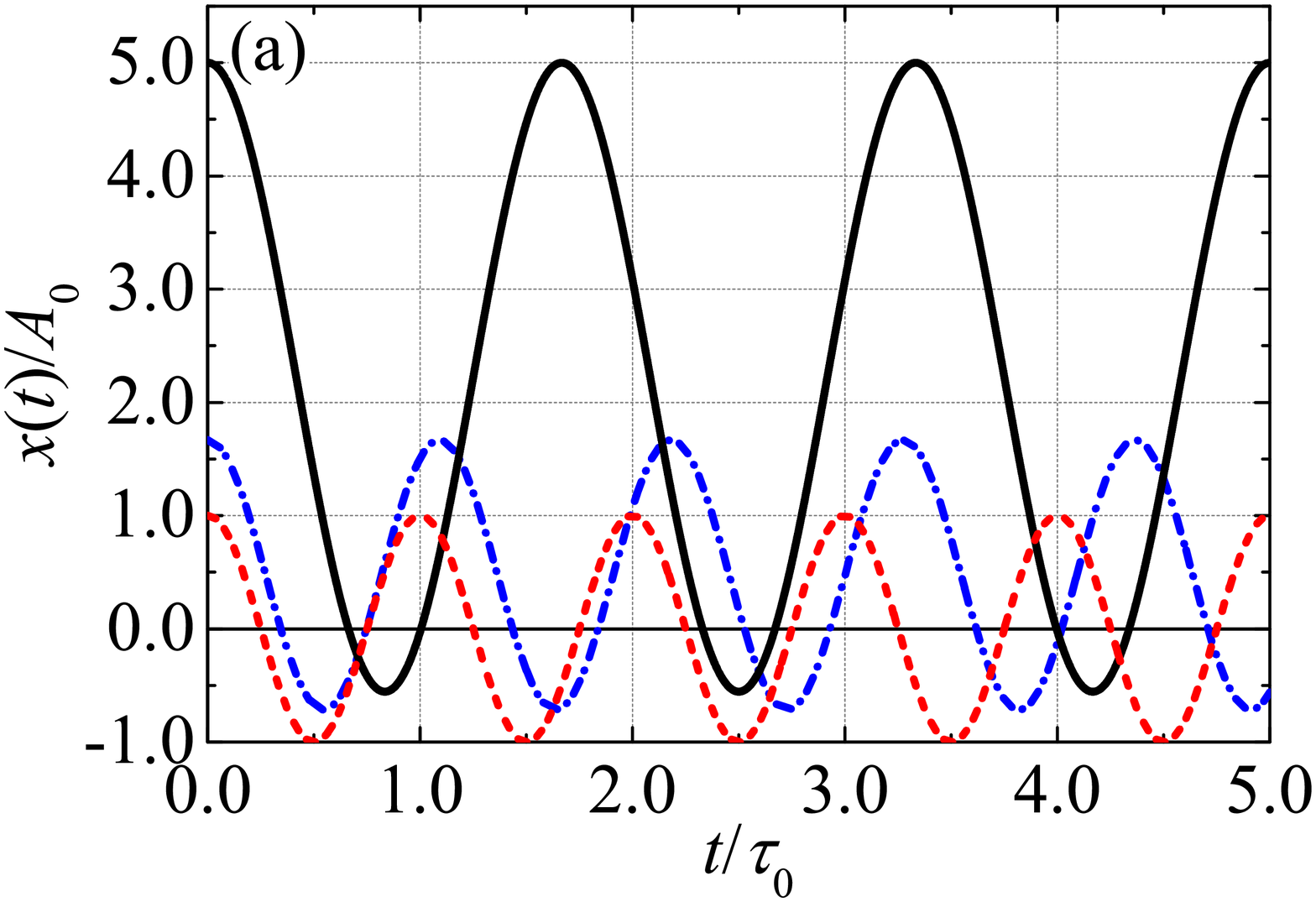}
\end{minipage}
\begin{minipage}[h]{0.32\linewidth}
\includegraphics[width=\linewidth ]{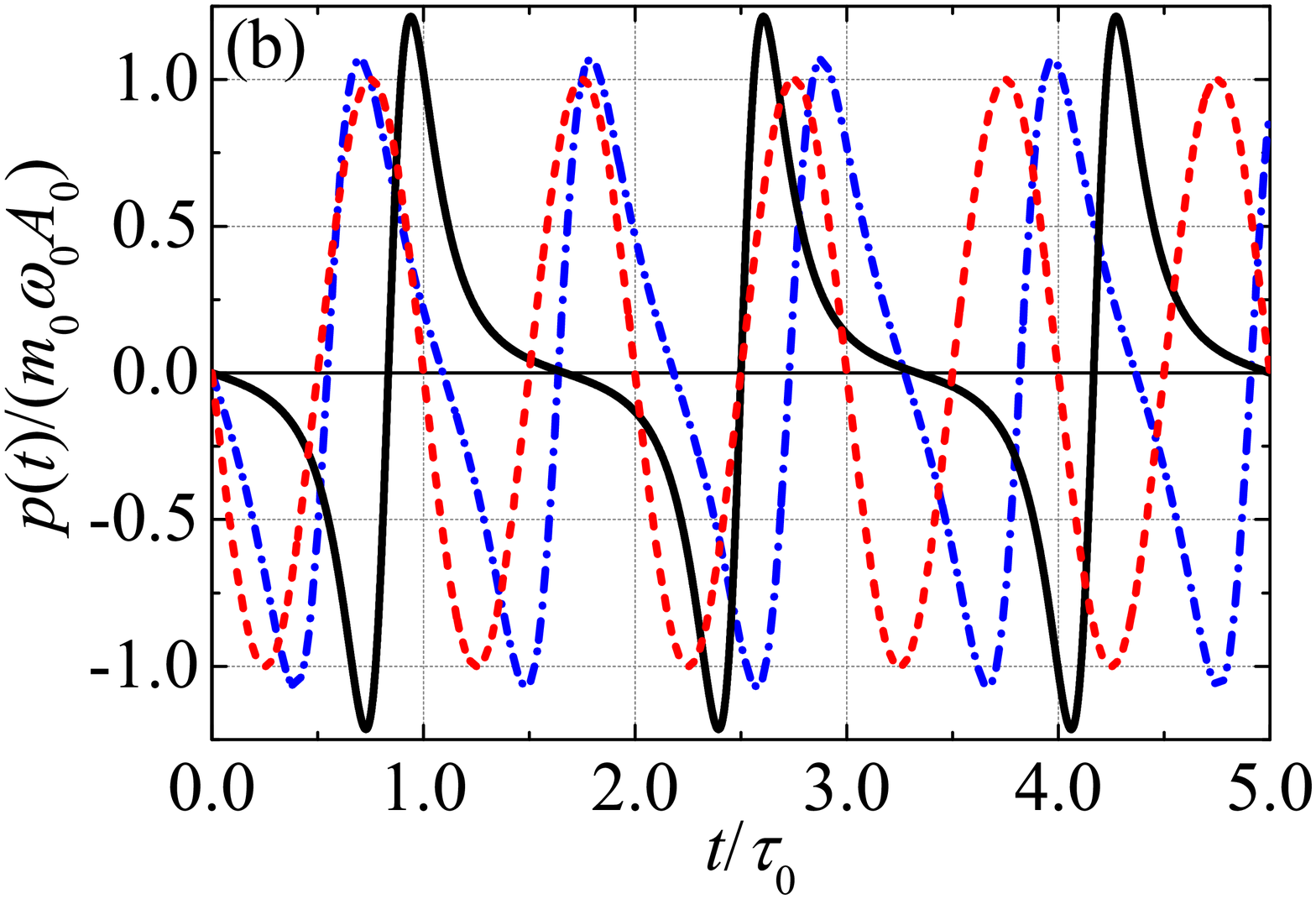}
\end{minipage}
\begin{minipage}[h]{0.32\linewidth}
\includegraphics[width=\linewidth]{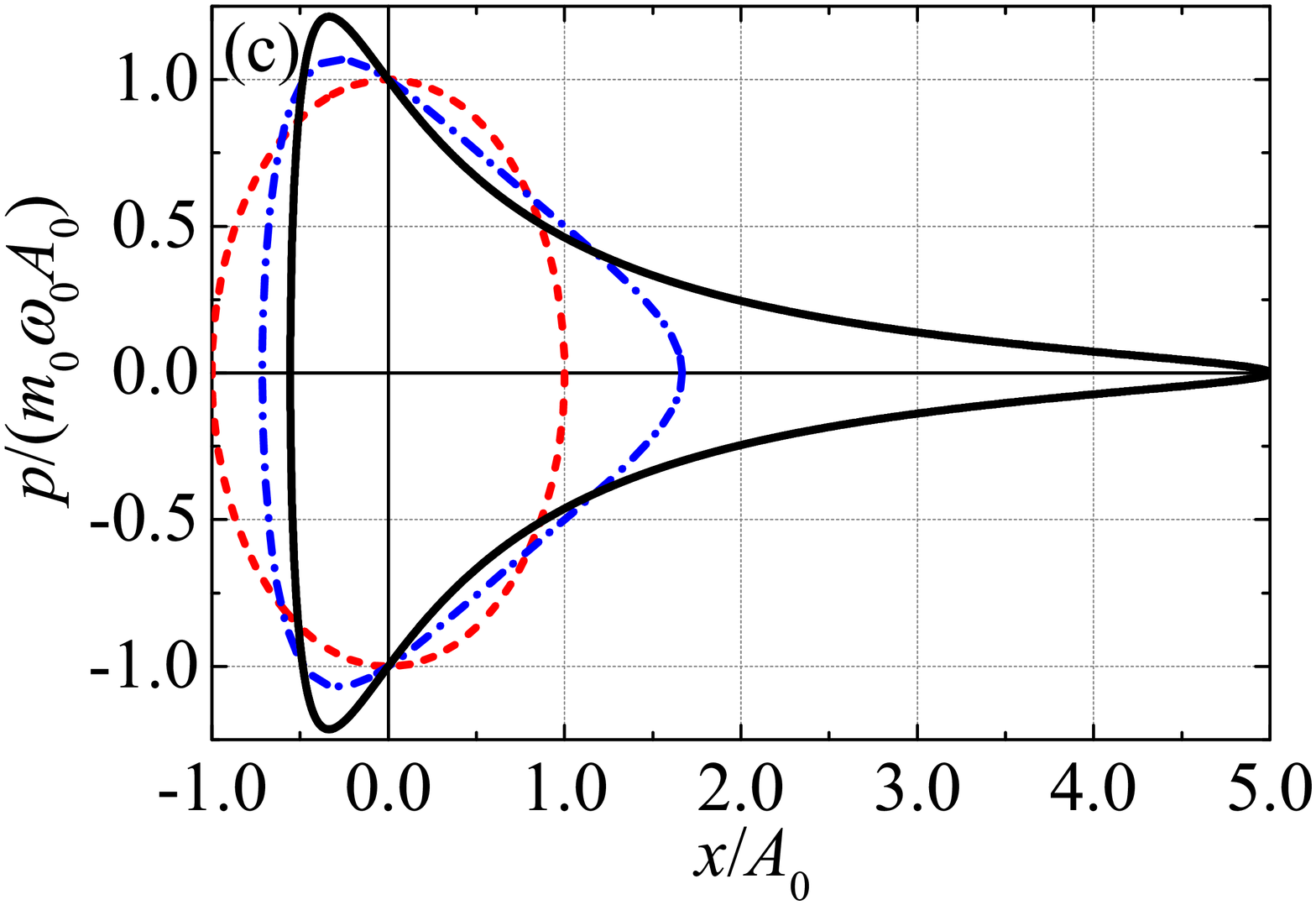}
\end{minipage}\\
\begin{minipage}[h]{0.32\linewidth}
\includegraphics[width=\linewidth]{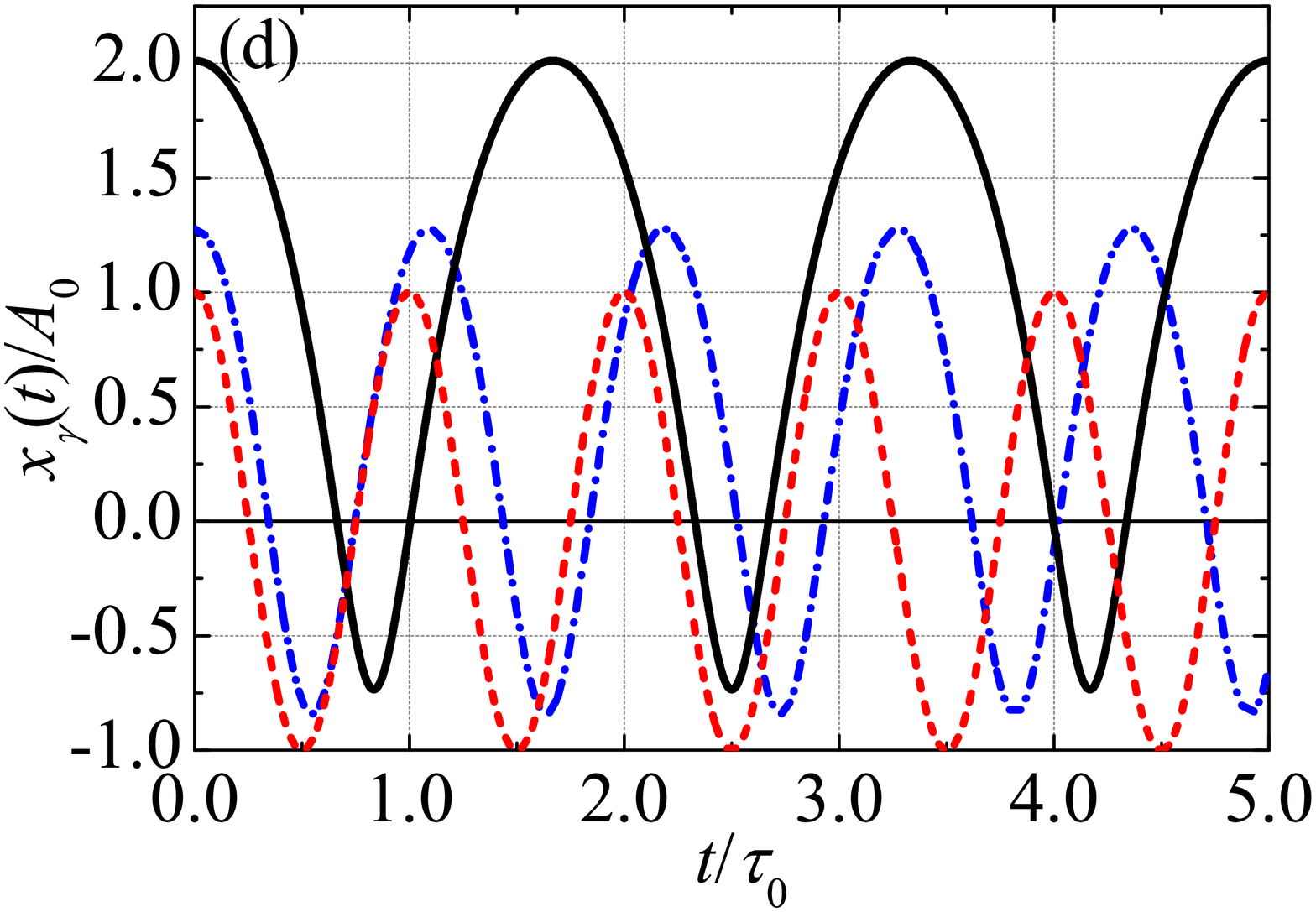}
\end{minipage}
\begin{minipage}[h]{0.32\linewidth}
\includegraphics[width=\linewidth ]{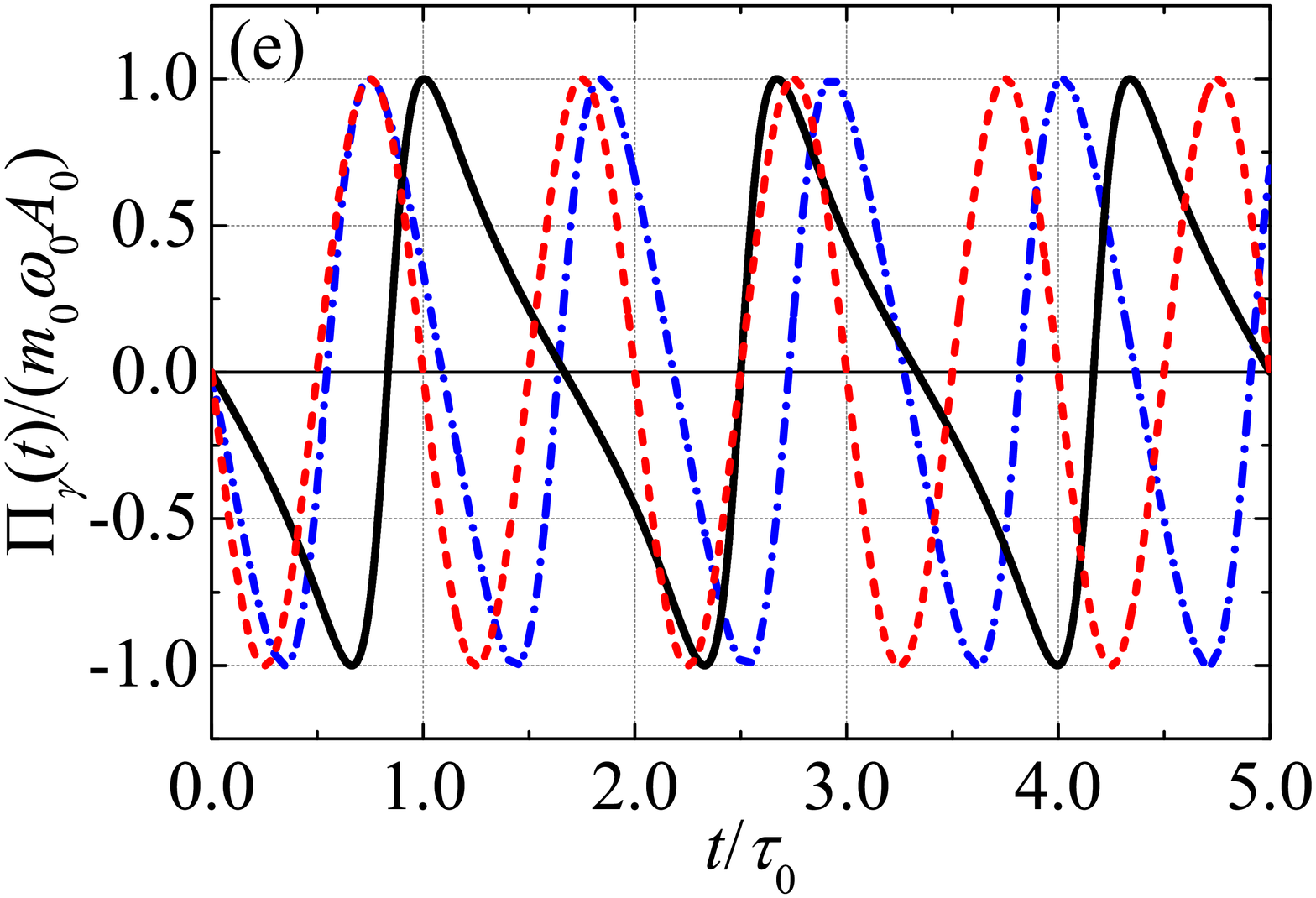}
\end{minipage}
\begin{minipage}[h]{0.32\linewidth}
\includegraphics[width=\linewidth]{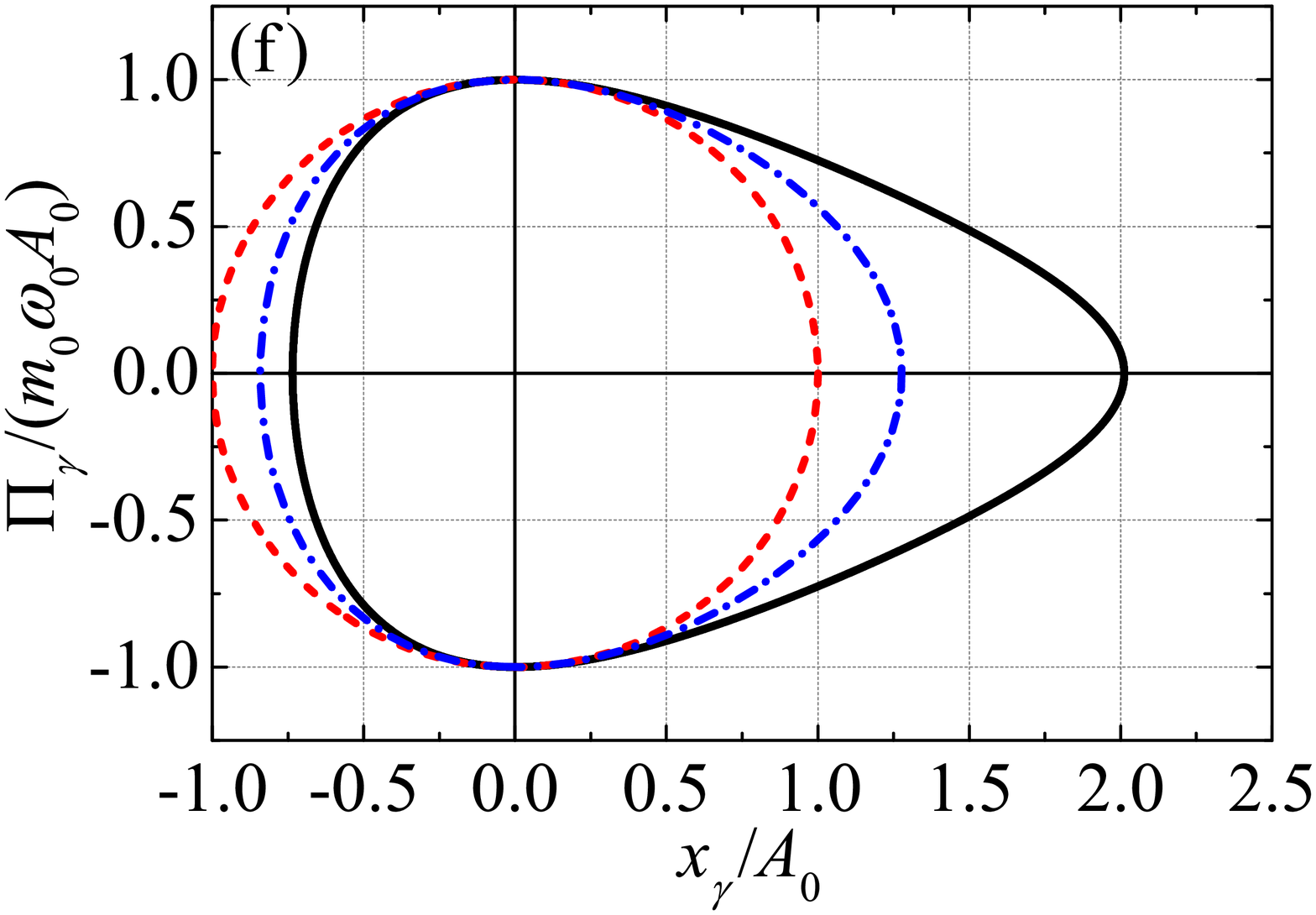}
\end{minipage}
\caption{
\label{fig:2}
Deformed oscillator dynamics with $m(x)$ 
given by Eq.~(\ref{eq:m(x)}) and potential 
$V(x)=\frac{1}{2}m(x) \omega_0^2 x^2$. 
Upper line: undeformed canonical coordinate representation --- 
time evolution of (a) the position $x(t)$ and 
(b) the linear momentum $p(t)$
and (c) paths in phase space $(x, p)$.
Bottom line: deformed canonical coordinate representation ---
time evolution of (d) the deformed position $x_\gamma(t)$ and 
(e) the linear pseudomomentum $\Pi_\gamma(t)$
and (f) paths in phase space $(x_\gamma, \Pi_\gamma)$.
The deformation parameter in graphics are 
$\gamma A_0$ = 0 (dashed red line), 
0.2 (dashed--dotted blue line), 
and 0.4 (solid black line).
}
\end{figure}

\begin{figure}[htb]
\begin{minipage}[h]{0.32\linewidth}
\includegraphics[width=\linewidth]{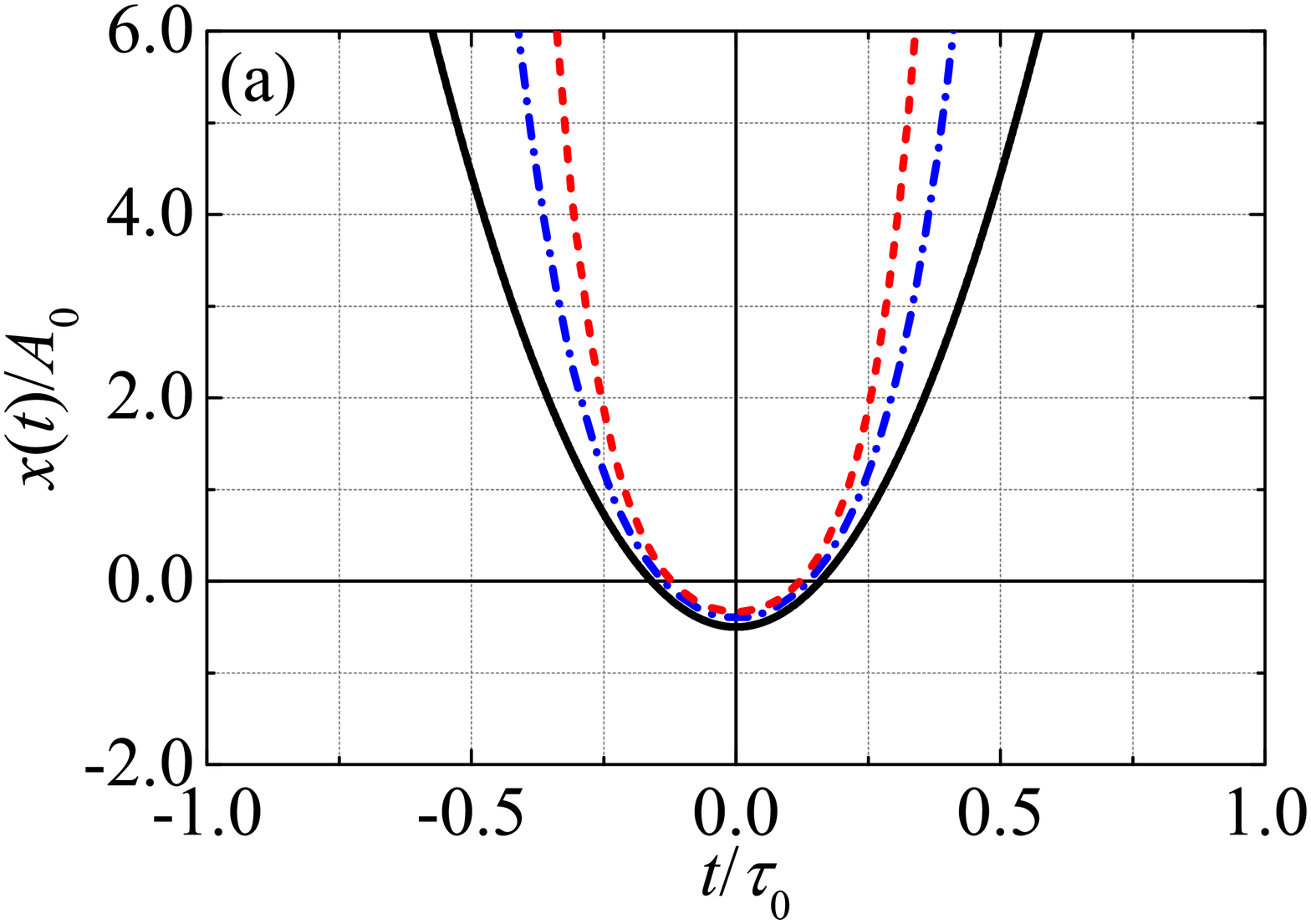}
\end{minipage}
\begin{minipage}[h]{0.32\linewidth}
\includegraphics[width=\linewidth ]{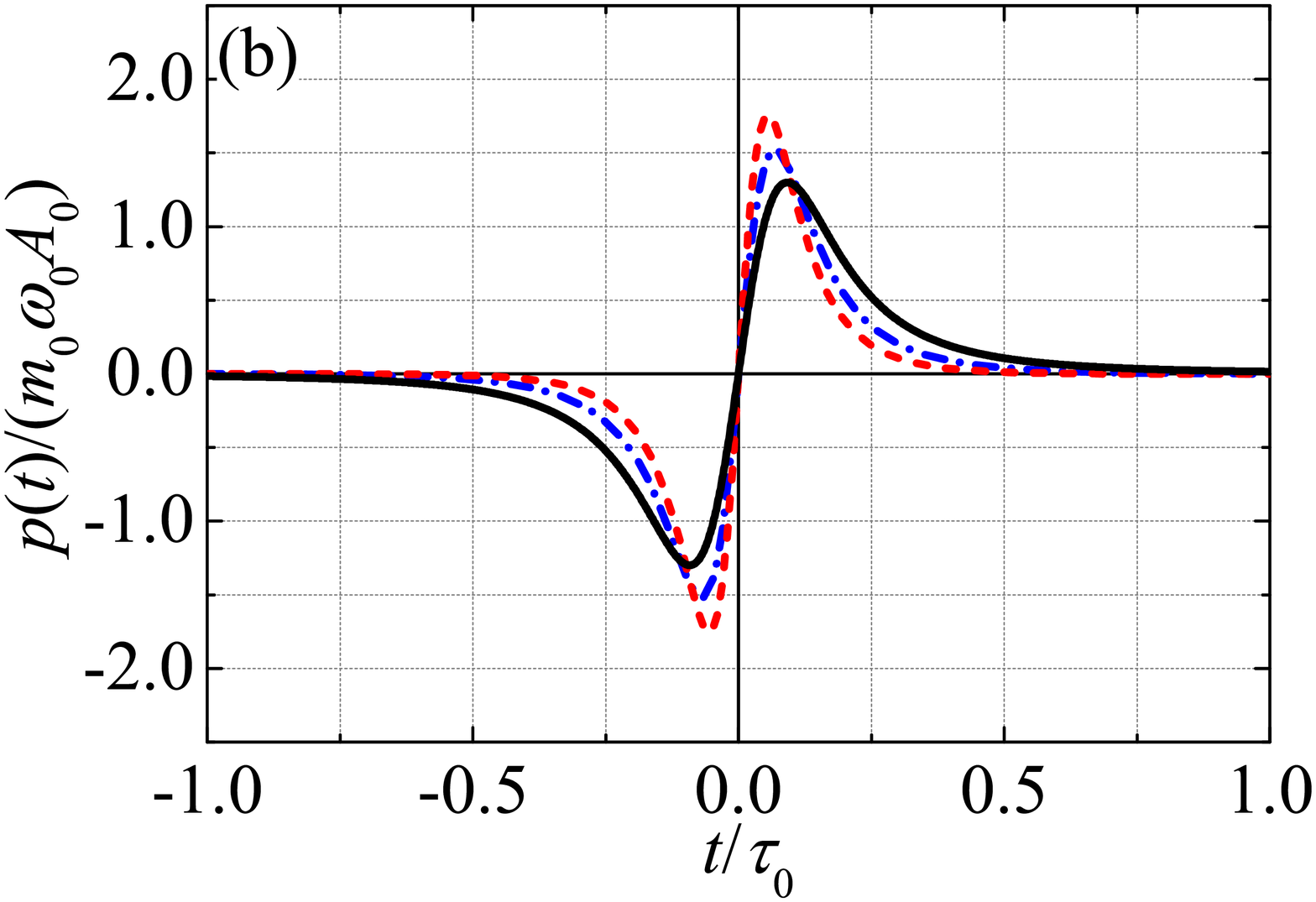}
\end{minipage}
\begin{minipage}[h]{0.32\linewidth}
\includegraphics[width=\linewidth]{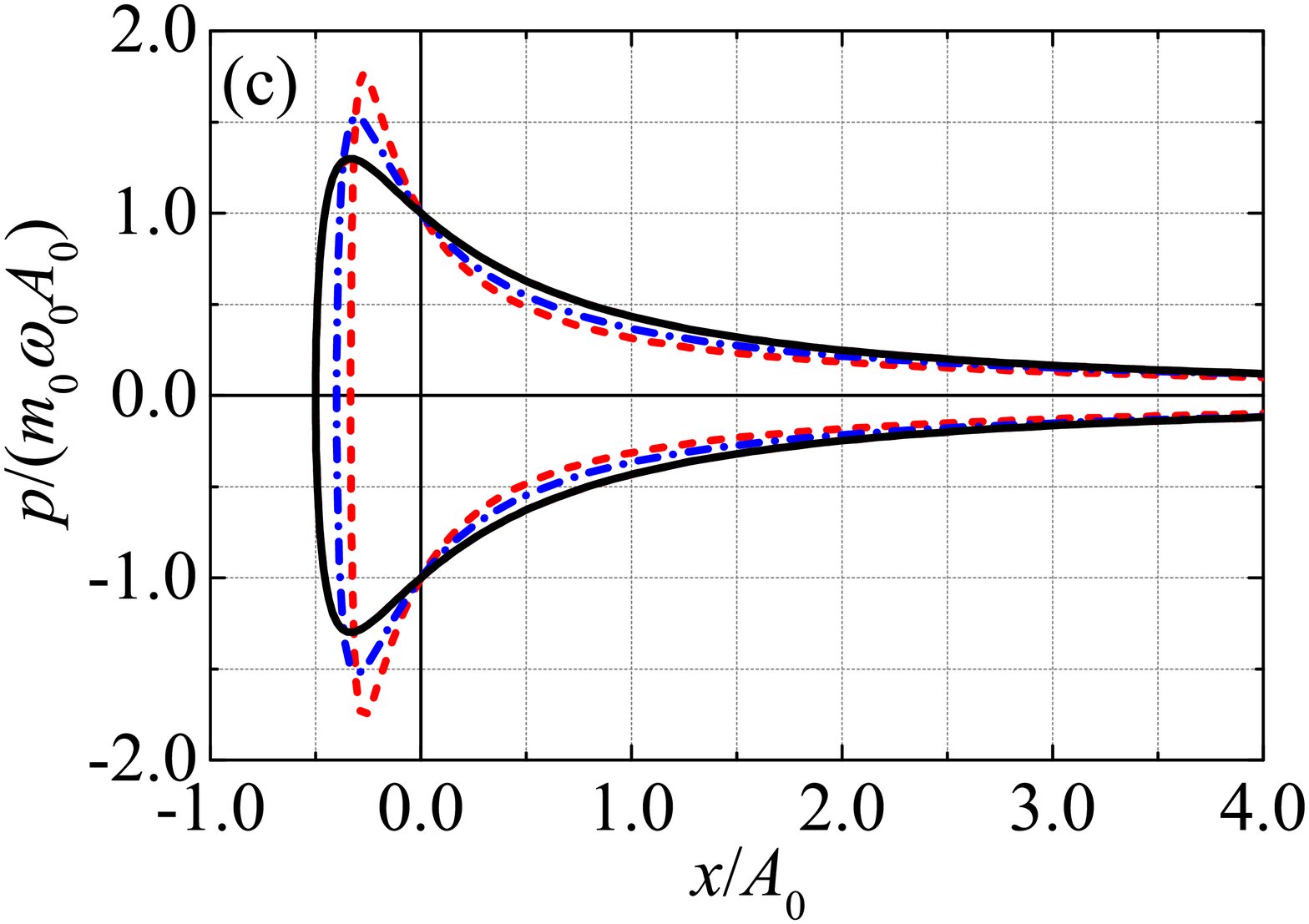}
\end{minipage}\\
\begin{minipage}[h]{0.32\linewidth}
\includegraphics[width=\linewidth]{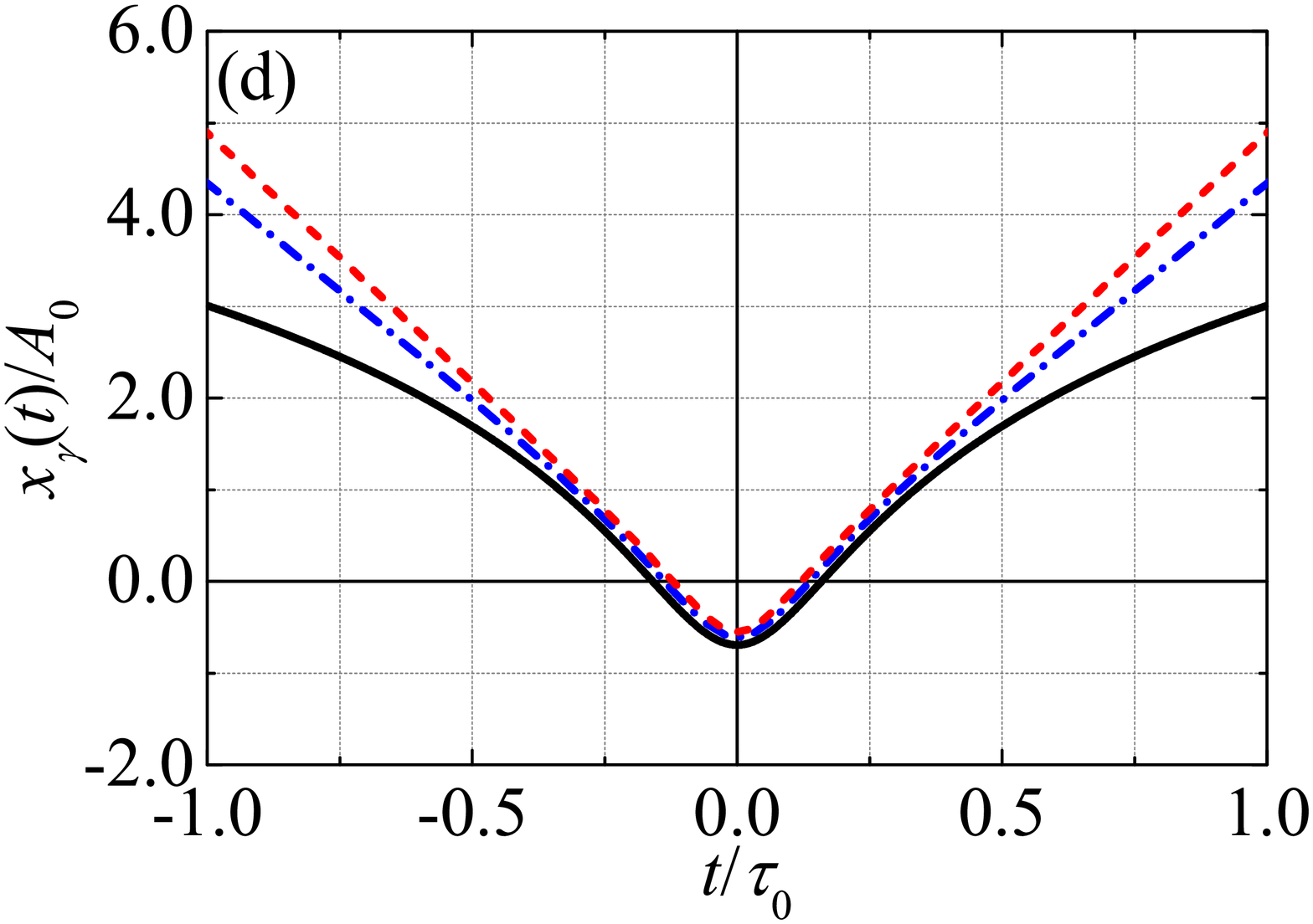}
\end{minipage}
\begin{minipage}[h]{0.32\linewidth}
\includegraphics[width=\linewidth ]{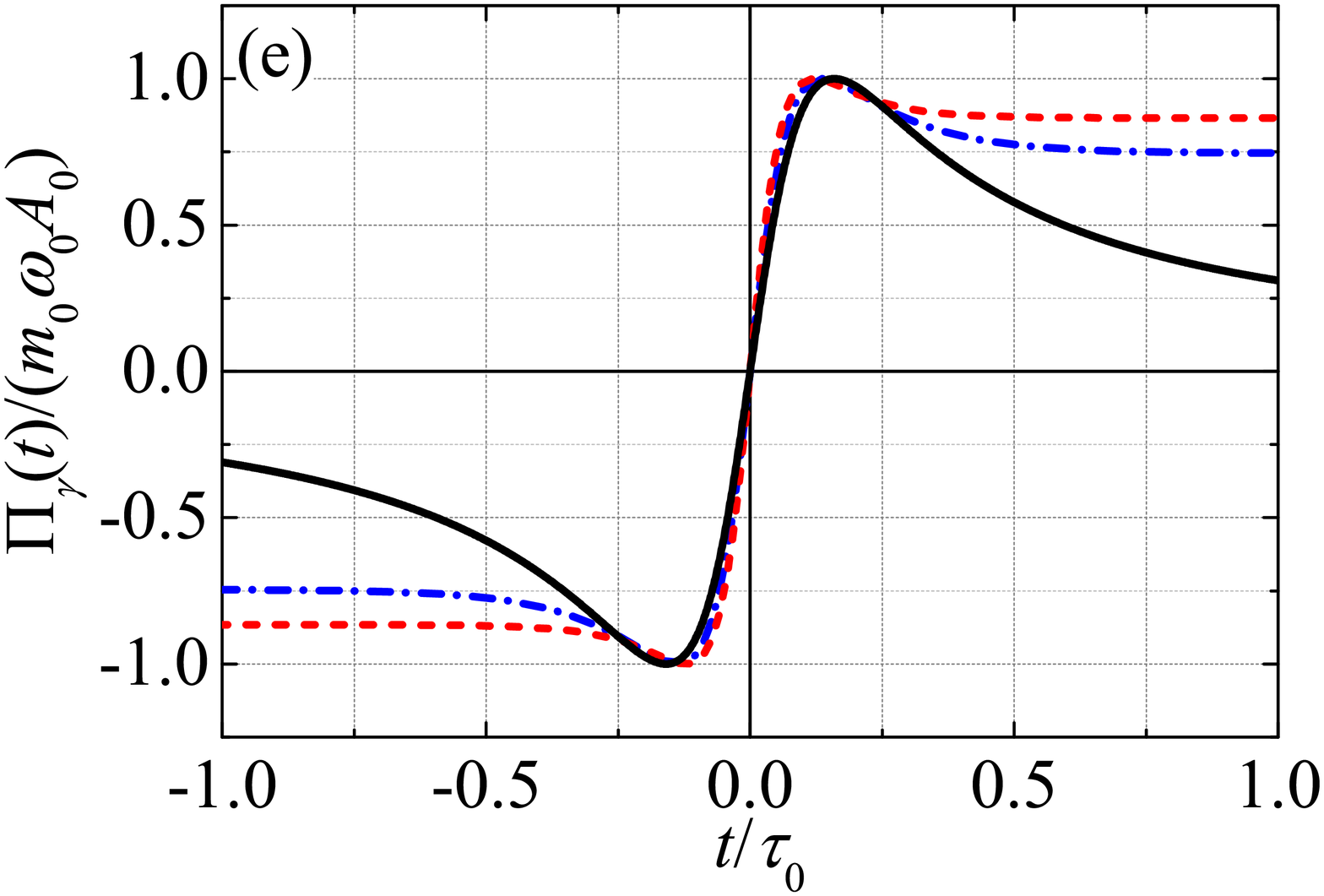}
\end{minipage}
\begin{minipage}[h]{0.32\linewidth}
\includegraphics[width=\linewidth]{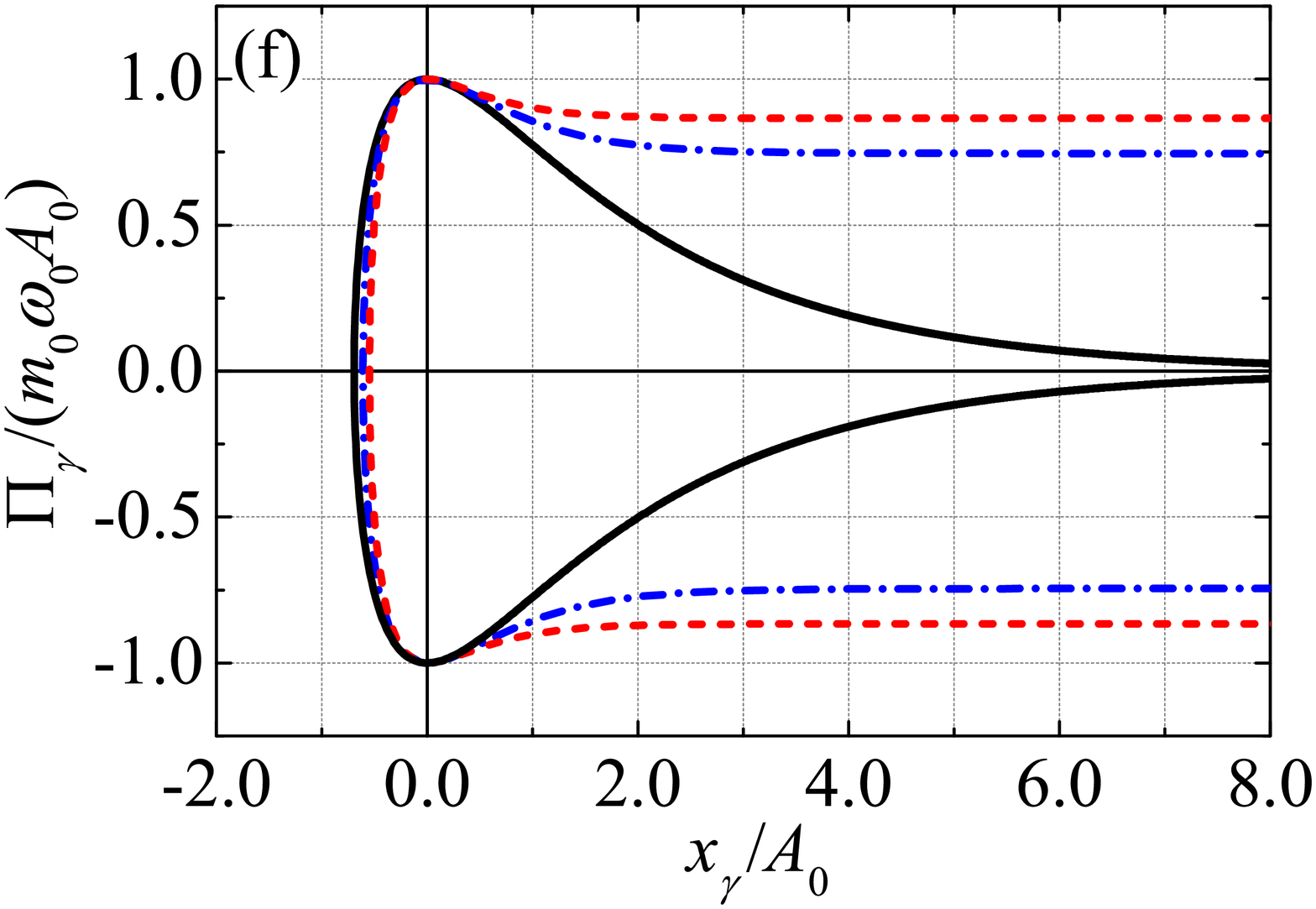}
\end{minipage}
\caption{
\label{fig:3}
Phase space for open trajectories of the
deformed oscillator dynamics with $m(x)$ 
given by Eq.~(\ref{eq:m(x)}) and potential 
$V(x)=\frac{1}{2}m(x) \omega_0^2 x^2$. 
Upper line: undeformed canonical coordinates representation --- 
time evolution of (a) the position $x(t)$ and 
(b) the linear momentum $p(t)$
and (c) paths in phase space $(x, p)$.
Bottom line: deformed canonical coordinates representation --- 
time evolution of (d) the deformed position $x_\gamma(t)$ and 
(e) the linear pseudomomentum $\Pi_\gamma(t)$
and (f) paths in phase space $(x_\gamma, \Pi_\gamma)$.
The deformation parameter in graphics are 
$\gamma A_0$ = 1.0 (solid black line),
1.5 (dashed--dotted blue line), 
and 2.0 (dashed red line).
}
\end{figure}

Figure \ref{fig:2} shows time evolution
$x(t)$ and $p(t)$ as well as $x_\gamma(t)$ and $\Pi_\gamma(t)$
for $0 \leq \gamma A_0 < 1$.
The trajectories in both $(x, p)$ and $(x_\gamma, \Pi_\gamma)$ 
phase spaces are also shown.
Figure \ref{fig:3} 
illustrates these same results, but now in the case
$\gamma A_0 > 1$.

For $0 < \gamma A_0 < 1$, the probability density for finding particle 
in the interval between $x$ and $x+\textrm{d}x$ is given by
\begin{equation}
\label{eq:rho_classic}
\rho_{\textrm{classic}} (x) = \frac{1}{\pi \sqrt{A_\gamma^2 - (x-\gamma A_0 A_\gamma)^2}},
\end{equation}
so the first moments of position and linear momentum are
\begin{subequations}
\label{eq:mean_values_x-p}
\begin{align}
\overline{x} &= \frac{\gamma A_0^2}{1-\gamma^2 A_0^2}, \\
\overline{x^2} &= \frac{A_0^2}{2}
				  \frac{1+2\gamma^2 A_0^2}{(1-\gamma^2 A_0^2)^2}, \\
\overline{p} &= 0, \\
\overline{p^2} &= \frac{1}{2} m_0^2 \omega_0^2 A_0^2
				  \sqrt{1-\gamma^2 A_0^2}. 
\end{align}
\end{subequations}
Similarly, the first moments of the linear pseudomomentum are
\begin{subequations}
\begin{align}
\overline{\, \Pi_\gamma} &=0,\\
\overline{\Pi_\gamma^2} &=m_0^2 \omega_0^2 A_0^2
		\left(
			1-\frac{1-\sqrt{1-\gamma^2 A_0^2}}{\gamma^2 A_0^2}
		\right).
\end{align}
\end{subequations}
It is straightforward to check that 
the mean value of the mass function is
$\overline{m(x)}=m_0 \sqrt{1-E/W_\gamma}$.
The mean values of the kinetic and potential 
energies satisfy the curious relation 
$
\overline{T(x)} = \overline{V(x)}\, \overline{m(x)}/ m_0,
$
with $\overline{T(x)} \geq \overline{V(x)}$
since for closed path, we have $E \leq W_\gamma$.
The standard virial theorem ($\overline{T} = \overline{V}$)
is only valid for $\gamma=0$ ($W_\gamma \rightarrow \infty$).


Alternatively, we can use the factorization method 
for obtaining the solution of the deformed oscillator.
For this purpose, we consider 
the conjugate complex variables
\begin{subequations}
\begin{align}
\label{eq:alpha_gamma}
& \alpha_\gamma (x,p) = \sqrt{\frac{m_0 \omega_0}{2\hbar}}
				\left[ \frac{x}{1+\gamma x} + \frac{i}{m_0 \omega_0} (1+\gamma x) p \right],
\\
& \alpha_\gamma^{\ast} (x,p) = \sqrt{\frac{m_0 \omega_0}{2\hbar}}
				\left[ \frac{x}{1+\gamma x} - \frac{i}{m_0 \omega_0} (1+\gamma x) p \right].
\end{align}
\end{subequations}
Hence, the classical Hamiltonian (\ref{eq:H(x,p)})
can be written as
$
\mathcal{H} = 
\hbar \omega_0 \alpha_{\gamma}^{\ast} \alpha_{\gamma},
$
with
$\alpha_\gamma$, $\alpha_\gamma^{\ast}$ and $\mathcal{H}$ 
satisfying the Poisson brackets
$\{ \alpha_\gamma, \alpha_\gamma^{\ast} \} = \frac{1}{i\hbar} \frac{1}{1+\gamma x}$,
$\{ \alpha_\gamma, \mathcal{H} \} = - \frac{i \omega_0 \alpha_\gamma}{1+\gamma x}$ 
and
$\{ \alpha_\gamma^{\ast}, \mathcal{H} \} = \frac{i \omega_0 \alpha_\gamma^{\ast}}{1+\gamma x}$,
as well as the  Jacobi identify 
$\{ \{ \alpha_\gamma, \alpha_\gamma^{\ast} \}, \mathcal{H} \} +
\{ \{ \mathcal{H}, \alpha_\gamma \}, \alpha_\gamma^{\ast} \} +
\{ \{ \alpha_\gamma^{\ast}, \mathcal{H} \}, \alpha_\gamma \} = 0.
$
In terms of dynamic variables 
$\alpha_\gamma (x,p)$ and $\alpha_\gamma^{\ast} (x,p)$,
the Poisson brackets assume the deformed algebraic structure
\begin{subequations}
\begin{align}
& \{ \alpha_\gamma, \alpha_\gamma^{\ast} \} = 
\frac{1}{i\hbar} 
\left[ 1- \frac{\gamma \sigma_0}{\sqrt{2}} \left( \alpha^{\ast}_\gamma + \alpha_\gamma \right) \right],
\\
& \{ \alpha_\gamma, \mathcal{H} \} =  
-i\omega_0 \alpha_\gamma 
\left[ 1- \frac{\gamma \sigma_0}{\sqrt{2}} \left( \alpha^{\ast}_\gamma + \alpha_\gamma \right) \right],
\\
& \{ \alpha_\gamma^{\ast}, \mathcal{H} \} = 
i\omega_0 \alpha_\gamma^{\ast} 
\left[ 1- \frac{\gamma \sigma_0}{\sqrt{2}} \left( \alpha^{\ast}_\gamma + \alpha_\gamma \right) \right].
\end{align}
\end{subequations}
Thus, the dynamical variables $\alpha_\gamma$, $\alpha_\gamma^{\ast}$ and $\mathcal{H}$ 
lead to a deformed algebraic structure
$\{\alpha_\gamma, \alpha_\gamma^{\ast}\} = \mathcal{G}_\gamma (\alpha_\gamma, \alpha_\gamma^{\ast})$,
$\{\mathcal{H} , \alpha_\gamma\}
 = i\omega_0 \mathcal{G}_\gamma(\alpha_\gamma, \alpha_\gamma^{\ast})\alpha_\gamma$
and
$\{\mathcal{H} ,\alpha_\gamma^{\ast} \}
 = -i \omega_0  \mathcal{G}_\gamma(\alpha_\gamma, \alpha_\gamma^{\ast})$
with deformation established by the function
$\mathcal{G}_\gamma(\alpha_\gamma, \alpha_\gamma^{\ast}) =
1 - \frac{\gamma \sigma_0}{\sqrt{2}} 
	(\alpha_\gamma^{\ast} + \alpha_{\gamma})$.
The standard case is recovered in the limit 
$\gamma \rightarrow 0$:
$\lim_{\gamma \rightarrow 0} i\hbar \{ \alpha_\gamma, \alpha_\gamma^{\ast} \} = 1$,
$\lim_{\gamma \rightarrow 0} \{ \mathcal{H}, \alpha_\gamma \} = i\omega_0 \alpha_0$,
and $\lim_{\gamma \rightarrow 0} \{ \mathcal{H}, \alpha_\gamma^{\ast} \} = -i\omega_0 \alpha_0^{\ast}$.

Of course, we have that the position, the linear momentum, 
and the linear pseudomomentum are, respectively,
\begin{subequations}
\begin{align}
x(t) &= 
\frac{\sqrt{2} \sigma_0 \textrm{Re} [\alpha_\gamma (t)]}{
		1-\sqrt{2} \sigma_0 \gamma \textrm{Re} [\alpha_\gamma (t)]}
\\
p(t) &= \sqrt{2} \frac{\hbar}{\sigma_0} \textrm{Im} [\alpha_\gamma (t)]
		\left\{ 1-\sqrt{2} \gamma \sigma_0 \textrm{Re} [\alpha_\gamma (t)]\right\},
\\
\Pi_\gamma (t) &= \sqrt{2} \frac{\hbar}{\sigma_0} \textrm{Im} [\alpha_\gamma (t)]
\end{align}
\end{subequations}
with 
$\sigma_0 = \sqrt{\frac{\hbar}{m_0 \omega_0}}$ 
being a characteristic length of the usual oscillator ($\gamma=0$) and
$\alpha_\gamma (t) = |\alpha_\gamma| e^{-i\theta_\gamma (t)}$.
The complex number $\alpha_\gamma$ characterizes the state 
of the deformed harmonic oscillator.

To end this section, it is important to mention that
the mass function $m(x)$ given by Eq.~(\ref{eq:m(x)}), 
and consequently the canonical transformation
(\ref{eq:classical-dynamical-variables}), 
are classical analog of the deformed space 
and linear momentum operators that emerge 
from the displacement operator method introduced 
by Costa Filho et al. 
\cite{CostaFilho-Almeida-Farias-AndradeJr-2011,
      CostaFilho-Alencar-Skagerstam-AndradeJr-2013,Aguiar-2020,
      Costa-Borges-2014,
      Costa-Borges-2018,
      Costa-Silva-Gomez-2021,
      Costa-Gomez-Portesi-2020,
      Jamshir-Lari-Hassanabadi-2021}
to describe quantum systems 
with effective mass dependent on position,
and thus, a Morse oscillator in the deformed space
$x_\gamma =\gamma^{-1}\ln (1+\gamma x)$.
According to Ref.~\onlinecite{Bravo-PRD-2016},
other different choices of the mass function 
and potential term can also lead 
to a Morse potential in a deformed space $\eta$.
For example, consider a family classical Hamiltonians written in the form
$
\mathcal{H}_\chi (x, p) = \frac{p^2}{2\mathcal{M}(x)} + \mathcal{V}(x),
$
with $\mathcal{V}(x) = \frac{1}{2} m_0 \omega_0^2 [\chi(x)]^2$
being a quadratic potential of the function $\chi(x)$
and $\mathcal{M}(x)$ being a generic mass function.
Hence, the classical Hamiltonians $\mathcal{H}_\chi (x, p)$ are mapped 
into the Morse oscillator in the deformed phase space
with constant mass $m_0$ and potential term
$U_{\textrm{Morse}}(\eta) = \frac{m_0 \omega_0^2}{2\gamma^2} (e^{-\gamma \eta} - 1)^2,$
since
\begin{subequations}
\label{eq:chi_M}
\begin{align}
\eta (x) &= -\frac{1}{\gamma} \ln [1 \mp \gamma \chi (x)],\\
\mathcal{M} (x) &= 
m_0 \left[ \frac{\chi' (x)}{1 \mp \gamma \chi (x)} \right]^2.
\end{align}
\end{subequations}
The upper signal is equivalent to 
the lower one in Eq.~(\ref{eq:chi_M}) 
through the transformation
$(\gamma, \chi, \eta) \rightarrow (-\gamma, -\chi, -\eta)$, 
which does not change the potential $\mathcal{V}(x)$ 
in the usual phase space $(x, p)$.
Table \ref{table:1} presents some
different choices for the mass functions $\mathcal{M}(x)$
and the potential terms $\mathcal{V}(x)$ that lead
to the Morse oscillator in deformed space.
Hamiltonians $\mathcal{H}_a$ and $\mathcal{H}_b$ 
have the same mass function (unless replacement
$\gamma \rightarrow -\gamma$)
but different potentials $\mathcal{V}(x)$: 
asymmetric for $\mathcal{H}_a$ and symmetric for
$\mathcal{H}_b$.
The Hamiltonian $\mathcal{H}_b$ has been investigated in
Refs.~\onlinecite{Costa-Borges-2018,CostaFilho-Alencar-Skagerstam-AndradeJr-2013}.
Hence, we choose to develop factorization methods, 
coherent states, and other topics in quantum mechanics on
the asymmetric system $\mathcal{H}_a$, 
which has not yet been discussed in the literature.
The other systems can be found in Ref.~\onlinecite{Cruz-2019}.

In addition, according Ref.~\onlinecite{Bravo-PRD-2016}, 
systems with PDM can be obtained 
from a particle with position-independent mass 
in Euclidean and Minkowski spaces.
The Lagrangian family
$\mathcal{L} =\frac{1}{2} \mathcal{M}(x)\dot{x}^2 - \mathcal{V}(x)$
for classical systems shown in table \ref{table:1} 
can be obtained by reduction from
a particle in two-dimensional hyperbolic space 
$(X, Y) = (\sinh [\eta(x)/l_0], \cosh [\eta(x)/l_0])$
[with $\eta(x)$ depending on the mass function
$\mathcal{M}(x)$ for each case].
The corresponding Lagrangian becomes
$\mathcal{L} = \frac{m_0 l_0^2}{2} ( \dot{X}^2 - \dot{Y}^2) - U(Y)$,
with $l_0$ being a characteristic length ($X$ and $Y$ are dimensionless)
and the potential term being a modified Morse potential
$U(Y)=\frac{m_0 \omega_0^2}{2\gamma^2} 
\left[ e^{-\gamma l_0 \cosh^{-1} (Y)} -1\right]^2$.
The mass function in this approach is given by 
$\mathcal{M}(x) = \frac{m_0 l_0^2}{1+X^2} \left( \frac{\textrm{d}X}{\textrm{d}x}\right)^2$. 
Similarly, in a two-dimensional Euclidean space
$(X, Y) = (\sin [\eta(x)/l_0], \cos [\eta(x)/l_0])$,
the systems in table \ref{table:1}
lead to a Lagrangian $\mathcal{L} = \frac{m_0 l_0^2}{2} ( \dot{X}^2 + \dot{Y}^2) - U(Y)$,
with modified Morse potential
$U(Y)=\frac{m_0 \omega_0^2}{2\gamma^2} 
\left[ e^{-\gamma l_0 \cos^{-1} (Y)} -1\right]^2$.
The mass function is now given by
$\mathcal{M}(x) = \frac{m_0 l_0^2}{1-X^2} \left( \frac{\textrm{d}X}{\textrm{d}x}\right)^2 $.

\begin{table}[h]
\caption{
Family of classical Hamiltonians 
$\mathcal{H}_\chi(x, p)= \frac{p^2}{2\mathcal{M}(x)} 
+ \frac{1}{2} m_0 \omega_0^2 [\chi(x)]^2$
mapped into Morse oscillator 
$\mathcal{K}_{\textrm{Morse}} (\eta, \Pi)
= \frac{1}{2m_0}\Pi^2 
+ \frac{m_0 \omega_0^2}{2\gamma^2} (e^{-\gamma \eta} - 1)^2$
by means of the canonical transformation
(\ref{eq:classical-dynamical-variables-Bravo}),
with $\chi(x)$ and $\mathcal{M}(x)$ given by
Eqs.~(\ref{eq:chi_M}).
The deformation parameter $\gamma$
has dimension inverse length.}
\small
\label{table:1}
\begin{tabular}{ccccccccc}
\toprule
\toprule
Case 
&  \quad $\chi(x)$ 
&  \quad $\mathcal{M}(x)$ 
&  \quad $\mathcal{V}(x)$ 
&  \quad $(x_1, x_2)$
&  \quad $\eta(x)$  
\\
\midrule
  \quad $\mathcal{H}_a$
&  \quad $ \frac{x}{1+\gamma x}$  	    
&  \quad $ \frac{m_0}{(1+\gamma x)^2}$ 
&  \quad $ \frac{m_0 \omega_0^2 x^2}{2(1+\gamma x)^2}$ 
&  \quad $(-\frac{1}{\gamma}, +\infty)$
&  \quad $ \frac{1}{\gamma} \ln (1+\gamma x)$  
\\
\\[5pt]
 \quad $\mathcal{H}_b$
&  \quad $x$  	    
&  \quad $ \frac{m_0}{(1-\gamma x)^2}$ 
&  \quad $ \frac{1}{2}m_0 \omega_0^2 x^2$ 
&  \quad $(-\infty, \frac{1}{\gamma})$
&  \quad $ -\frac{1}{\gamma} \ln (1-\gamma x)$  
\\[5pt]
  \quad $\mathcal{H}_c$
&  \quad $ \frac{\sqrt{1+\gamma^2 x^2} - 1 - \gamma x }{\gamma}$ 
&  \quad $ \frac{m_0}{1+\gamma^2 x^2}$  
&  \quad $ \frac{m_0 \omega_0^2}{2\gamma^2}
          \left( \sqrt{1+\gamma^2 x^2} - 1 - \gamma x \right)^2$
&  \quad $(-\infty, +\infty)$
&  \quad $ \frac{1}{\gamma} \sinh^{-1} (\gamma x)$   
\\[5pt]
  \quad $\mathcal{H}_d$
&  \quad $ \displaystyle \frac{1-e^{-\tan^{-1} (\gamma x)}}{\gamma}$ 
&  \quad $ \frac{m_0}{(1+\gamma^2 x^2)^2}$  
&  \quad $ \displaystyle \frac{m_0 \omega_0^2}{2\gamma^2}
         \left[e^{-\tan^{-1} (\gamma x)}-1 \right]^2$ 
&  \quad $(-\infty, +\infty)$
&  \quad $ \frac{1}{\gamma} \tan^{-1} (\gamma x)$   
\\
  \quad $\mathcal{H}_e$
&  \quad $ \displaystyle \frac{1-e^{-\tan^{-1} (\sinh (\gamma x))}}{\gamma}$ 
&  \quad $ m_0\, \textrm{sech}^2 (\gamma x)$  
&  \quad $ \displaystyle \frac{m_0 \omega_0^2}{2\gamma^2}
           \left[e^{-\tan^{-1} (\sinh (\gamma x))} - 1\right]^2$
&  \quad $(-\infty, +\infty)$
&  \quad $ \frac{1}{\gamma} \tan^{-1}(\sinh (\gamma x))$   
\\
  \quad $\mathcal{H}_f$
&  \quad $ \displaystyle \frac{1-e^{-\sinh^{-1} (\tan (\gamma x))}}{\gamma}$ 
&  \quad $ m_0\, \textrm{sec}^2 (\gamma x)$  
&  \quad $ \displaystyle \frac{m_0 \omega_0^2}{2\gamma^2}
           \left[e^{-\sinh^{-1} (\tan (\gamma x))} - 1\right]^2$
&  \quad $(-\frac{\pi}{2\gamma}, \frac{\pi}{2\gamma})$
&  \quad $ \frac{1}{\gamma} \sinh^{-1}(\tan (\gamma x))$   
\\
\bottomrule
\bottomrule
\end{tabular}
\end{table}

\section{\label{sec:quantum-deformed-formalism}
		 Deformed quantum oscillator with position-dependent mass}

As a consequence of the non-commutative structure 
of the mass function $m(\hat{x})$ 
and the linear momentum $\hat{p}$,
the quantization of systems with PDM leads
to the problem of ordering ambiguity in the definition 
of the kinetic energy operator.
A general form for a Hermitian kinetic energy operator
of a particle with PDM
in one-dimensional was introduced by von Roos
\cite{vonroos_1983} 
\begin{equation}
\label{eq:general-kinetic-operator-pdm}
\hat{T}(\hat{x}, \hat{p}) =
		\frac{1}{4} \left\{
			[m(\hat{x})]^{-\delta_1}\hat{p}
			[m(\hat{x})]^{-1+\delta_1+\delta_2}
			\hat{p}[m(\hat{x})]^{-\delta_2} 
		+[m(\hat{x})]^{-\delta_2}\hat{p}
		[m(\hat{x})]^{-1+\delta_1+\delta_2}
		\hat{p}[m(\hat{x})]^{-\delta_1} 
		\right\},
\end{equation}
with $\delta_1$ and $\delta_2$ named ambiguity parameters.
Several proposals for the kinetic energy operator
are particular case of (\ref{eq:general-kinetic-operator-pdm}),
such as,
Ben Daniel and Duke ($\delta_1 = \delta_2 = 0$),\cite{BenDaniel-Duke-1966} 
Gora and Williams ($\delta_1 = 1$, $\delta_2 = 0$),\cite{Gora-Williams-1969} 
Zhu and Kroemer ($\delta_1 = \delta_2 = \frac{1}{2}$),\cite{Zhu-Kroemer-1983} 
Li and Kuhn ($\delta_1 = 0, \delta_2 = \frac{1}{2}$).\cite{Li-Kuhn-1993} 
Morrow and Brownstein\cite{Morrow-Brownstein-1984} investigated 
that the case $\delta_1 = \delta_2$ obeys conditions of continuity 
of the wave function at the boundaries of 
a heterojunction in crystals.
In particular, Mustafa and Mazharimousavi 
\cite{Mustafa-Mazharimousavi-2007}
showed that the case $\delta_1  = \delta_2 = \frac{1}{4}$  
allows the mapping of a quantum Hamiltonian with PDM
into another Hamiltonian with constant mass
through a point canonical transformation,
regardless of the potential $V(\hat{x})$ 
to which the particle is subjected.
For classical Hamiltonians with kinetic terms including 
PDM, Bravo and Plyushchay\cite{Bravo-PRD-2016} 
introduced  a fictitious gauge transformation,
which leads to a more general ordering 
that includes all these particular cases pointed out
in Refs.~\onlinecite{BenDaniel-Duke-1966,
                      Gora-Williams-1969,
                      Zhu-Kroemer-1983,
                      Li-Kuhn-1993,
  					  Morrow-Brownstein-1984,
                      Mustafa-Mazharimousavi-2007}.

Considering the quantum Hamiltonian
\begin{equation}
\label{eq:hamiltonian-MM}
 \hat{H}(\hat{x}, \hat{p}) = \frac{1}{2}
	[m(\hat{x})]^{-\frac{1}{4}}
	\hat{p} [m(\hat{x})]^{-\frac{1}{2}} \hat{p}
	[m(\hat{x})]^{-\frac{1}{4}}
 	+ V(\hat{x}),
\end{equation}
the time-independent Schr\"odinger equation
$ \hat{H}| \psi \rangle = E | \psi \rangle$
in the position representation $\{ |\hat{x}\rangle \}$ is
\begin{equation}
\label{eq:SE-MM-m(x)}
 \left( -\frac{\hbar^2}{2m_0} \sqrt[4]{\frac{m_0}{m(x)}}
 \frac{\textrm{d}}{\textrm{d} x} \sqrt{\frac{m_0}{m(x)}}
 \frac{\textrm{d}}{\textrm{d}x}\sqrt[4]{\frac{m_0}{m(x)}}
 + V(x) \right) \psi (x) = E \psi (x),
\end{equation}
where $E$ is the eigenvalue corresponding to
the eigenfunction $\psi(x)$ of $\hat{H}$.

Let the linear pseudomomentum  
introduced previously by Mustafa and Mazharimousavi be
\cite{Mustafa-Mazharimousavi-2007}
\begin{equation}
\label{eq:Pi_momentum}
\hat{\Pi} \equiv \sqrt{g(\hat{x})}\,\hat{p}\,\sqrt{g(\hat{x})}
   = \frac{1}{2} (\hat{P}_g^{\dagger} + \hat{P}_g)
\end{equation}
with $g(\hat{x}) = \sqrt{m_0/m(\hat{x})}$
being a spatial metric term
and $\hat{P}_g \equiv g(\hat{x}) \hat{p}$
being a non-Hermitian linear momentum.
The linear pseudomomentum (\ref{eq:Pi_momentum}) 
has also been investigated recently
by Plyushchay\cite{Plyushchay-2017}
from the anomaly-free prescription so that the structure is
used to get quantization of particles in curved spaces. 
Hamiltonian (\ref{eq:hamiltonian-MM}) can be written 
in the simplified form
$\hat{H} (\hat{x}, \hat{p}) 
= \frac{1}{2m_0}\hat{\Pi}^2 (\hat{x}, \hat{p})
+ V(\hat{x}).
$
Note that the linear pseudomomentum $\hat{\Pi}$ is 
the physical quantity conserved for a free particle with PDM
($V(\hat{x})=0$).
Moreover, $\hat{x}$ and $\hat{\Pi}$ 
satisfy the commutation relation
$[\hat{x}, \hat{\Pi}] = i\hbar g(\hat{x})$,
so this algebraic structure results in
a generalized uncertainty principle,
\begin{equation}
\label{eq:GUP}
\Delta x \Delta \Pi \geq 
	\frac{\hbar}{2}
	| \langle g(\hat{x}) \rangle |.
\end{equation}
Alternatively, we can rewrite Eq.~(\ref{eq:SE-MM-m(x)}) 
by means of the transformation
$\varphi (x) = \sqrt{g(x)} \psi (x)$,
such that
\begin{equation}
\label{eq:m(x)-SE}
\left[ 
-\frac{\hbar^2}{2m_0} \left( g(x)
\frac{\textrm{d}}{\textrm{d} x} \right)^2 
+V(x)\right] \varphi (x) = E \varphi (x),
\end{equation}
which is associated with the Hamiltonian operator
$
\hat{H}_g (\hat{x}, \hat{P}_g) = \frac{1}{2m_0}\hat{P}_g^2 + V(\hat{x}).
$
From deformed space 
$\hat{\eta} \equiv 
\int^{\hat{x}} [g(y)]^{-1}\textrm{d}y = 
\int^{\hat{x}} \sqrt{m(y)/m_0} \textrm{d}y$,
we obtain the standard commutation relation 
$[\hat{\eta}, \hat{\Pi}] = i\hbar\hat{1}$.
Therefore, we have that
$(\hat{x}, \hat{p}) \rightarrow (\hat{\eta}, \hat{\Pi})$
constitutes a point canonical transformation
that maps the Hamiltonian (\ref{eq:hamiltonian-MM}) with PDM
into another Hamiltonian with constant mass
$\hat{K} (\hat{\eta}, \hat{\Pi}) = \frac{1}{2m_0}\hat{\Pi}^2 + U(\hat{\eta}),$
whose potential in the deformed space is
$U(\hat{\eta}) = V(\hat{x}(\hat{\eta}))$.
The corresponding Schr\"odinger equation in the deformed space
is denoted by $\hat{K}\phi( \eta) = E\phi(\eta)$,
with $\phi(\eta) = \varphi (x(\eta))$ being
the wave function in representation $\{|\hat{\eta}\rangle\}$.

For the mass function (\ref{eq:m(x)}), 
the spatial metric term is 
$g(\hat{x}) = \hat{1} + \gamma \hat{x}$,
and the deformed space and linear pseudomomentum
are given, respectively, by
\begin{subequations}
\label{eq:hat-x_k-Pi_k}
\begin{align}
\hat{x}_\gamma & =
\frac{\ln (1+\gamma \hat{x})}{\gamma}, \\
\hat{\Pi}_\gamma & = 
     (\hat{1}+\gamma \hat{x})^{1/2}\hat{p}
     (\hat{1}+\gamma \hat{x})^{1/2}
     = \frac{1}{2}(\hat{p}_\gamma^{\dagger} + \hat{p}_\gamma),
\end{align}
\end{subequations}
with $\hat{p}_\gamma = -i\hbar D_\gamma$ and 
$D_\gamma \varphi(x) = 
(1+\gamma x)\frac{\textrm{d}\varphi(x)}{\textrm{d}x}$
being a deformed derivative [dual to 
$\mathcal{D}_\gamma x(t) = 
\frac{1}{1+\gamma x} \frac{\textrm{d}x}{\textrm{d}t}$ ---
see Ref.~\onlinecite{Costa-Borges-2018} for more details].

Considering the potential 
$V(x) = \frac{1}{2} m(x) \omega_0^2 x^2$
with mass function~(\ref{eq:m(x)}), 
one can write Eq.~(\ref{eq:m(x)-SE}) 
as a deformed Schr\"odinger equation for the problem of 
the asymmetric oscillator with PDM 
\cite{CostaFilho-Almeida-Farias-AndradeJr-2011,Costa-Borges-2018}
\begin{equation}
\label{eq:deformed-schrodinger-equation-and-derivative}
-\frac{\hbar^2}{2m_0} D_{\gamma}^2 \varphi (x)
+ \frac{m_0 \omega_0^2 x^2}{2(1+\gamma x)^2}  
 \varphi (x) =  E \varphi (x).
\end{equation}
In the deformed space $x_\gamma$,
Eq.~(\ref{eq:deformed-schrodinger-equation-and-derivative})
can be expressed in terms of the new field 
$\phi(x_\gamma) = \varphi(x(x_\gamma))$,
and it becomes the Schr\"odinger equation 
for the quantum Morse oscillator,\cite{CostaFilho-Alencar-Skagerstam-AndradeJr-2013,Costa-Borges-2018}
\begin{equation}
\label{eq:SE-QMO}
-\frac{\hbar^2}{2m_0} \frac{\textrm{d}^2{\phi}(x_\gamma)}{\textrm{d}x_\gamma^2}
+W_\gamma (e^{-\gamma x_\gamma}-1)^2 \phi ({x_\gamma})
= E \phi ({x_\gamma}).
\end{equation}
The eigenfunctions of Eq.~(\ref{eq:SE-QMO})
are
\begin{equation}
\phi_n (x_{\gamma}) = 
	(-1)^n \mathcal{N}_n e^{-\frac{1}{2} \xi (x_\gamma)}
	[\xi (x_\gamma)]^{\frac{\nu_n}{2}}
	L_n^{(\nu_{n})} (\xi(x_\gamma )),	
\end{equation}
where
$\xi (x_\gamma) = 2se^{-\gamma x_\gamma},$ 
$s = \frac{1}{\gamma^2 \sigma_0^2},$
$\nu_n=2s - 2n - 1>0,$
$\mathcal{N}_n^2 = \nu_n \gamma \Gamma(n+1)/\Gamma(\nu_n+n+1),$
and
$L_n^{(\nu)}(z)$ are the associated Laguerre polynomials.
From the solutions of (\ref{eq:SE-QMO}), 
we obtain the normalized eigenfunctions 
of the deformed oscillator,
\begin{align}
\label{eq:egeinfucntions-osc}
\psi_n (x)  &= \frac{\varphi_n (x)}{\sqrt{1+\gamma x}} 
			=(-1)^n  \frac{\mathcal{N}_n}{\sqrt{2s}} e^{-\frac{\zeta(x)}{2}} 
	          [\zeta (x)]^{\frac{\nu_n+1}{2}} L_n^{(\nu_{n})} (\zeta(x))
\end{align}
for $x > -1/\gamma$ and $\psi_n (x)=0$  otherwise,
where $\zeta(x) = \frac{2s}{1+\gamma x}$.
The energy eigenvalues are
\begin{equation}
\label{eq:E_n}
E_n({\gamma}) = \hbar \omega_0 \left( n+\frac{1}{2} \right)
	- \frac{\hbar^2 \gamma^2}{2m_0} \left( n+\frac{1}{2} \right)^2.
\end{equation}
Curiously, this energy spectrum is identical to 
the case of the quadratic potential $V_0 (x) = \frac{1}{2}m_0 \omega_0^2 x^2$
previously studied in Ref.~\onlinecite{Costa-Silva-Gomez-2021}.
From the condition $E_n < W_\gamma$, 
the number of bound states is restricted by the condition
$0 \leq n + \frac{1}{2} < \frac{m_0 \omega_0}{\hbar \gamma^2}$
(i.e., $\nu_n > 0$).

In particular, the eigenfunction for $n=0$ is
\begin{equation}
\label{eq:ground-state}
\psi_0 (x) 	= 
     \sqrt{\frac{\gamma}{2s \Gamma (2s-1)}}
	  e^{-\frac{\zeta(x)}{2}} [\zeta(x)]^{s}.
\end{equation}
The probability density 
$\rho_0 (x) = |\psi_0(x)|^2$
behaves like a Gamma distribution,
\begin{equation}
\label{eq:Gamma_distribuition}
\rho_0 (x) = 
	\frac{\gamma}{2s} \frac{1}{\Gamma (\lambda)}
	e^{-\zeta(x)} [\zeta(x)]^{\lambda+1},
\end{equation}
with $\lambda = \frac{2}{\gamma^2 \sigma_0^2} -1$ 
being the shape parameter.
The ground state energy is
$
E_0 ({\gamma}) = \frac{\hbar \omega_0}{2} 
                 - \frac{\hbar^2 \gamma^2}{8m_0}.
$

For $E> W_\gamma$, the
eigenstates become unbound and the eigenfunctions 
are the non-normalized wavefunction
\begin{equation}
\psi (x) = 
\frac{e^{-\frac{\zeta(x)}{2}} }{\sqrt{2s}} 
	          [\zeta (x)]^{w+  \frac{1}{2}} 
\left[C_1 \,L_{s-1-w}^{(2w)} (\zeta (x))+
C_2\,{_{1}}F_1 \left( 1 + w -s; 1 + 2w; \zeta (x) \right)
\right]\!,
\end{equation}
where $_{1}F_1( \,\cdot\, ; \,\cdot\, ; \,\cdot\, )$ represent 
the confluent hypergeometric functions of the first kind,
$C_{1}$ and $C_2$ are constants, and
$w^2 = 2m_0(E-W_\gamma)/\hbar^2 \gamma^2$.
For the sake of simplicity, we will continue in this work discussing 
the properties of the bound states for the deformed oscillator.

\begin{figure}[!htb]
\centering
\begin{minipage}[h]{0.32\linewidth}
\includegraphics[width=\linewidth]{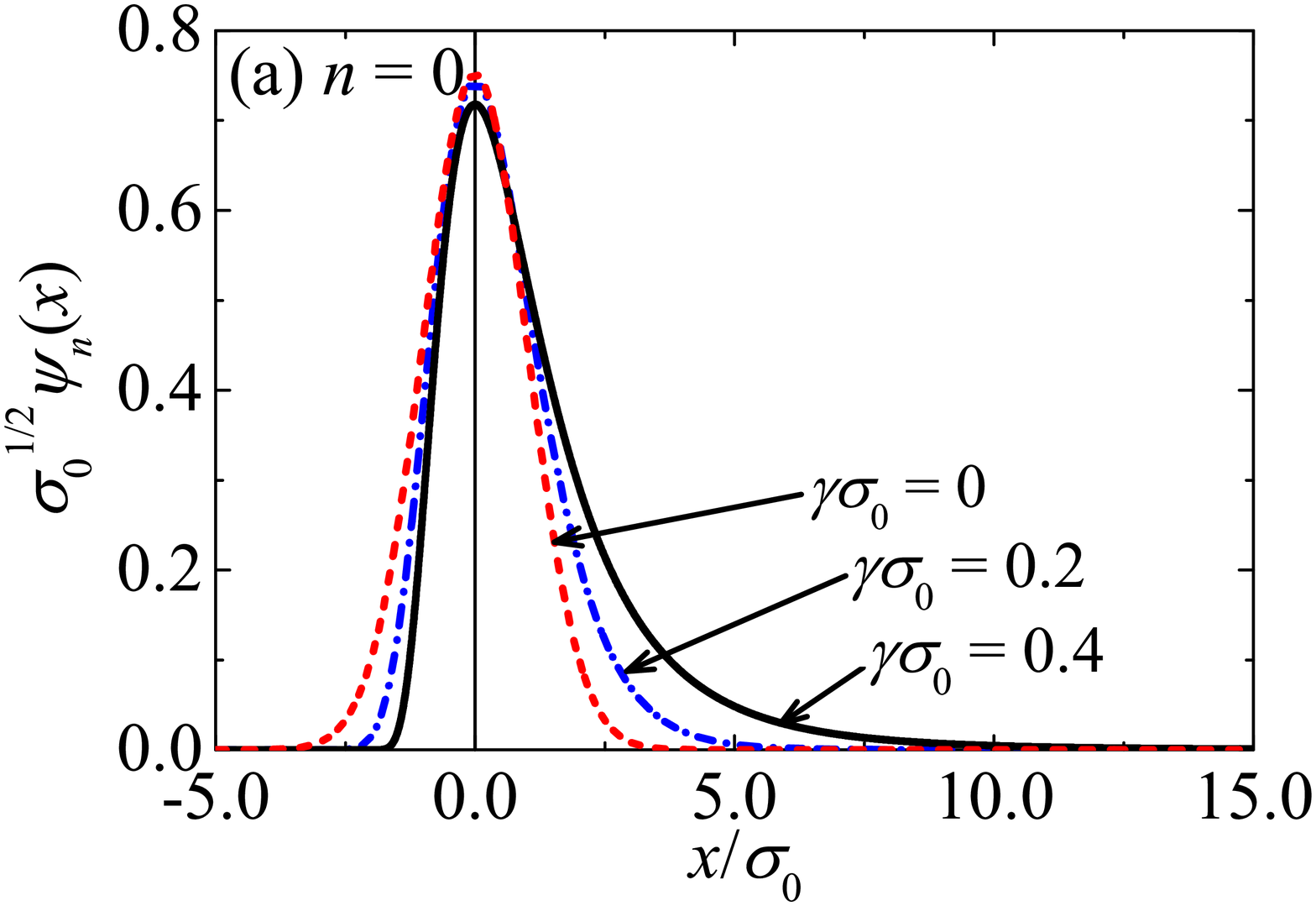}
\end{minipage}
\begin{minipage}[h]{0.32\linewidth}
\includegraphics[width=\linewidth ]{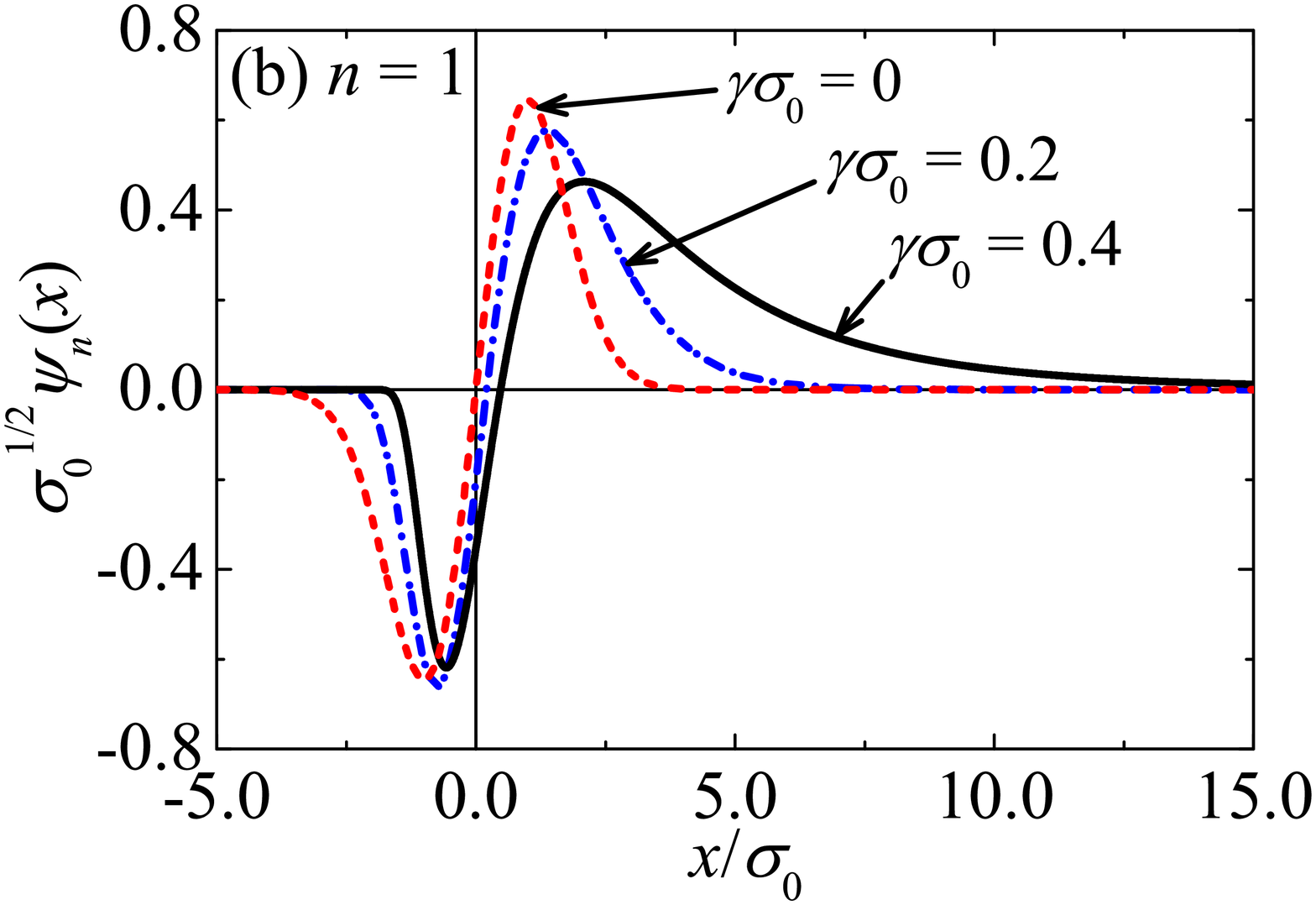}
\end{minipage}
\begin{minipage}[h]{0.32\linewidth}
\includegraphics[width=\linewidth]{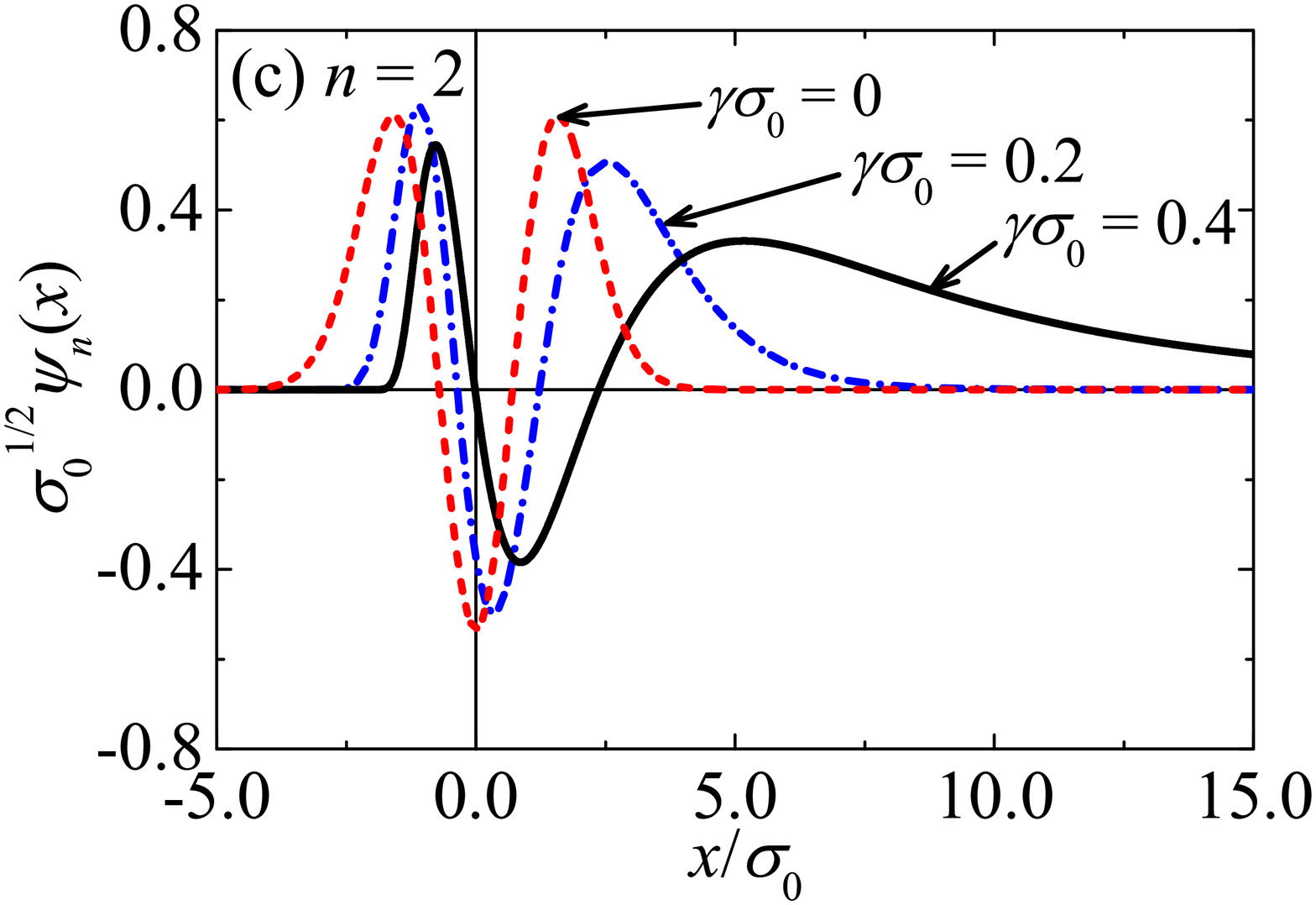}
\end{minipage}
\\
\begin{minipage}[h]{0.32\linewidth}
\includegraphics[width=\linewidth]{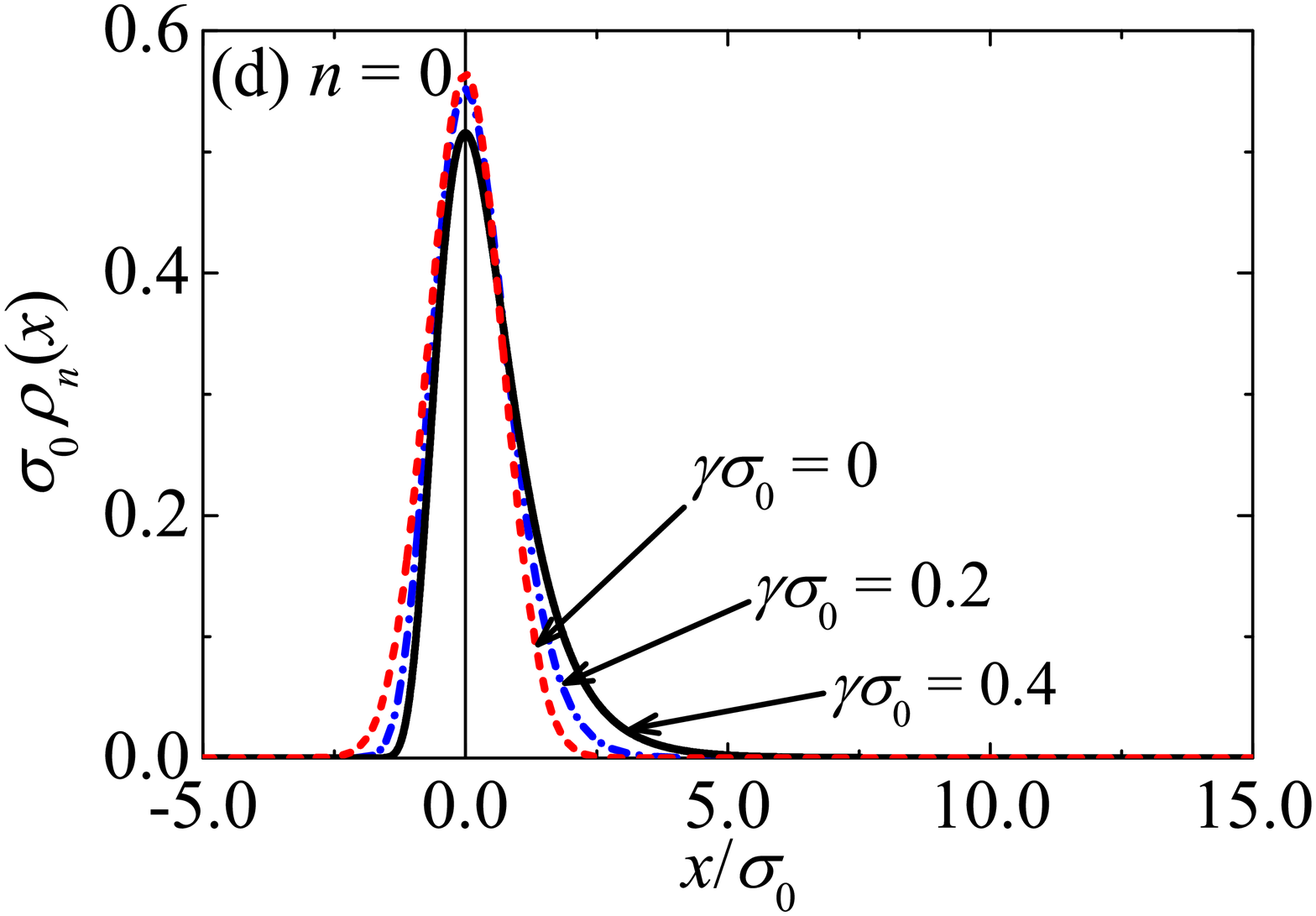}
\end{minipage}
\begin{minipage}[h]{0.32\linewidth}
\includegraphics[width=\linewidth ]{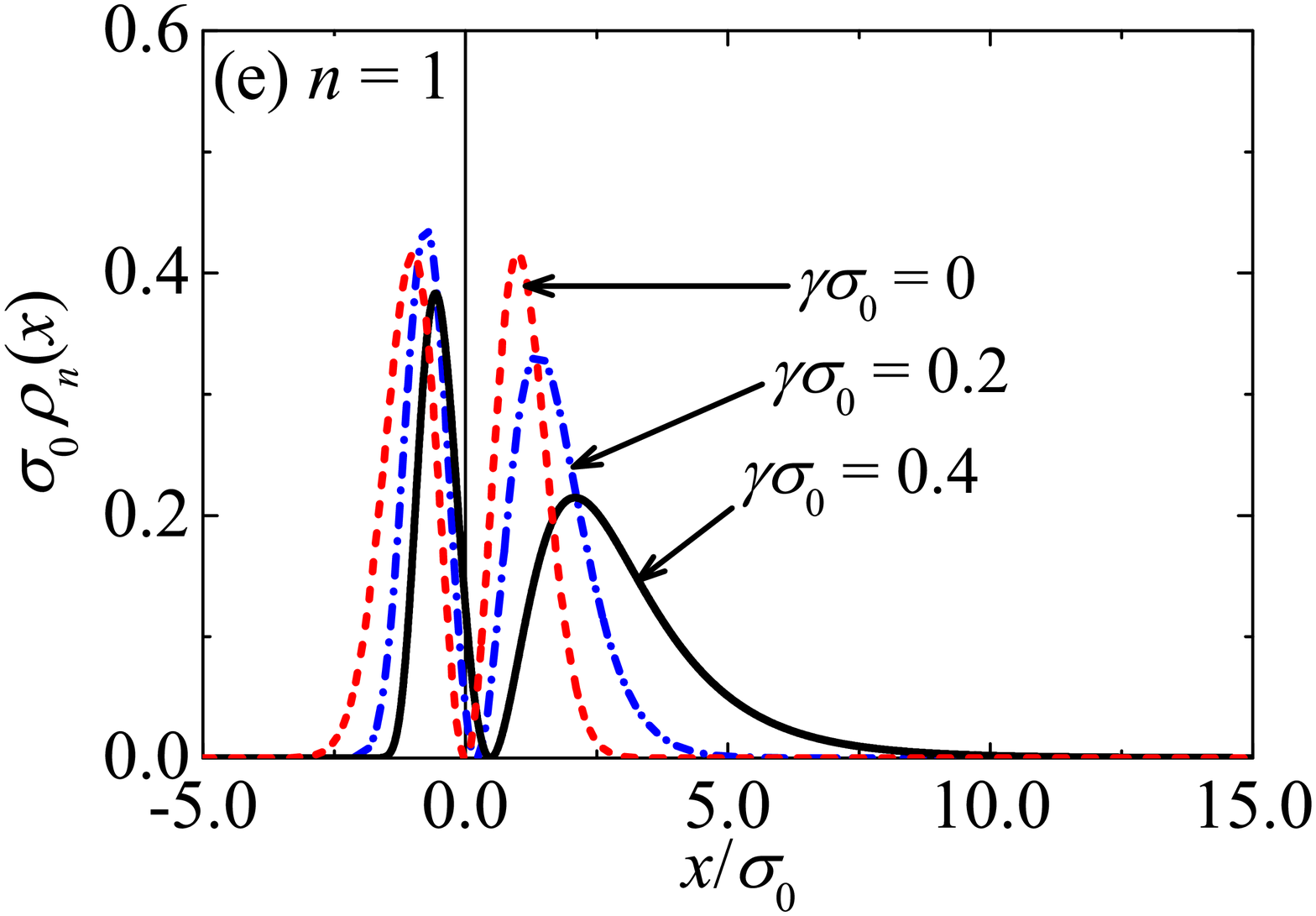}
\end{minipage}
\begin{minipage}[h]{0.32\linewidth}
\includegraphics[width=\linewidth]{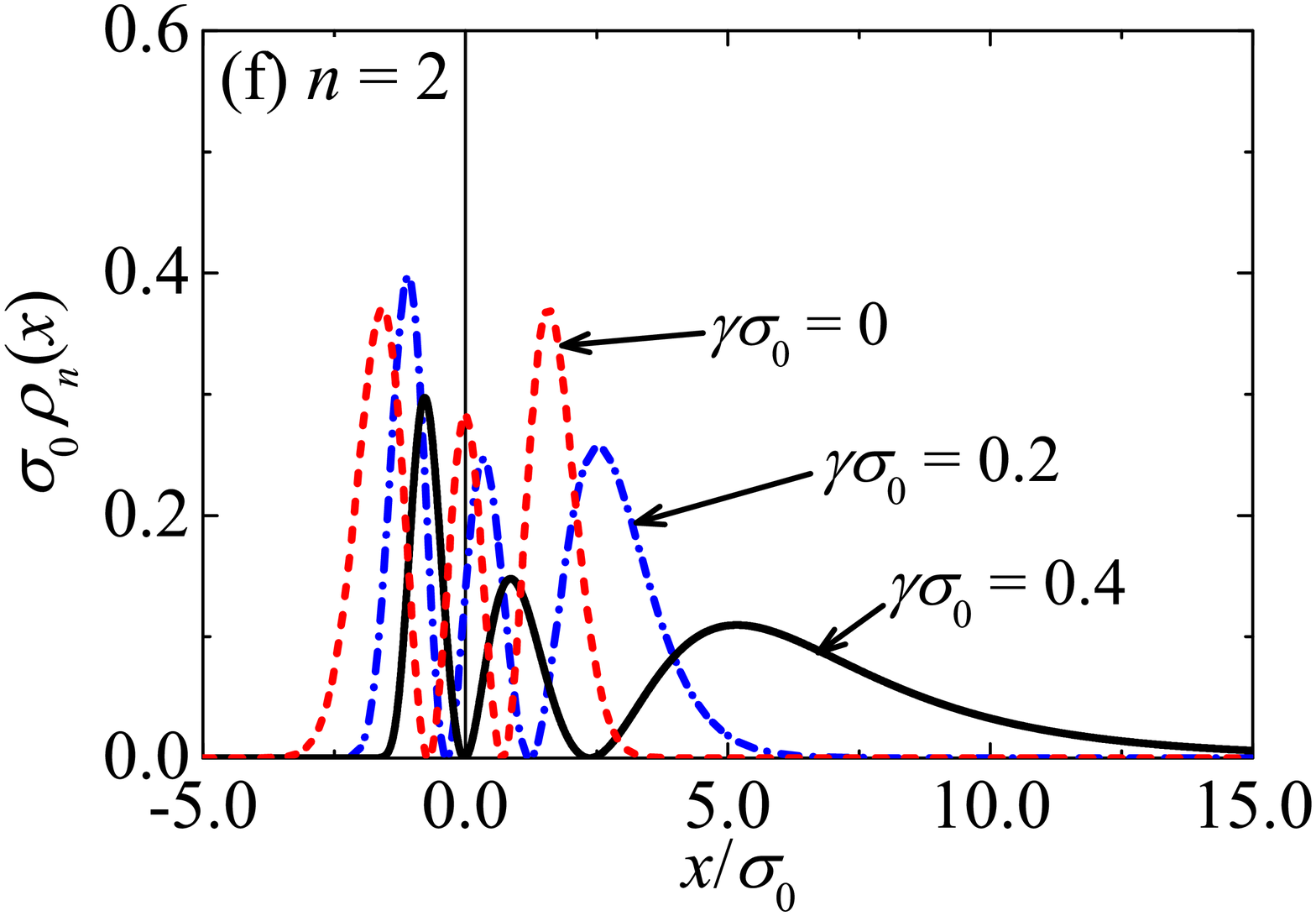}
\end{minipage}
\caption{\label{fig:4}
Wave functions $\psi_n(x)$ (upper line)
and probability densities $\rho_n(x)$ (bottom line)
for deformed oscillator with PDM and
values of $\gamma \sigma_0 = 0$ 
(standard case for comparison, dashed red line), 
$0.2$ (dashed--dotted blue line), 
and $0.4$ (solid black line).
[(a) and (d)] $n = 0$ (ground state), 
[(b) and (e)] $n = 1$ (first excited state), and
[(c) and (f)] $n = 2$ (second excited state).
}
\end{figure}

\begin{figure}[!htb]
\centering
\begin{minipage}[h]{\linewidth}
\includegraphics[width=0.35\linewidth]{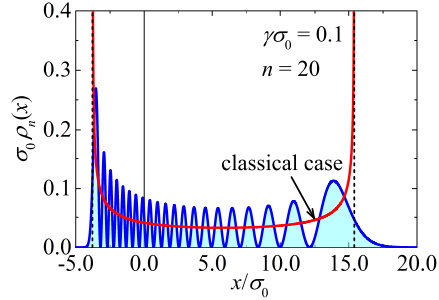}
\end{minipage}
\caption{\label{fig:5}
Probability density for a deformed oscillator with PDM
given by Eq.~(\ref{eq:m(x)}) and potential $V(x)=\frac{1}{2}m(x)\omega_0^2x^2$
with $\gamma \sigma_0 = 0.1$ and $n=20$.
The classical case [Eq.~(\ref{eq:rho_classic})]
(with $A_{\gamma,n}^2=2E_n(\gamma)/m_0\omega_0^2$ 
replaced by $A_0^2$) is shown for comparison.
Dotted vertical lines correspond to 
$x_{\textrm{min}} = -A_{\gamma, n}/(1+\gamma A_{\gamma, n})$ and
$x_{\textrm{max}} = A_{\gamma, n}/(1-\gamma A_{\gamma, n})$.
}
\end{figure}

In Fig.~\ref{fig:4}, we plot eigenfunctions $\psi_n(x)$
and density probabilities $\rho_n (x) = |\psi_n(x)|^2$ 
for the ground state and the first two excited states
with different values of the deformation parameter
$\gamma \sigma_0$.
Figure \ref{fig:5} shows that for large quantum numbers, 
exemplified here with $n = 20$, the mean quantum probability density 
$\rho_n(x)$ tends to approach the classical probability density 
$\rho_{\textrm{classic}}(x)$ [Eq.~(\ref{eq:rho_classic})] with 
$A_{\gamma,n}^2 = 2E_n(\gamma)/m_0\omega_0^2$ in place of $A_0^2$. 

\begin{figure}[!htb]
\begin{minipage}[h]{\linewidth}
\includegraphics[width=0.35\linewidth]{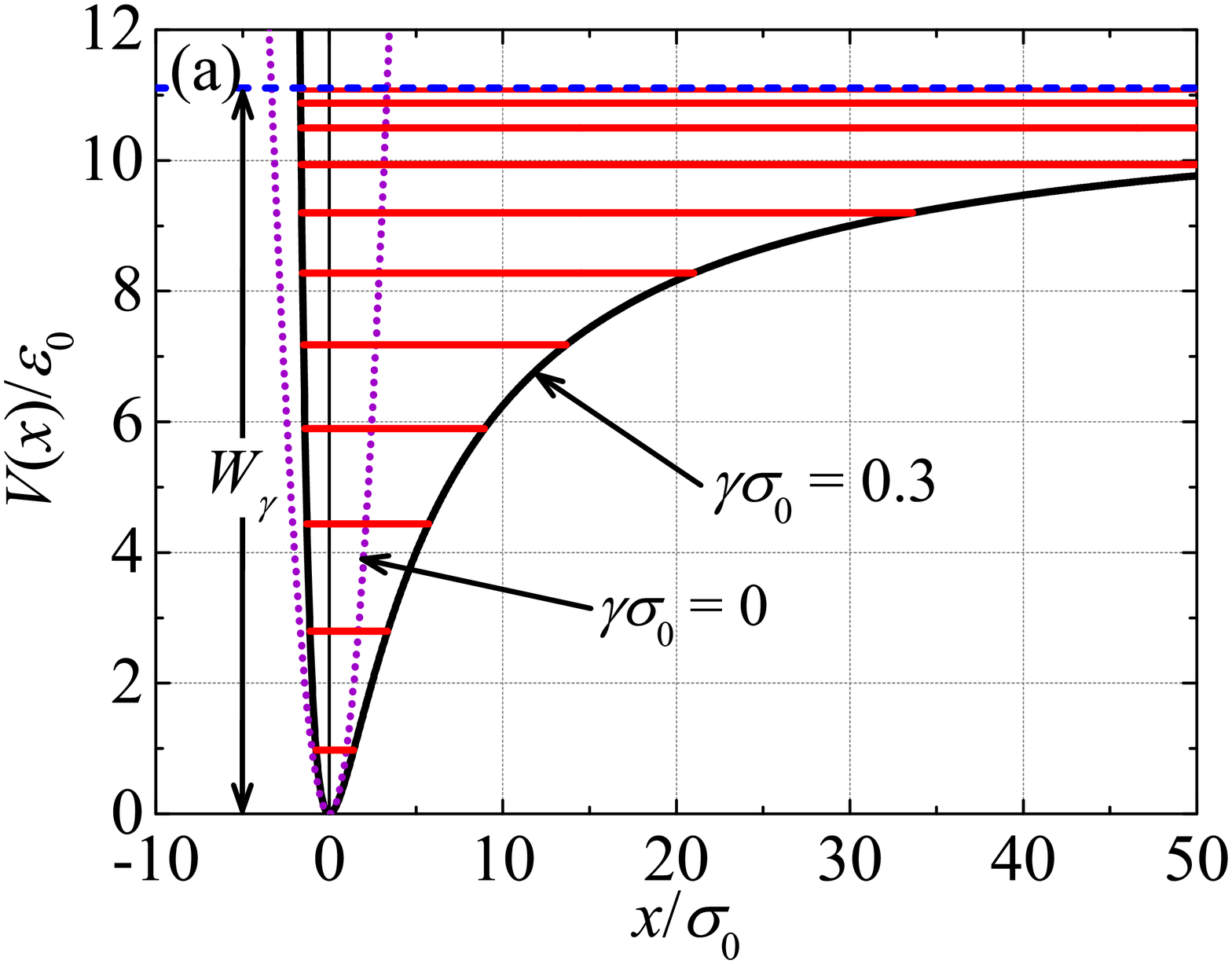}
\includegraphics[width=0.355\linewidth]{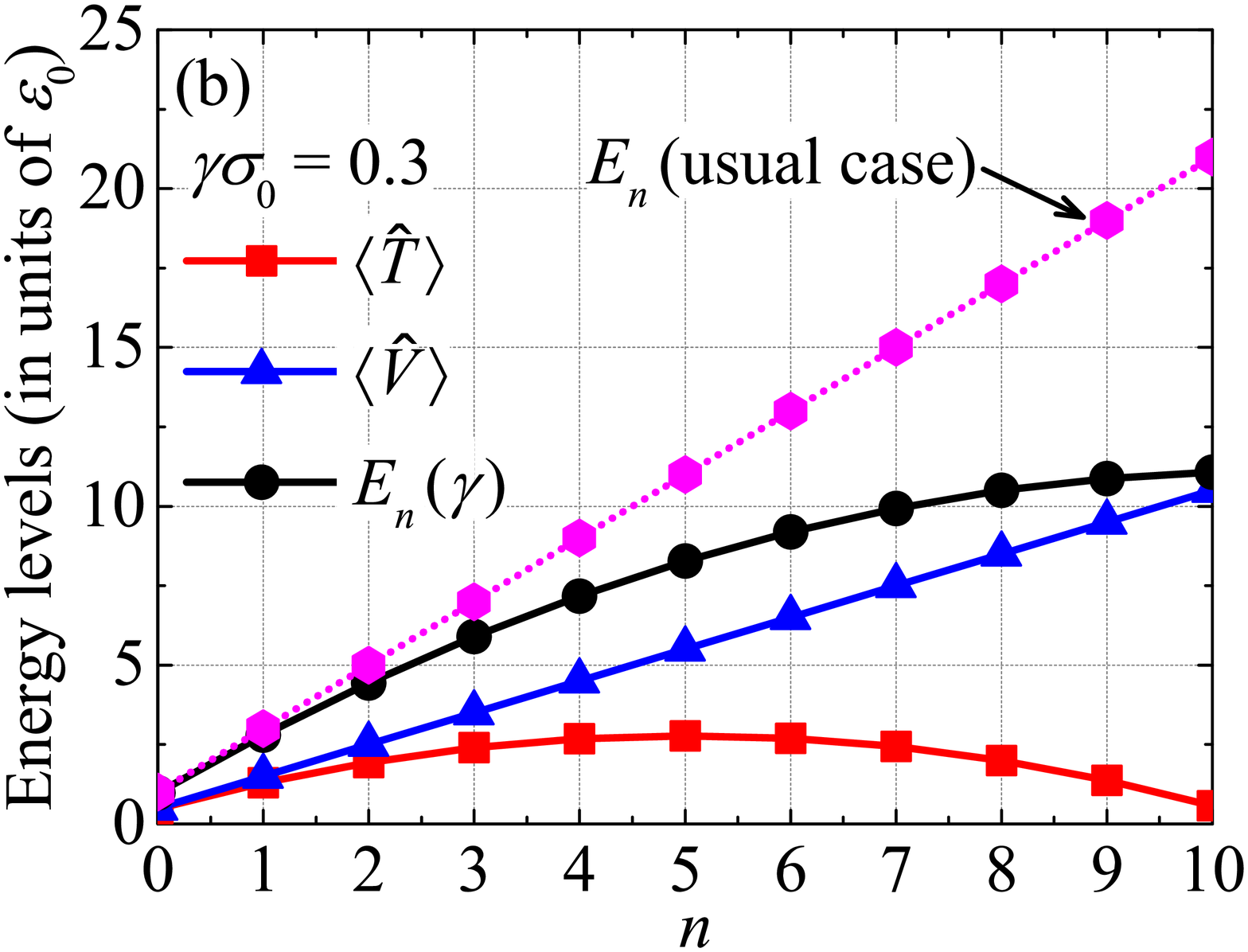}
\end{minipage}
\caption{\label{fig:6}
(a) Energy levels of the bound states for the oscillator with PDM $m(x)$ 
[Eq.~\eqref{eq:m(x)}] and potential $V(x) = \frac{1}{2}m(x) \omega_0^2 x^2$.
The deformation parameter is $\gamma \sigma_0 = 0.3$, with
$\varepsilon_0 = \frac{1}{2}\hbar \omega_0$. 
(b) Expected values $\langle \hat{T} \rangle$, $\langle \hat{V} \rangle$,
and $E_n$ as a function of the quantum number $n$
(the usual case $\gamma\sigma_0 = 0$ is shown for comparison).
}
\end{figure}

We obtain that the expected values of the potential
and kinetic energies are given respectively by
\begin{subequations}
\begin{align}
\label{eq:V-expected-value}
\langle \hat{V} \rangle &=
\frac{\hbar \omega_0}{2} \left( n + \frac{1}{2} \right),
\\
\label{eq:T-expected-value}
\langle \hat{T} \rangle &=
\frac{\hbar \omega_0}{2} \left( n + \frac{1}{2} \right)
\left[1 -\gamma^2 \sigma_0^2 \left( n + \frac{1}{2} \right) \right].
\end{align}
\end{subequations}
Figure \ref{fig:6} illustrates the potential $V(x)$
(in terms of the standard ground energy $\varepsilon_0 = \frac{1}{2} \hbar \omega_0$) 
and the energy levels (\ref{eq:E_n}) of the bound states.
Together, a comparison between $E_n$, $\langle \hat{T} \rangle$ and 
$\langle \hat{V} \rangle$ is illustrated.

Similarly, we calculate that 
the expected value of mass operator $m(\hat{x})$ is
$
\langle m(\hat{x}) \rangle =
m_0 \left[1 -\gamma^2 \sigma_0^2 \left( n + \frac{1}{2} \right) \right],
$
in which it leads to the relationship
$ \langle\hat{T} \rangle =  {\langle\hat{V} \rangle \langle m(\hat{x}) \rangle}/{m_0}$.
Note that $\langle \hat{T} \rangle \geq \langle \hat{V} \rangle$,
and the equality is satisfied only for the usual case ($\gamma=0$, i.e., 
$\langle m(\hat{x}) \rangle  = {m_0}$).

After some careful calculations, we obtain that the expected values 
$ \langle \hat{x} \rangle $,
$ \langle \hat{x}^2 \rangle $,
$ \langle \hat{p} \rangle $ and
$ \langle \hat{p}^2 \rangle $
for the $n$th eigenstate are given by
\begin{subequations}
\label{eq:x-p-x^2-p^2-expec-values}
\begin{align}
\label{eq:expected-value-x-oscillator}
\langle \hat{x} \rangle &= \frac{1}{\gamma} \left\{
	\frac{1}{\left[ 1-\gamma^2 \sigma_0^2 \left( n + \frac{1}{2} \right) \right]^2
	-\frac{\gamma^2 \sigma_0^2}{4}}
	-1 \right\},\\
\label{eq:expected-value-x^2-oscillator}
\langle \hat{x}^2 \rangle &= \frac{1}{\gamma^2}
	\frac{1 + \gamma^2 \sigma_0^2 \left( n + \frac{1}{2} \right) 
		  - \frac{1}{2}\gamma^4 \sigma_0^4  n(n+1)}{
		\left\{ 
			\left[ 1 - \gamma^2 \sigma_0^2 \left( n + \frac{1}{2} \right) \right]^2 
			- \frac{\gamma^4 \sigma_0^4}{4} 
		\right\}
		\left\{ 
			\left[ 1 - \gamma^2 \sigma_0^2 \left( n + \frac{1}{2} \right) \right]^2 
			- \gamma^4 \sigma_0^4 
		\right\}
		}
	\nonumber \\
	& \quad
	- \frac{2}{\gamma^2} 
		\frac{1}{\left[ 1 - \gamma^2 \sigma_0^2 \left( n + \frac{1}{2} \right) \right]^2 
				- \frac{\gamma^4 \sigma_0^4}{4}}
	+ \frac{1}{\gamma^2},
	\\
\label{eq:expected-value-p-oscillator}
 \langle \hat{p} \rangle &= 0,\\
\label{eq:expected-value-p^2-oscillator}
\langle \hat{p}^2 \rangle &= 
\frac{\hbar^2}{\sigma_0^2} \left[ \left( n + \frac{1}{2} \right) 
-\frac{\gamma^2 \sigma_0^2}{2} (n^2 + n - 1) \right]
\left[1 -\gamma^2 \sigma_0^2 \left( n + \frac{1}{2} \right) \right].
\end{align}
\end{subequations}
Similarly, it is straightforward to verify that
the expected values of the linear pseudomomentum are
\begin{subequations}
\label{eq:Pi-expected-value-psi_n}
\begin{align}
\label{eq:Pi-expected-value}
\langle \hat{\Pi}_\gamma \rangle &=0,
\\
\label{eq:Pi^2-expected-value}
\langle \hat{\Pi}_\gamma^2 \rangle &=
\frac{\hbar^2}{\sigma_0^2} \left( n + \frac{1}{2} \right)
\left[1 -\gamma^2 \sigma_0^2 \left( n + \frac{1}{2} \right) \right].
\end{align}
\end{subequations}
The expected values $\langle \hat{p}^2 \rangle$
and  $\langle \hat{\Pi}_\gamma^2 \rangle$ 
different for $\gamma \neq 0$.
We can clearly see that in the limit
$\gamma =0$ the usual cases are recovered: 
$\lim_{\gamma \rightarrow 0}\langle \hat{x} \rangle = 0$,
$\lim_{\gamma \rightarrow 0} \langle \hat{x}^2 \rangle = 
\sigma_0^2 \left( n+\frac{1}{2} \right)$
and
$\lim_{\gamma \rightarrow 0} \langle \hat{p}^2 \rangle = 
\lim_{\gamma \rightarrow 0} \langle \hat{\Pi}_\gamma^2 \rangle = 
\frac{\hbar^2}{\sigma_0^2} \left( n+\frac{1}{2} \right)$.

Since $A_{\gamma, n}^2 = 2 \sigma_0^2 E_n({\gamma})/\hbar \omega_0$,
according to the principle of correspondence, 
in the limit of large quantum numbers (or $\hbar \rightarrow 0$), 
we have $E_n \gg \hbar \omega_0$, and consequently,
Eqs.~(\ref{eq:x-p-x^2-p^2-expec-values}) can be written as
\begin{subequations}
\begin{align}
\langle \hat{x} \rangle &= 
	\frac{\gamma A_{\gamma, n}^2}{1-\gamma^2 A_{\gamma, n}^2}, \\
\langle \hat{x}^2 \rangle &= 
	\frac{A_{\gamma, n}^2}{2}
	\frac{1+2\gamma^2 A_{\gamma, n}^2}{(1-\gamma^2 A_{\gamma, n}^2)^2}, \\
\langle \hat{p}^2 \rangle &= \frac{1}{2} m_0^2 \omega_0^2 A_{\gamma, n}^2
				  \sqrt{1-\gamma^2 A_{\gamma, n}^2},
\end{align}
\end{subequations}
which coincide with the corresponding mean values 
of the classic oscillator [Eq.~(\ref{eq:mean_values_x-p})].

\begin{figure}[htb]
\centering
\includegraphics[width=0.32\linewidth]{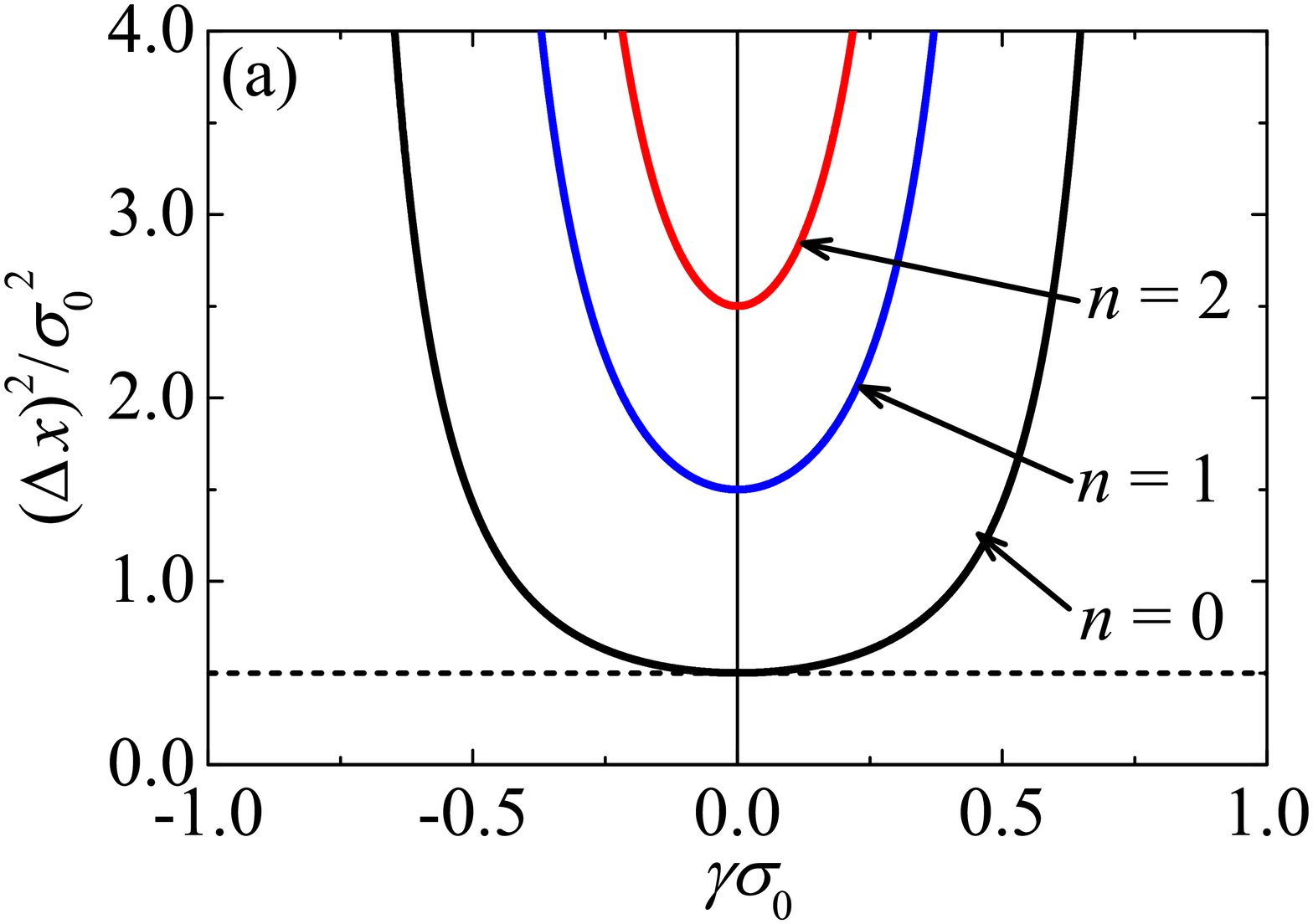}
\includegraphics[width=0.32\linewidth]{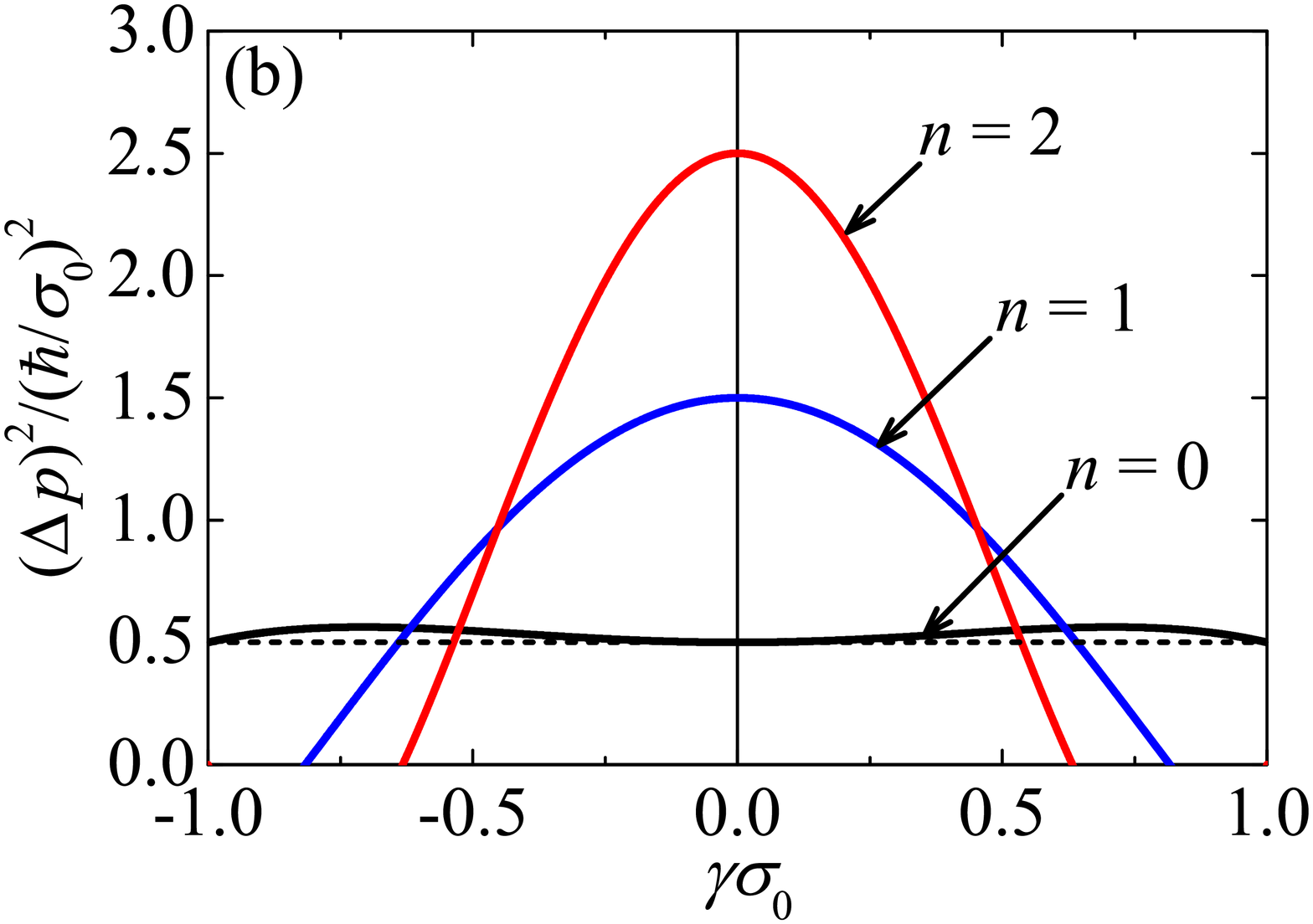}\\
\includegraphics[width=0.32\linewidth]{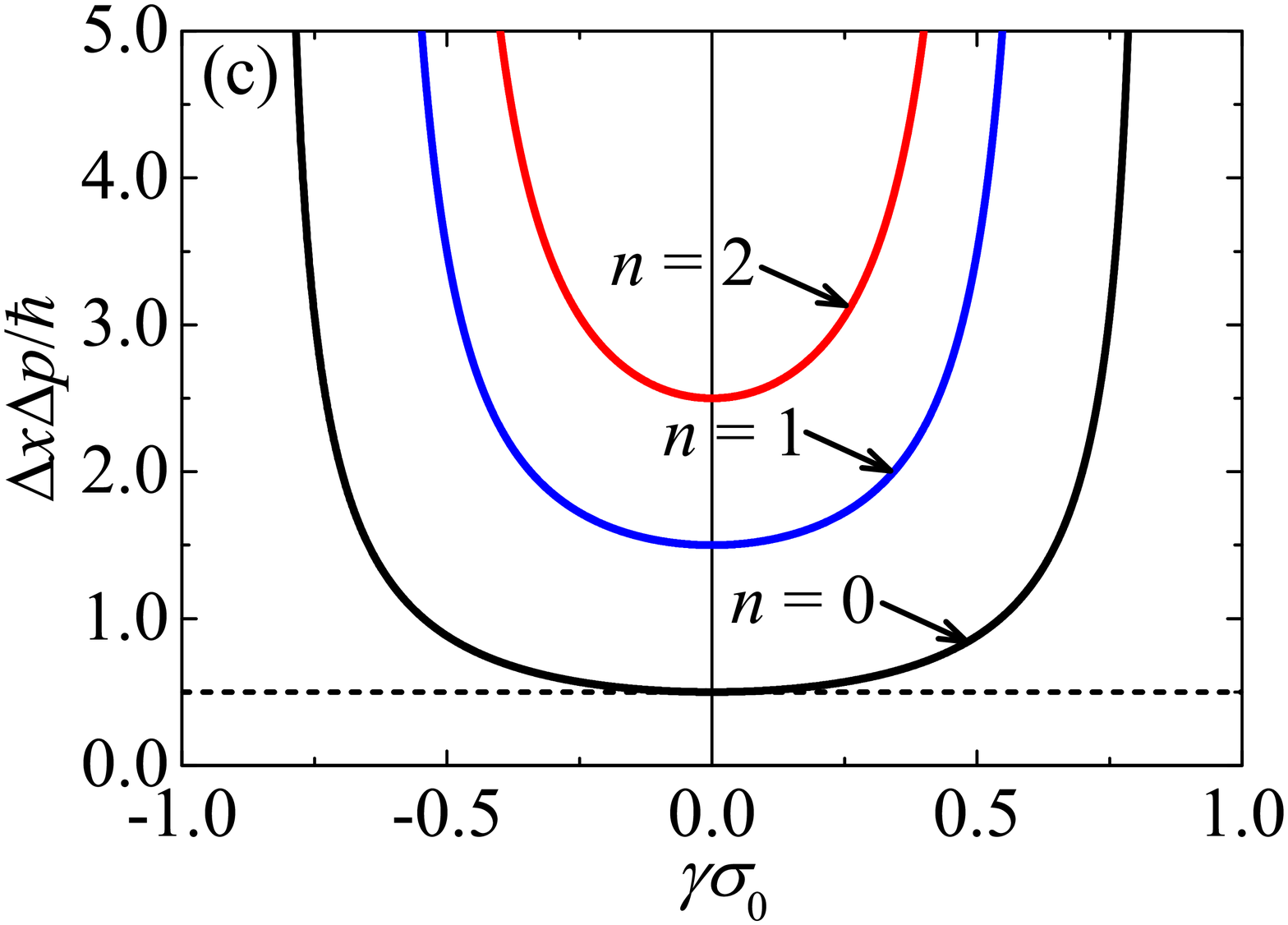}
\includegraphics[width=0.32\linewidth]{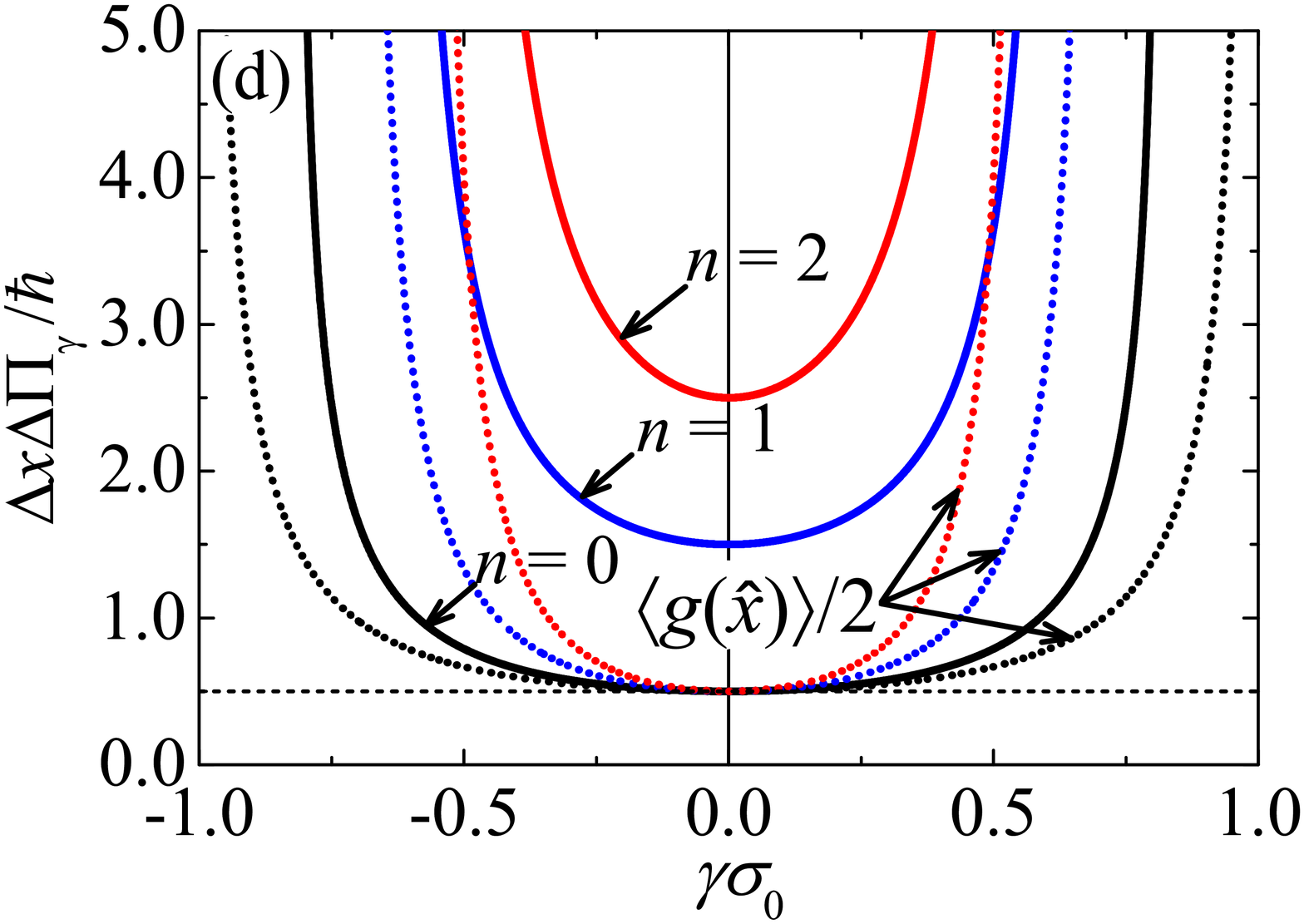}
\caption{\label{fig:7}
Mean square deviation of (a) position $x$ 
and (b) linear momentum $p$
and products (c) $\Delta x \Delta p$
and (d) $\Delta x \Delta \Pi_\gamma$ 
of the deformed oscillator as a function 
of the dimensionless parameter $\gamma \sigma_0$ 
for the states $n=0$, $1$ and $2$.
}
\end{figure}

In Fig.~\ref{fig:7}, we plot 
the square of the uncertainties of $x$ and $p$, 
$(\Delta x)^2$ and $(\Delta p)^2$, 
along with the uncertainty relation $\Delta x \Delta p$ 
for states $n=0$, $1$ and $2$.
As the dimensionless parameter $|\gamma \sigma_0|$ approaches 1, 
the position uncertainty increases and 
the linear momentum one decreases
(except for state $n=0$ which has a peculiar behavior), 
but as expected, the standard uncertainty relation 
$\Delta x \Delta p \geq \frac{\hbar}{2}$ is kept.
In addition, the uncertainty product
is symmetric around $\gamma \sigma_0 = 0$ and 
grows with the quantum number $n$.
On the other hand, for the mass function
(\ref{eq:m(x)}), we have the spatial metric term
$g(\hat{x}) = 1+\gamma \hat{x}$ so that
the observables $\hat{x}$ and $\hat{\Pi}_\gamma$ 
satisfy the generalized uncertainty principle
$\Delta x \Delta \Pi_\gamma \geq 
\frac{\hbar}{2} (1 + \gamma \langle \hat{x} \rangle)$,
as shown in the figure \ref{fig:7}(d). 
Despite the different expressions for 
expected values of 
$\langle \hat{p}^2 \rangle$
and $\langle \hat{\Pi}_\gamma^2 \rangle$, 
the products of uncertainty are quite similar
$\Delta x \Delta p$ and $\Delta x \Delta \Pi_\gamma$.

\section{\label{CS-deformed-oscillator} 
        Coherent states for deformed oscillator}

\subsection{Factorization method for quantum deformed oscillator}

For the potential 
$V(\hat{x}) = \frac{1}{2}m(x) \omega_0^2 \hat{x}^2,$
the Hamiltonian operator 
(\ref{eq:hamiltonian-MM})
can be factorized as
\begin{equation}
\label{eq:H-factorized}
\hat{H} = \hbar \omega_0 \hat{a}_\gamma^{\dagger} \hat{a}_\gamma
		  + E_0 ({\gamma}), 
\end{equation}
with the annihilation and the creation operators given by
\begin{subequations}
\label{eq:a_gamma-and-a_gamma-dagger}
\begin{align}
\label{eq:a_gamma}
\hat{a}_\gamma 
	&= \sqrt{\frac{m_0 \omega_0}{2\hbar}}	
		\left[ \frac{\hat{x}}{\hat{1} + \gamma \hat{x}} 
		+ \frac{i}{m_0\omega_0} (\hat{1} + \gamma \hat{x}) \hat{p} \right] 
		\nonumber \\
	&= \sqrt{\frac{m_0 \omega_0}{2\hbar}}	
		\left[ \hat{\Phi}_\gamma (\hat{x})
		+ \frac{i}{m_0\omega_0} \hat{\Pi}_\gamma (\hat{x}, \hat{p}) \right]
\end{align}
and
\begin{align}
\label{eq:a_gamma_dagger}
\hat{a}_\gamma^{\dagger}
	&= \sqrt{\frac{m_0 \omega_0}{2\hbar}}	
		\left[ \frac{\hat{x}}{\hat{1} + \gamma \hat{x}} 
		- \frac{i}{m_0\omega_0} \hat{p} (\hat{1} + \gamma \hat{x})  \right]
			\nonumber \\	&= \sqrt{\frac{m_0 \omega_0}{2\hbar}}	
		\left[ \hat{\Phi}_\gamma (\hat{x})
		- \frac{i}{m_0\omega_0} \hat{\Pi}_\gamma (\hat{x}, \hat{p}) \right],
\end{align}
\end{subequations}
where 
$\hat{\Phi}_\gamma (\hat{x}) =
\frac{\hat{x}}{\hat{1} + \gamma \hat{x}} - \frac{\hbar \gamma}{2m_0 \omega_0}
$ 
corresponds to superpotential\cite{Ruby-Senthilvelan-2010}
and satisfies along linear pseudomomentum $\hat{\Pi}_\gamma$
the relation
$[\hat{\Phi}_\gamma, \hat{\Pi}_\gamma] 
= \frac{i\hbar}{\sqrt{m(\hat{x})/m_0}} \frac{\textrm{d}\Phi_\gamma (\hat{x})}{\textrm{d} \hat{x}}
= \frac{i\hbar}{1+\gamma \hat{x}}$.
It is easy to verify that $\hat{a}_\gamma \psi_0(x) = 0$, i.e.,
the ground state [Eq.~(\ref{eq:ground-state})] 
is annihilated by the operator $\hat{a}_\gamma$.
The ground state can also be obtained 
from the superpotential through the relation
\begin{equation}
\psi_0 (x)  \propto \sqrt[4]{\frac{m(x)}{m_0}}
			\exp \left[
			-\frac{1}{\sigma_0^2} \int^x
			\sqrt{\frac{m(y)}{m_0}}
			\Phi_\gamma (y) \textrm{d}y
			\right].
\end{equation}

The PDM effect leads to a commutator
between $\hat{a}_\gamma$ and $\hat{a}_\gamma^{\dagger}$ 
dependent on the spatial coordinate given by
$
[\hat{a}_\gamma, \hat{a}_\gamma^{\dagger}] = ( \hat{1}+\gamma \hat{x} )^{-1}.
$
Additionally, it is straightforward to show that the operators
(\ref{eq:a_gamma}) and (\ref{eq:a_gamma_dagger})
satisfy the commutation relations
$
[\hat{a}_\gamma, \hat{a}_\gamma^{\dagger} \hat{a}_\gamma ]
= ( \hat{1}+\gamma \hat{x} )^{-1} \hat{a}_\gamma
$
and
$
[\hat{a}_\gamma^{\dagger}, \hat{a}_\gamma^{\dagger} \hat{a}_\gamma ]
= -\hat{a}_\gamma^{\dagger} ( \hat{1} + \gamma \hat{x} )^{-1}.
$
In terms of the annihilation and creation operators
\begin{subequations}
\begin{align}
[\hat{a}_\gamma, \hat{a}_\gamma^{\dagger}]
& = \hat{1} - \frac{\gamma \sigma_0}{\sqrt{2}} 
	(\hat{a}_\gamma^{\dagger} + \hat{a}_{\gamma}) 
	- \frac{\gamma^2 \sigma_0^2}{2},
	\\
[\hat{a}_\gamma, \hat{a}_\gamma^{\dagger} \hat{a}_\gamma ]
	&= \left[ \hat{1} - \frac{\gamma \sigma_0}{\sqrt{2}} 
	(\hat{a}_\gamma^{\dagger} + \hat{a}_{\gamma}) 
	- \frac{\gamma^2 \sigma_0^2}{2} \right] \hat{a}_\gamma,
	\\
[\hat{a}_\gamma^{\dagger}, \hat{a}_\gamma^{\dagger} \hat{a}_\gamma ]
	&= -\hat{a}_\gamma^{\dagger}
	\left[ \hat{1} - \frac{\gamma \sigma_0}{\sqrt{2}} 
	(\hat{a}_\gamma^{\dagger} + \hat{a}_{\gamma}) 
	- \frac{\gamma^2 \sigma_0^2}{2} \right].
\end{align}
\end{subequations}
In compact form, the operators $\hat{a}_\gamma$, $\hat{a}_\gamma^{\dagger}$ and $\hat{H}$ 
constitute the deformed algebraic structure
$[\hat{a}_\gamma, \hat{a}_\gamma^{\dagger}] = G_\gamma (\hat{a}_\gamma, \hat{a}_\gamma^{\dagger})$,
$[\hat{H} , \hat{a}_\gamma]
 = - \hbar \omega_0 G_\gamma(\hat{a}_\gamma, \hat{a}_\gamma^{\dagger})\hat{a}_\gamma$
and
$[\hat{H} ,\hat{a}_\gamma^{\dagger} ]
 = \hbar \omega_0  \hat{a}_\gamma^{\dagger} G_\gamma(\hat{a}_\gamma, \hat{a}_\gamma^{\dagger})$
with deformation established by the function
$G_\gamma(\hat{a}_\gamma, \hat{a}_\gamma^{\dagger}) =
\hat{1} - \frac{\gamma \sigma_0}{\sqrt{2}} 
	(\hat{a}_\gamma^{\dagger} + \hat{a}_{\gamma}) 
	- \frac{\gamma^2 \sigma_0^2}{2}$.
Similar to the classical formalism, 
the operators $\hat{a}_\gamma$, $\hat{a}_\gamma^{\dagger}$ and $\hat{H}$
satisfy the Jacobi identify 
$
[[\hat{a}_\gamma, \hat{a}_\gamma^{\dagger}], \hat{H}]+
[[\hat{H}, \hat{a}_\gamma ],\hat{a}_\gamma^{\dagger}]+
[[\hat{a}_\gamma^{\dagger}, \hat{H}],\hat{a}_\gamma] =0.
$

It is possible to observe that the operators 
(\ref{eq:a_gamma-and-a_gamma-dagger}) are not bosonic,
since in terms of the anticommutator 
$\{ \hat{a}_\gamma,\hat{a}_\gamma^{\dagger} \}$
the Hamiltonian $\hat{H}$ becomes
\begin{align}
\label{eq:H_uniforme_field}
\hat{H} &= \frac{\hbar \omega_0}{2} 
\{ \hat{a}_\gamma,\hat{a}_\gamma^{\dagger} \}
-\frac{\hbar \omega_0}{2} \frac{1}{\hat{1}+ \gamma \hat{x}} + E_0(\gamma)
\nonumber \\
&=\frac{\hbar \omega_0}{2} \{ \hat{a}_\gamma,\hat{a}_\gamma^{\dagger} \}
  +\frac{\hbar \omega_0 \gamma \sigma_0}{2\sqrt{2}}
  (\hat{a}_\gamma^{\dagger}+\hat{a}_\gamma) 
   + \frac{\hbar^2 \gamma^2}{8m_0}.
\end{align}
The Hamiltonian operator written in the form 
of Eq. (\ref{eq:H_uniforme_field}) has the same structure of 
a quantum harmonic oscillator subjected to a uniform static field,
i.e.,
$\hat{H} = \hat{H}_\gamma - \mathcal{F}_\gamma \hat{\Phi}_\gamma + \frac{\mathcal{F}_\gamma^2}{2m_0\omega_0^2},$
with 
$\hat{H}_\gamma = \frac{\hbar \omega_0}{2} 
\{ \hat{a}_\gamma,\hat{a}_\gamma^{\dagger} \}$ being a bosonic Hamiltonian,
$\hat{\Phi}_\gamma =\frac{\sigma_0}{\sqrt{2}} (\hat{a}_\gamma^{\dagger} + \hat{a}_\gamma)$
being a generalized coordinate  and 
$\mathcal{F}_\gamma = -\frac{1}{2}\hbar \omega_0 \gamma$ a constant force.
However, considering the displaced annihilation operator
$\hat{b}_\gamma 
= \hat{a}_\gamma + \frac{\gamma \sigma_0}{2\sqrt{2}} \hat{1}
= \frac{1}{\sqrt{2} \sigma_0} 
\left(\frac{\hat{x}}{\hat{1}+\gamma \hat{x}} 
+ \frac{i}{m_0 \omega_0} \hat{\Pi}_\gamma \right)$
and its self-adjoint operator $\hat{b}_\gamma^{\dagger}$, 
we can rewrite $\hat{H}$ in bosonic form, 
\begin{equation}
\label{eq:bosonic-Hamiltonian}
\hat{H} = \frac{\hbar \omega_0}{2} 
\{ \hat{b}_\gamma , \hat{b}_\gamma^{\dagger} \},
\end{equation}
or more explicitly,
\begin{equation}
\label{eq:H-bosonic-operators}
\hat{H} =\hbar \omega_0
\left( \hat{b}_\gamma^{\dagger} \hat{b}_\gamma + \frac{1}{2} \right)
- \frac{\hbar \omega_0 \gamma \sigma_0 }{2\sqrt{2}} 
( \hat{b}_\gamma^{\dagger}  + \hat{b}_\gamma ).
\end{equation}
with 
$[\hat{b}_\gamma, \hat{b}_\gamma^{\dagger}] = \hat{1}-\frac{\gamma \sigma_0}{\sqrt{2}}
(\hat{b}_\gamma^{\dagger}+\hat{b}_\gamma).$
The first term in (\ref{eq:H-bosonic-operators}) 
corresponds to the Hamiltonian operator
of a deformed oscillator, i.e., 
$\hat{h}_\gamma = \hbar \omega_0 
\left( \hat{b}_\gamma^{\dagger} \hat{b}_\gamma 
+ \frac{1}{2} \right)$,
while the second term is equivalent to an interaction potential 
of a uniform electric field  due to spatial deformation, 
$\hat{\mathcal{V}}_\gamma = 
-\frac{\hbar \omega_0 \gamma \sigma_0 }{2\sqrt{2}} 
( \hat{b}_\gamma^{\dagger}  + \hat{b}_\gamma )$.

Considering the Pauli matrices 
$\widehat{\sigma}_{\pm}$
and the supercharges\cite{Plyushchay-2017}
\begin{equation}
\widehat{\mathcal{Q}}_\gamma = \widehat{\sigma}_{-} \hat{b}_\gamma =
\left(
\begin{array}{cc}
0 & 0 \\
\hat{b}_\gamma & 0
\end{array}
\right)
\qquad 
\textrm{and}
\qquad
\widehat{\mathcal{Q}}_\gamma^\dagger = \widehat{\sigma}_{+} \hat{b}_\gamma^{\dagger} =
\left(
\begin{array}{cc}
0 & \hat{b}_\gamma^{\dagger} \\
0 & 0
\end{array}
\right),
\end{equation}
the supersymmetric Hamiltonian is
\begin{align}
\hat{H}_{\textrm{ss}} 
&= \hbar \omega_0
\{ \widehat{\mathcal{Q}}_\gamma , \widehat{\mathcal{Q}}_\gamma^{\dagger} \}
\nonumber \\
&= \hbar \omega_0  \left(
\begin{array}{cc}
\hat{b}_\gamma^{\dagger}\hat{b}_\gamma & 0\\
0 & \hat{b}_\gamma \hat{b}_\gamma^{\dagger}	
\end{array}
\right) \nonumber \\
&=
\left(
\begin{array}{cc}
\hat{H} -\frac{\hbar \omega_0}{2} \frac{1}{\hat{1}+ \gamma \hat{x}} & 0\\
0 & \hat{H} + \frac{\hbar \omega_0}{2} \frac{1}{\hat{1} + \gamma \hat{x}}
\end{array}
\right),
\end{align}
or more compactly,
$
\hat{H}_{\textrm{ss}} = \hat{H}
- \widehat{\sigma}_z \left(\frac{\hbar \omega_0}{2} 
+ \hat{\mathcal{V}}_\gamma \right),
$
where $\widehat{\sigma}_z$ is 
the diagonal Pauli matrix.

In order to apply the supersymmetric quantum mechanics, 
the annihilation and the creation operators
(\ref{eq:a_gamma-and-a_gamma-dagger}) lead to partners Hamiltonian 
$
	\hat{H}_{\scriptscriptstyle +} 
	= \hbar \omega_0 \hat{a}_{\gamma}^{\dagger} \hat{a}_\gamma
$
and
$
	\hat{H}_{\scriptscriptstyle -} 
	= \hbar \omega_0 \hat{a}_\gamma\hat{a}_{\gamma}^{\dagger},
$
which can be written in the form	
$
	\hat{H}_{\scriptscriptstyle \pm} (\hat{x}, \hat{p}) 
	= \frac{1}{2m_0} \hat{\Pi}_\gamma^2 (\hat{x}, \hat{p}) + 
	V_{\scriptscriptstyle \pm}(\hat{x}),
$
with partner potentials given respectively by 
$
V_{\scriptscriptstyle +} (x) = 
\frac{1}{2}m_0\omega_0^2 
\left( \frac{x}{1+\gamma x} \right)^2 
- E_0(\gamma)
$
and
$
V_{\scriptscriptstyle -} (x) = 
\frac{1}{2}m_0\omega_0^2 
\left( \frac{x}{1+\gamma x} - \frac{\hbar \gamma}{m_0 \omega_0} \right)^2 
- E_0(\gamma) + \hbar \omega_0  - \frac{\hbar^2 \gamma^2}{2m_0}.
$
Both potentials have a shift equal to ground state energy 
in binding energy, i.e., 
$\lim_{x \rightarrow \infty} V_{\scriptscriptstyle \pm} (x) = W_\gamma - E_0(\gamma)$.
The equilibrium positions 
for $V_{\scriptscriptstyle +}(x)$ and $V_{\scriptscriptstyle -}(x)$ 
are respectively
$x_{\textrm{eq.}}^{{\scriptscriptstyle (+)}} = 0$ 
and 
$x_{\textrm{eq.}}^{{\scriptscriptstyle (-)}} 
= \gamma \sigma_0^2/(1-\gamma^2 \sigma_0^2).$

The time-independent Schr\"odinger equation for
partners operators are denoted as
$\hat{H}_{\scriptscriptstyle +} \psi_n^{\scriptscriptstyle (+)}(x) 
= E_n^{\scriptscriptstyle (+)}\psi_n^{\scriptscriptstyle (+)}(x)$
and 
$\hat{H}_{\scriptscriptstyle -} \psi_n^{\scriptscriptstyle (-)}(x) 
= E_n^{\scriptscriptstyle (-)}\psi_n^{\scriptscriptstyle (-)}(x)$.
The partner operators $\hat{H}_{\scriptscriptstyle -}$ and
$\hat{H}_{\scriptscriptstyle +}$ are intertwined as follows
\begin{subequations}
\begin{align}
\hat{H}_{\scriptscriptstyle -} (\hat{a}_\gamma \psi_n^{\scriptscriptstyle (+)}) 
= \hat{a}_\gamma (\hat{H}_{\scriptscriptstyle +} \psi_n^{\scriptscriptstyle (+)})
= E_n^{\scriptscriptstyle (+)} (\hat{a}_\gamma \psi_n^{\scriptscriptstyle (+)}),
\\
\hat{H}_{\scriptscriptstyle +} (\hat{a}_\gamma^{\dagger} \psi_n^{\scriptscriptstyle (-)}) 
= \hat{a}_\gamma^{\dagger} (\hat{H}_{\scriptscriptstyle -} \psi_n^{\scriptscriptstyle (-)})
= E_n^{\scriptscriptstyle (-)} (\hat{a}_\gamma^{\dagger} \psi_n^{\scriptscriptstyle (-)} ),
\end{align}
\end{subequations}
and so, $\hat{a}_\gamma \psi_n^{\scriptscriptstyle (+)}$ 
($\hat{a}_\gamma^{\dagger} \psi_n^{\scriptscriptstyle (-)}$)
is eigenfunction of $\hat{H}_{\scriptscriptstyle -}$ 
($\hat{H}_{\scriptscriptstyle +}$).
Consequently, the energy eigenvalues and eigenfunctions of 
the partner Hamiltonians are related as
\begin{subequations}
\label{eq:a_gamma-and-adjunct-eigenfunctions}
\begin{align}
\label{eq:E_n^+andE_n^-}
E_{n}^{\scriptscriptstyle (+)} 
		  & = E_{n-1}^{\scriptscriptstyle (-)} 
		    = E_n - E_0({\gamma})
\\
\label{eq:a_gamma-eigenfunctions}
\hat{a}_\gamma \psi_n^{\scriptscriptstyle (+)}(x)
	&= \left[\frac{E_n^{\scriptscriptstyle (+)}}{\hbar \omega_0}\right]^{1/2} 
	  \psi_{n-1}^{\scriptscriptstyle (-)} (x)
\\
\label{eq:a_gamma-dagger-eigenfunctions}
\hat{a}_\gamma^{\dagger} \psi_n^{\scriptscriptstyle (-)}(x)
	&= \left[\frac{E_n^{\scriptscriptstyle (-)}}{\hbar \omega_0}\right]^{1/2}
	   \psi_{n+1}^{\scriptscriptstyle (+)} (x).
\end{align}
\end{subequations}

Since $\hat{H}_{\scriptscriptstyle +} = \hat{H} - E_0(\gamma)$,
we immediately obtain $\psi_n^{\scriptscriptstyle (+)}(x) = \psi_n (x)$.
From the wave function
$
 \varphi_n^{\scriptscriptstyle (-)}(x) = 
 \sqrt{1+\gamma x} \psi_n^{\scriptscriptstyle (-)}(x)
$
and 
$
 \epsilon_n = E_{n}^{\scriptscriptstyle (-)} 
			  - \hbar \omega_0 + \frac{\hbar^2 \gamma^2}{2m_0} + E_0 (\gamma),
$
we can write the following deformed Schr\"odinger equation
from the potential $V_{\scriptscriptstyle -}(x)$
\begin{equation}
\label{eq:deformed-SE-electric-field}
	-\frac{\hbar^2}{2m_0} D_{\gamma}^2 \varphi_n^{\scriptscriptstyle (-)} (x)
	+ \frac{1}{2}m_0\omega_0^2 \left( 
	\frac{x}{1+\gamma x} - \gamma \sigma_0^2 \right)^2 
	\varphi_n^{\scriptscriptstyle (-)} (x) 
	= \epsilon_n \varphi_n^{\scriptscriptstyle (-)} (x).
\end{equation}
In deformed space $x_\gamma$, Eq.~(\ref{eq:deformed-SE-electric-field}) 
is also turned into a Morse oscillator
\begin{equation}
\label{eq:deformed-SE-electric-field2}
-\frac{\hbar^2}{2m_0} 
\frac{\textrm{d}^2 \phi_n^{\scriptscriptstyle (-)}(x_\gamma)}{\textrm{d} x_\gamma^2}
+ \mathcal{W}_\gamma  (e^{-\gamma (x_\gamma - \mu_{\gamma})} - 1)^2 
\phi_n^{\scriptscriptstyle (-)} (x_\gamma) 
= \epsilon_n \phi_n^{\scriptscriptstyle (-)} (x_\gamma),
\end{equation}
but now with a shifted binding energy
$\mathcal{W}_\gamma = m_0 \omega_\gamma^2/2\gamma^2$,
a frequency of small oscillations around the equilibrium position
$\omega_\gamma = \omega_0 (1 - \gamma^2 \sigma_0^2)$,
equilibrium position
$\mu_\gamma = -\frac{1}{\gamma} \ln (1 - \gamma^2 \sigma_0^2)$,
and
$
\phi_n^{\scriptscriptstyle (-)} (x_\gamma) =
\varphi_n^{\scriptscriptstyle (-)} (x(x_\gamma)).
$
From the solution of the above equation, we arrive at the eigenfunctions
\begin{equation}
\label{eq:psi_n^-} 
\psi_{n}^{\scriptscriptstyle (-)}(x) =
    \frac{\varphi_n^{\scriptscriptstyle (-)} (x)}{\sqrt{1+\gamma x}} \nonumber \\
    =(-1)^{n}\frac{ \widetilde{\mathcal{N}}_n}{\sqrt{2s}} e^{-\frac{\zeta (x)}{2}} 
	[\zeta (x)]^{\frac{\widetilde{\nu}_{n}+1}{2}} 
	L_n^{(\widetilde{\nu}_{n})} (\zeta (x)),
\end{equation}
with $\widetilde{\nu}_{n} = \nu_{n+1}$ and 
$\widetilde{\mathcal{N}}_n^2 = 
  \widetilde{\nu}_n\gamma \Gamma(n+1)/\Gamma(\widetilde{\nu}_n+n+1).$

\subsection{Shape invariance}

Let us consider the shape invariance technique 
(see Ref.~\onlinecite{Plastino-etal-1999,Amir-Iqbal-2016} for more details).
The partner Hamiltonians satisfy the integrability condition 
\begin{equation}
\label{eq:integrability-condition}
	\hat{H}_{\scriptscriptstyle -} ({\beta}_{j})
	- \hat{H}_{\scriptscriptstyle +}({\beta}_{j+1}) 
	= R({\beta}_{j}),
\end{equation}
where the set of parameters is 
related by a function $f$ such that 
$\beta_{j+1} = f(\beta_j)$, 
and the remainder term $R(\beta_j)$ is independent 
of the position and linear momentum operators.
Since the partner operators $\hat{H}_{\scriptscriptstyle \pm}$ 
differ only by an additive constant, 
their energy spectra and eigenstates are related respectively as
\begin{subequations}
\label{eq:energy_and_states_SI}
\begin{align}	
	E_n^{\scriptscriptstyle (-)} ({\beta}_{j})
	& = E_n^{\scriptscriptstyle (+)}({\beta}_{j+1}) 
	+ R({\beta}_{j}), \\
	|\psi_n^{\scriptscriptstyle (-)} (\beta_{j}) \rangle 
	& = |\psi_n^{\scriptscriptstyle (+)} (\beta_{j+1}) \rangle
	  = |\psi_n (\beta_{j+1}) \rangle.
\end{align}
\end{subequations}

The shape invariance method applied to 
the deformed oscillator leads to change 
the intertwining operators
(\ref{eq:a_gamma-and-a_gamma-dagger}) so that
\begin{subequations}
\label{eq:a_gamma-and-a_gamma-dagger_beta}
\begin{align}
\label{eq:a_gamma_beta}
\hat{a}_\gamma ({\beta})
	&= \sqrt{\frac{m_0 \omega_0}{2\hbar}}		
		\left[ \Phi_\gamma (\hat{x})
		- \frac{(\beta -1) \gamma \hbar }{2 m_0 \omega_0} \hat{1}
		+ \frac{i}{m_0\omega_0} \hat{\Pi}_\gamma 
		\right]
\end{align}
and
\begin{align}
\label{eq:a_gamma_dagger_beta}
\hat{a}_\gamma ({\beta})
	&= \sqrt{\frac{m_0 \omega_0}{2\hbar}}		
		\left[ \Phi_\gamma (\hat{x})
		- \frac{(\beta -1) \gamma \hbar}{2 m_0 \omega_0} \hat{1}
		- \frac{i}{m_0\omega_0} \hat{\Pi}_\gamma 
		\right].
\end{align}
\end{subequations}
The creation and annihilation operators (\ref{eq:a_gamma-and-a_gamma-dagger})
are recovered as $\beta \rightarrow 1$.
The supersymmetric partner Hamiltonians 
$
\hat{H}_{\scriptscriptstyle +} ({\beta}) =
\hbar \omega_0 \hat{a}_\gamma^{\dagger} ({\beta}) \hat{a}_\gamma ({\beta})
$
and
$
\hat{H}_{\scriptscriptstyle -} ({\beta}) = 
\hbar \omega_0 \hat{a}_\gamma ({\beta}) \hat{a}_\gamma^{\dagger} ({\beta}) 
$
are
$
\hat{H}_{\scriptscriptstyle \pm}  = 
\frac{1}{2m_0}\hat{\Pi}_\gamma^2 + V_{\scriptscriptstyle \pm}(\hat{x}, \beta )
$
with potentials 
$
V_{\scriptscriptstyle \pm}(\hat{x}, \beta ) =  V(\hat{x})
-\frac{\hbar \omega_0 (\beta \mp 1)}{2} \frac{\gamma \hat{x}}{\hat{1}+\gamma \hat{x}}
\mp \frac{\hbar \omega_0}{2} +\frac{\beta^2 \hbar^2 \gamma^2}{8m_0}.
$

The integrability condition 
(\ref{eq:integrability-condition}) for the deformed oscillator is
\begin{equation}
\hbar \omega_0 \hat{a}_\gamma ({\beta}_j) \hat{a}_\gamma^{\dagger} ({\beta}_j)
- \hbar \omega_0 \hat{a}_\gamma^{\dagger} ({\beta_{j+1}}) \hat{a}_\gamma({\beta_{j+1}})
= R(\beta_{j}),
\end{equation}
where the set $\beta$-parameters satisfy 
the translational shape invariance
$\beta_{j+1} = \beta_j + \varsigma$ and $\varsigma=2$
such that $\beta_n = \beta + 2(n-1)$
and $\beta_1 \equiv \beta$.
The remainder term is
$ 
R(\beta) = \hbar \omega_0 \left[ 
			1-\frac{1}{2}\gamma^2 \sigma_0^2 (\beta+1) 	
			\right].
$
The energy levels of the operator 
$\hat{H}_{\scriptscriptstyle +}({\beta})$ 
are given by
$E_n^{\scriptscriptstyle (+)}({\beta}) 
= \sum_{j=1}^{n} R({\beta}_j),$
with
$E_0^{\scriptscriptstyle (+)}({\beta}) = 0.$
It is straightforward to verify that
\begin{equation}
E_n^{\scriptscriptstyle (+)}({\beta}) = 
		\hbar \omega_0 n 
		\left[ 
			1 - \frac{\gamma^2 \sigma_0^2}{2} ( n+\beta ) 
		\right],
\end{equation}
and therefore,
the operator for deformed oscillator
$\hat{H}(\beta) = \hbar \omega_0 
\hat{a}_\gamma^{\dagger}(\beta) \hat{a}_\gamma(\beta) + E_0(\gamma)$ 
has the energy spectrum
$E_n = E_n^{\scriptscriptstyle (+)}({\beta}) + E_0 (\gamma).$

From Eq.~(\ref{eq:a_gamma-eigenfunctions}), 
the eigenstates 
$|\psi_n ({\beta}_j)\rangle$
satisfy the recurrence relation
\begin{equation}
|\psi_n ({\beta}_j)\rangle =
	\left[ 
		\frac{E_n^{\scriptscriptstyle (+)}}{\hbar \omega_0} 
	\right]^{-1/2}	
		\hat{a}_\gamma^{\dagger} ({\beta}_{j+1}) 
		|\psi_{n-1} (\beta_{j+1})\rangle.
\end{equation}
Applying $n$ interactions, we get
\begin{equation}
\label{eq:psi_n_a_dagger_n}
	|\psi_n (\beta_{1}) \rangle =
	\frac{1}{\sqrt{[n_\gamma ({\beta})]!}} 
	\hat{a}_\gamma^{\dagger} ({\beta}_1)
	\hat{a}_\gamma^{\dagger} ({\beta}_2)
	...
	\hat{a}_\gamma^{\dagger} ({\beta}_n)
	|\psi_0 (\beta_{n+1})\rangle
\end{equation}
with the deformed factorial given by
$[n_\gamma ({\beta})]! 
	= \frac{n!}{(2s)^n} 
	\frac{\Gamma (2s + 1 - \beta - n)}{\Gamma (2s + 1 - \beta - 2n)}.
$

The operator (\ref{eq:a_gamma_dagger_beta}) can be recasted as
\begin{align}
 \hat{a}_\gamma^{\dagger} ({\beta}) y (\zeta)
		&= \frac{\gamma \sigma_0}{\sqrt{2}}
		\left[ \frac{1}{\gamma^2 \sigma_0^2}
		-\frac{\beta +1}{2} - \frac{\zeta}{2} + \zeta \frac{\textrm{d}}{\textrm{d} \zeta}
		\right] y (\zeta)
		\nonumber \\
		&= \frac{\gamma \sigma_0}{\sqrt{2}} 
			\left[ \frac{1}{h(\zeta)}
			\left( \zeta \frac{\textrm{d}}{\textrm{d}\zeta} \right)
			h (\zeta) \right] y (\zeta)
\end{align}
with $y(\zeta)$ being a generic function and
$h (\zeta (x)) = 
e^{-\frac{\zeta (x)}{2}} [\zeta(x)]^{\frac{2s-\beta+1}{2}}.$
From that, we have
\begin{equation}
\label{eq:product_a_dagger}
\prod_{j=1}^n \hat{a}_\gamma^{\dagger} ({\beta}_j) y (\zeta)
	= \left( \frac{\gamma \sigma_0}{\sqrt{2}} \right)^{n}
	\left[ \frac{1}{h(\zeta)} \left(
	\zeta^n \frac{\textrm{d}^n}{\textrm{d}\zeta^n} 
	\right) h(\zeta)  \right] y (\zeta).
\end{equation}
The condition 
$\hat{a}_\gamma ({\beta}) |\psi_{0} ({\beta}) \rangle = 0$ 
leads to the ground state
\begin{equation}
\label{eq:SI_ground_state}
\psi_{0,\beta} (x)  = \frac{1}{\sqrt{2s}}
		\sqrt{ \frac{\gamma }{\Gamma ( 2s - \beta )} }
		e^{-\frac{\zeta (x)}{2}} [\zeta (x)]^{\frac{2s-\beta+1}{2}}.
\end{equation}
Substituting (\ref{eq:SI_ground_state}) into (\ref{eq:psi_n_a_dagger_n}),
and using Rodrigues' formula $L_n^{(\nu)}(\zeta) = \frac{1}{n!}e^\zeta \zeta^{-\nu} 
\frac{\textrm{d}^n}{\textrm{d}\zeta^n} (\zeta^{\nu+n} e^{-\zeta}),$
the eigenfunctions for the Hamiltonian $\hat{H}(\beta)$ are given by
\textcolor{red}{
\begin{align}
\psi_{n,\beta} (x) 
	&=\frac{(-1)^n}{\sqrt{2s}}\left[ 
   \frac{(\nu_n + 1 - \beta ) \gamma \Gamma (n+1) }{
      \Gamma (\nu_n + 2 - \beta + n)} 
    \right]^{1/2}
   e^{-\frac{\zeta(x)}{2}} [\zeta (x)]^{\frac{\nu_n - \beta + 2}{2}}
   L_n^{( \nu_n + 1 - \beta )}(\zeta (x)),
\end{align}
}
where the expression above reduces to the 
eigenfunctions (\ref{eq:egeinfucntions-osc})
as $\beta \rightarrow 1$.

The commutator between the $\hat{a}_\gamma(\beta)$ and 
$\hat{a}_\gamma^{\dagger}(\beta)$ depends on the position,
so they can not be chosen as ladder operators.
Nonetheless, the Hamiltonian operators 
$\hat{H}_{\scriptscriptstyle \pm} ({\beta})$
are translational shape invariance, and in this case 
the ladder operators are defined as
\begin{equation}
\label{eq:ladder_operators}
\hat{L}_{\scriptscriptstyle -} (\beta) = 
\hat{\mathcal{T}}^{\dagger}({\beta})\hat{a}_{\gamma}({\beta})
\quad \textrm{and} \quad
\hat{L}_{\scriptscriptstyle +}(\beta) =  
\hat{a}_{\gamma}^{\dagger} ({\beta}) \hat{\mathcal{T}}({\beta})
\end{equation}
with $\hat{\mathcal{T}}^{\dagger}({\beta})$ 
and $\hat{\mathcal{T}}({\beta})$ 
being unitary translational operators 
on parameter $\beta$, 
which satisfy the reparameterization
$\hat{\mathcal{T}} (\beta) |\psi_n (\beta)\rangle = 
|\psi_n (\beta + \varsigma)\rangle.$ 
The translational operators 
$\hat{\mathcal{T}}({\beta})$ and
$\hat{\mathcal{T}}^{\dagger}({\beta})$ 
are given, respectively, by 
\cite{Amir-Iqbal-2016}
\begin{equation}
\hat{\mathcal{T}}({\beta}) = 
	\exp \left( \varsigma \frac{\partial}{\partial \beta} \right) 
\ \ \textrm{and} \ \
\hat{\mathcal{T}}^{\dagger}({\beta}) = 
	\exp \left( - \varsigma \frac{\partial}{\partial \beta} \right). 
\end{equation}

Once $\hat{\mathcal{T}}^{\dagger} (\beta) \hat{\mathcal{T}} (\beta) = \hat{1}$
the factorization for deformed oscillator preserves the form
$
\hat{H}_{\scriptscriptstyle +} ({\beta})= \hat{H}(\beta) -E_0(\gamma) =  
\hbar \omega_0 \hat{L}_{\scriptscriptstyle +}(\beta)
\hat{L}_{\scriptscriptstyle -}(\beta).
$
From Eqs.~(\ref{eq:energy_and_states_SI}) and
(\ref{eq:ladder_operators}),  
the action of the ladder operators on the ket vectors 
$|\psi_n (\beta) \rangle$ is given by 
\begin{subequations}
\label{eq:eq:ladder_operators_C_n}
\begin{align}
&\hat{L}_{\scriptscriptstyle -} (\beta) |\psi_n ({\beta})\rangle =
	\sqrt{n \left[ 1-\frac{\gamma^2 \sigma_0^2}{2} ( n+\beta ) \right]}
	|\psi_{n-1} ({\beta})\rangle,
\end{align}
and
\begin{align}
&\hat{L}_{\scriptscriptstyle +} (\beta) |\psi_n ({\beta})\rangle =
\sqrt{(n+1) \left[ 1-\frac{\gamma^2 \sigma_0^2}{2} ( n+1+\beta ) \right]}
	|\psi_{n+1} ({\beta})\rangle.
\end{align}
\end{subequations}
The ladder operators have the form
\begin{subequations}
\label{eq:eq:ladder_operators_explicitly}
\begin{align}
&\hat{L}_{\scriptscriptstyle -}  ({\beta}) = 
	e^{-\varsigma \frac{\partial}{\partial \beta}}
	\frac{1}{\sqrt{2} \sigma_0}
		\left[ \frac{x}{1+\gamma x}
		-\frac{(\beta - 1) \gamma \sigma_0^2}{2} 
		+ \sigma_0^2 (1+\gamma x) \frac{\textrm{d}}{\textrm{d}x}
		\right]
\\
&\hat{L}_{\scriptscriptstyle +}  ({\beta}) = 
	\frac{1}{\sqrt{2} \sigma_0}
		\left[ \frac{x}{1+\gamma x}
		-\frac{(\beta + 1) \gamma \sigma_0^2}{2} 
		- \sigma_0^2 (1+\gamma x) \frac{\textrm{d}}{\textrm{d}x}
		\right]
	e^{\varsigma \frac{\partial}{\partial \beta}},
\end{align}
\end{subequations}
and the effects of the ladder operators on
wavefunctions are 
\begin{align}
\hat{L}_{\scriptscriptstyle \pm} ({\beta})
\psi_{n,\beta} (x)
=& \frac{1}{\sqrt{2}\sigma_0} 
	\left[ 
		\sqrt{{\frac{(2s-2n -{\beta} \mp 2)(2s-n -{\beta} \mp 1)^{\pm 1}}{
		 			 (2s-2n-{\beta})(2s-n-{\beta})^{\pm 1}}}}
	\frac{2s-2n -{\beta} \mp 1}{2s}
	\right] 
	\nonumber \\ 
	& 
	\left\{ \frac{1}{\gamma} 
	\left[
		\frac{2s + 1 - \beta }{2s - 2n  -\beta \mp 1}
		- \frac{2s-2n -\beta \mp 1}{2s} \left( 1+\gamma x \right)
	\right]
	\pm \sigma_0^2 (1+\gamma x)^2 \frac{\textrm{d}}{ \textrm{d} x}
	\right\}
	\psi_{n,\beta} (x).
\end{align}
The wavefunction for ground state
(\ref{eq:SI_ground_state}) 
is obtained from 
$\hat{L}_{\scriptscriptstyle -} (\beta)\psi_{0,\beta}(x) = 0,$
and the $n$th excited state is expressed by
\begin{equation}
\psi_{n,\beta} (x) = 
	\frac{1}{\sqrt{[n_\gamma ({\beta})]!}}
	[\hat{L}_{\scriptscriptstyle +}({\beta})]^n \psi_{0,\beta}(x).
\end{equation}

The action of the commutator 
$[\hat{L}_{\scriptscriptstyle -}, \hat{L}_{\scriptscriptstyle +}]$ 
for the case $\beta=1$ 
on the eigenfunctions is
$[\hat{L}_{\scriptscriptstyle -}, 
\hat{L}_{\scriptscriptstyle +}] \psi_n(x) = 
\left[ 1 - \gamma^2 \sigma_0^2 (n+1) \right] \psi_n(x),
$
so that we can introduce the operator
$
\hat{L}_{0} =\frac{1}{2} [ 1 - \gamma^2 \sigma_0^2 (\hat{n}+1) ].
$
From the operators
$\hat{M}_{\scriptscriptstyle \pm} = \sqrt{2s} \hat{L}_{\scriptscriptstyle \pm}$
and
$\hat{M}_{0} = 2s \hat{L}_{0}$,
we obtain the following commutation relations
\begin{equation}
[\hat{M}_{\scriptscriptstyle +}, \hat{M}_{\scriptscriptstyle -}] =
2\hat{M}_{0}
\quad \textrm{and}\quad
[\hat{M}_{0}, \hat{M}_{\scriptscriptstyle \pm}] =
\pm \hat{M}_{\scriptscriptstyle \pm},
\end{equation}
which corresponds to a Lie algebra $SU(1,1)$ for
the deformed oscillator.

\subsection{Coherent states and minimum relation of uncertainty}

By means of the canonical transformation
$(\hat{x}, \hat{p}) \rightarrow (\hat{x}_\gamma, \hat{\Pi}_\gamma)$,
the asymmetric deformed oscillator with PDM is mapped into a Morse oscillator, 
and the coherent states for the latter are well known. 
Coherent states for the Morse oscillator 
were previously investigated in different approaches:
minimum-uncertainty,\cite{Nieto-SimmonsJr-1979}
Perelomov generalized definition,\cite{Perelomov-1986}
time dilatation operator,\cite{Kais-Levine-1990}
algebraic method,\cite{Copper-1992}
Gazeau-Klauder and Barut-Girardello approaches,\cite{Roy-Roy-2002,Popov-Dong-Pop-Sajfert-Simon-2013}
generalized and Gaussian coherent states,\cite{Angelova-Hussin-2008}
generalized Heisenberg algebra and su(2)-like approach,\cite{Belfakir-Hassouni-Curado-2020}
etc.

For the sake of simplicity,
following the concept of coherent states for 
the usual oscillator introduced by Glauber,\cite{Glauber-1963} 
the coherent states for the PDM oscillator 
are eigenstates of the annihilation operator\cite{Ruby-Senthilvelan-2010}
\begin{equation}
\hat{a}_\gamma |\alpha_\gamma \rangle 
= \alpha_\gamma| \alpha_\gamma \rangle. 
\end{equation}
Let the wavefunction $\psi_{\textrm{cs}}(x) = \langle{x}|{\alpha_\gamma} \rangle$
for the coherent states in the position representation $\{| \hat{x} \rangle \}$.
By means of Eq.~(\ref{eq:a_gamma}), we obtain
\begin{equation}
\label{eq:CS-differential-equation}
\frac{1}{\sqrt{2}\sigma_0} \left[
	\frac{x}{1+\gamma x}+\sigma_0^2 (1+\gamma{x})\frac{\textrm{d}}{\textrm{d}x}
	\right] \psi_{\textrm{cs}}(x) = \alpha_\gamma \psi_{\textrm{cs}}(x).
\end{equation}
Solving Eq.~(\ref{eq:CS-differential-equation}), 
we obtain that the wavefunctions $\psi_{\textrm{cs}}(x)$ are similar 
to the ground state (\ref{eq:ground-state}), i.e., 
\begin{equation}
\label{eq:psi_cs(x)}
\psi_{\textrm{cs}}(x) = 
		\frac{\mathcal{N}_{\textrm{cs}}}{\sqrt{2s}}
		e^{-\frac{\zeta(x)}{2}} [\zeta (x)]^{\frac{1}{\gamma^2\sigma_0^2}
		-\frac{\sqrt{2}\alpha_\gamma}{\gamma\sigma_0}},
\end{equation}
with 
$
\mathcal{N}_{\textrm{cs}}^2 =
 \gamma /\Gamma (\lambda_{\textrm{cs}} (\alpha_{\gamma}))
$
and
$\lambda_{\textrm{cs}} (\alpha_{\gamma}) 
	= \frac{2}{\gamma^2 \sigma_0^2} 
	  -\frac{\sqrt{2}}{\gamma \sigma_0} (\alpha_\gamma^{\ast}+\alpha_{\gamma}) -1.$
The probability density for coherent states in position representation
behaves like the Gamma distribution (\ref{eq:Gamma_distribuition}),
but withe the shape parameter $\lambda$ replaced by $\lambda_{\textrm{cs}} (\alpha_{\gamma})$, 
i.e.,
$
\rho_{\textrm{cs}} (x) =
	\frac{\gamma}{2s}
	\frac{1}{\Gamma (\lambda_{\textrm{cs}} (\alpha_{\gamma}))} 
    e^{-\zeta(x)} [\zeta(x)]^{\lambda_{\textrm{cs}} (\alpha_{\gamma})+1}.
$

\begin{figure}[hbt]
\centering
\includegraphics[width=0.32\linewidth]{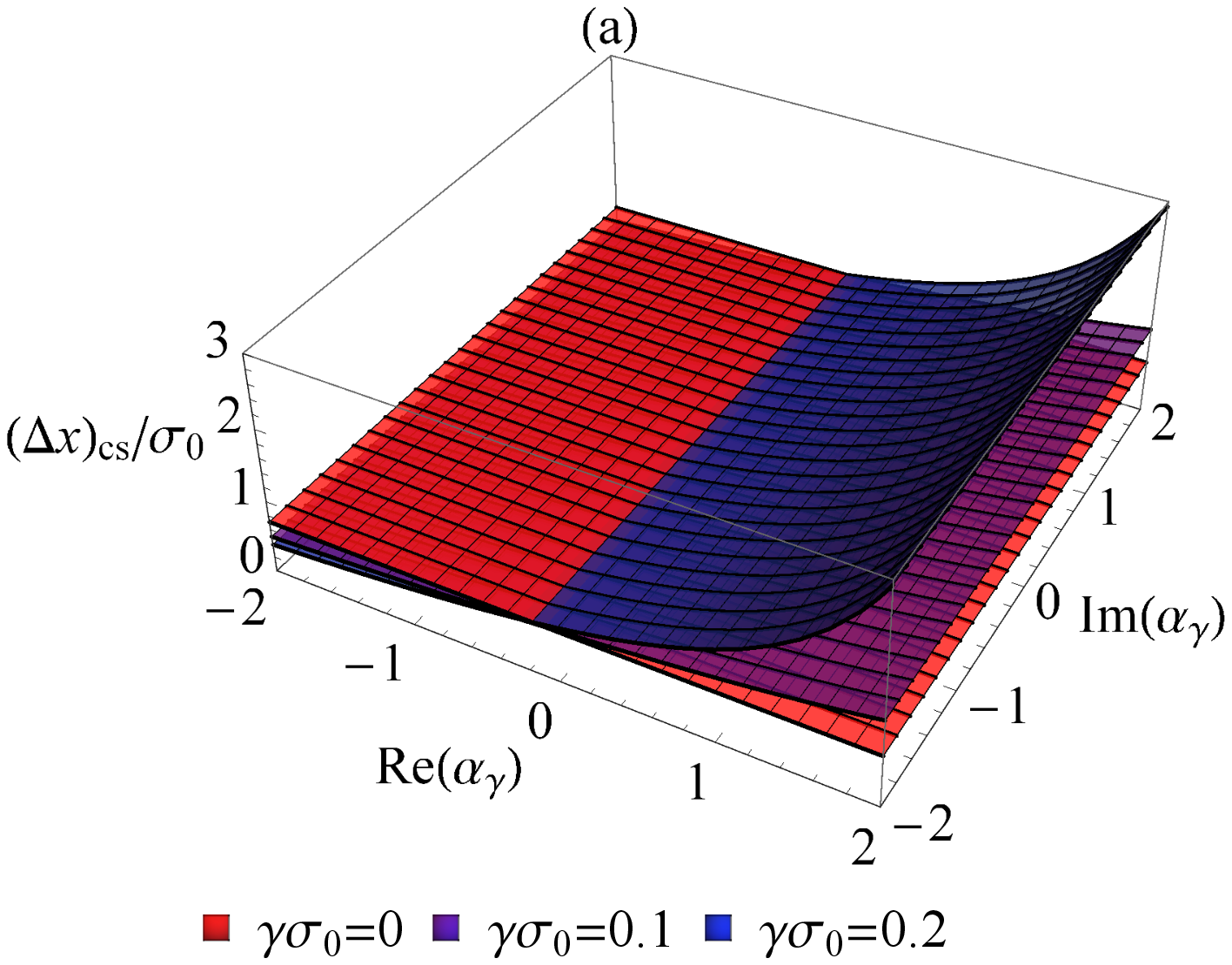}
\includegraphics[width=0.32\linewidth]{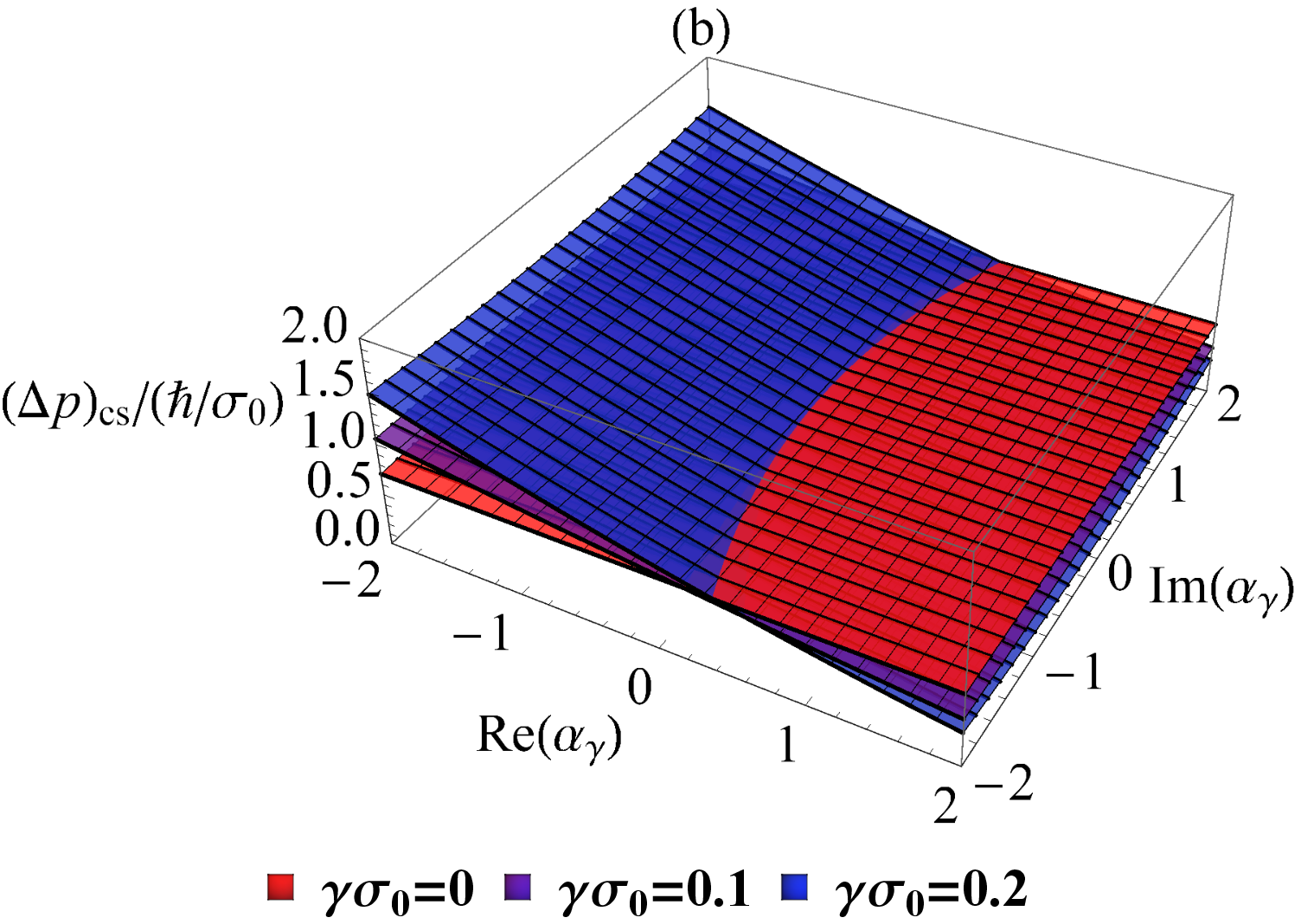}
\includegraphics[width=0.32\linewidth]{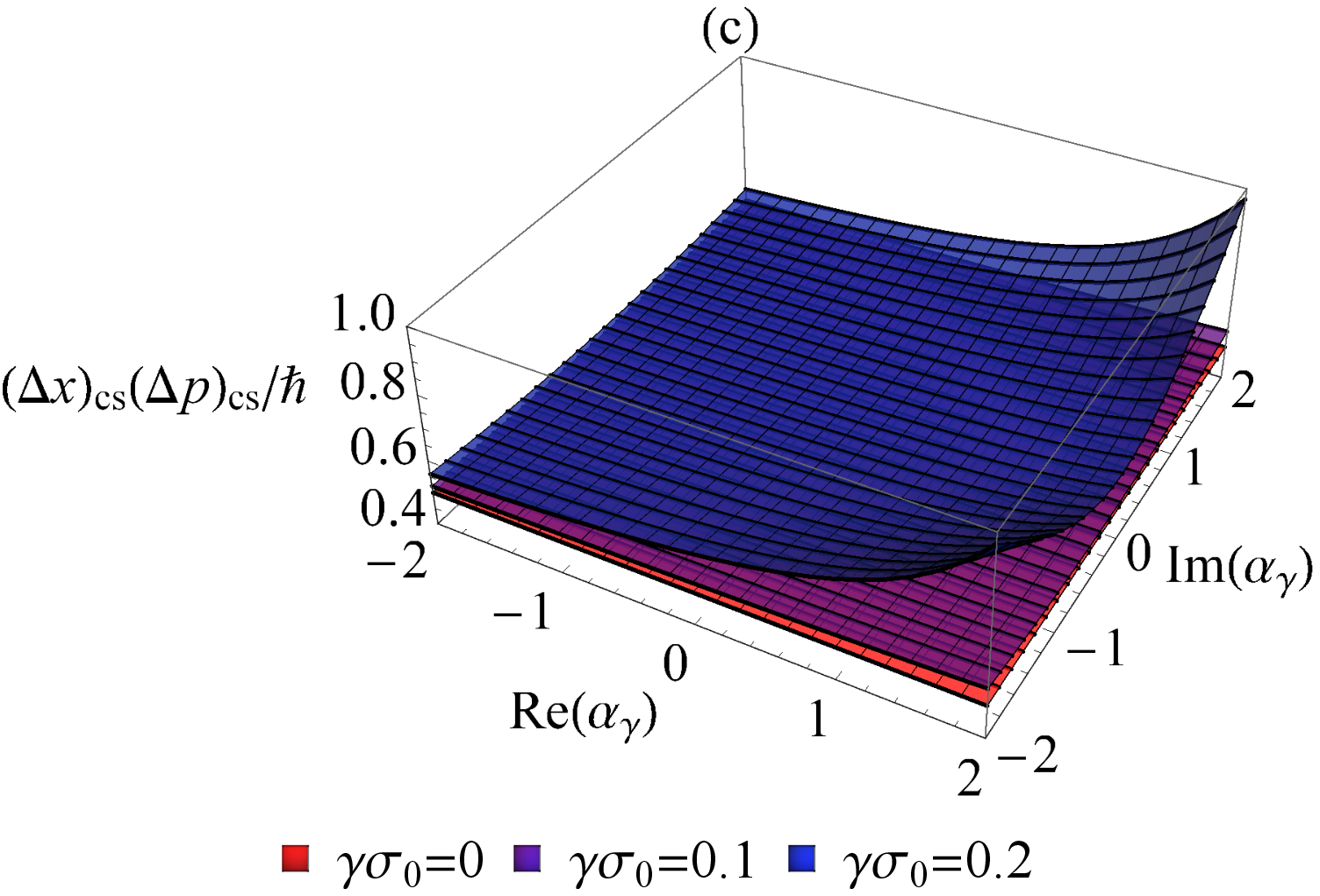}
\caption{\label{fig:8}
Uncertainties of 
(a) the position $(\Delta x)_{\textrm{cs}}$,
(b) the linear momentum $(\Delta p)_{\textrm{cs}}$ 
and (c) the product 
 $(\Delta x)_{\textrm{cs}} (\Delta p)_{\textrm{cs}}$ 
for the coherent states of the asymmetric deformed oscillator
as function of the real and imaginary parts of $\alpha_\gamma$,
with $\{ |\textrm{Re}(\alpha_\gamma)|,|\textrm{Im}(\alpha_\gamma)|<2 \}$.
The surfaces are plotted for deformation parameter values
$\gamma \sigma_0 = 0$ (usual case -- red planes),
$0.1$ (violet surfaces) and
$0.2$ (blue surfaces).
}
\end{figure}

The expected values 
$\langle \hat{x} \rangle_{\textrm{cs}}$,
$\langle \hat{x}^2 \rangle_{\textrm{cs}}$,
$\langle \hat{p} \rangle_{\textrm{cs}}$,
and $\langle \hat{p}^2 \rangle_{\textrm{cs}}$
for coherent states are given by
\begin{subequations}
\label{eq:x-p-x^2-p^2-cs}
\begin{align}
\label{eq:expected-value-x-cs}
\langle \hat{x} \rangle_{\textrm{cs}} & = \frac{1}{\gamma} 
\left[
	\frac{1}{1 - \frac{\gamma \sigma_0}{\sqrt{2}} (\alpha_\gamma^{\ast} + \alpha_\gamma) - \gamma^2 \sigma_0^2}
	-1 \right] ,\\
\label{eq:expected-value-x^2-cs}
\langle \hat{x}^2 \rangle_{\textrm{cs}} &= \frac{1}{\gamma^2} \left\{
	\frac{1}{\left[ 
			  1 - \frac{\gamma \sigma_0}{\sqrt{2}} (\alpha_\gamma^{\ast} + \alpha_\gamma) 
			  - \gamma^2 \sigma_0^2 
			\right] \left[ 
				1 - \frac{\gamma \sigma_0}{\sqrt{2}} (\alpha_\gamma^{\ast} + \alpha_\gamma)
			  - \frac{3}{2} \gamma^2 \sigma_0^2 
			\right]}
	-\frac{2}{1 
			- \frac{\gamma \sigma_0}{\sqrt{2}} (\alpha_\gamma^{\ast} + \alpha_\gamma)
			- \gamma^2 \sigma_0^2}
	+1 \right\},
	\\
\label{eq:expected-value-p-cs}
 \langle \hat{p} \rangle_{\textrm{cs}} &=
\frac{i\hbar}{\sqrt{2}\sigma_0} 
	(\alpha_{\gamma}^{\ast} - \alpha_{\gamma})
		\left[ 1- \frac{\gamma \sigma_0}{\sqrt{2}} (\alpha_\gamma^{\ast} + \alpha_\gamma) 
		- \frac{\gamma^2 \sigma_0^2}{2}\right],\\
\label{eq:expected-value-p^2-cs}
\langle \hat{p}^2 \rangle_{\textrm{cs}} & =
	\frac{\hbar^2}{2\sigma_0^2} 
		\left[ 1-\frac{\gamma \sigma_0}{\sqrt{2}} (\alpha_\gamma^{\ast} + \alpha_\gamma) \right] 
		\left[ 1-\frac{\gamma \sigma_0}{\sqrt{2}} (\alpha_\gamma^{\ast} + \alpha_\gamma)
			- \frac{\gamma^2 \sigma_0^2}{2}\right]
		\left[ 1 - (\alpha_{\gamma}^{\ast} - \alpha_{\gamma})^2
				- \frac{\gamma \sigma_0}{\sqrt{2}} (\alpha_\gamma^{\ast} + \alpha_\gamma)
				+ \gamma^2 \sigma_0^2 \right].
\end{align}
\end{subequations}
In figure~\ref{fig:8}, from Eqs.~(\ref{eq:x-p-x^2-p^2-cs})
we plot the uncertainties of the position $(\Delta x)_{\textrm{cs}}$ 
and of the linear momentum $(\Delta p)_{\textrm{cs}}$,
as well as the product $(\Delta x)_{\textrm{cs}}(\Delta p)_{\textrm{cs}}$
for the coherent states $|\alpha_\gamma \rangle$
within the region 
$|\textrm{Re}(\alpha_\gamma)|,|\textrm{Im}(\alpha_\gamma)|<2$.
It can be seen that these three quantities are symmetric around
$\textrm{Im}(\alpha_\gamma)=0$.
This feature stems from the spatial deformation caused by PDM, 
where the real part of the variable $\alpha_\gamma$ 
is a function only of the position.
The surfaces for the usual case ($\gamma \sigma_0 = 0$) 
correspond to planes, which become curved surfaces 
due to the effect of spatial deformation.
While $(\Delta x)_\textrm{cs}$ increases as $\textrm{Re}(\alpha_\gamma)$ varies 
within the interval $[-2, 2]$, $(\Delta p)_\textrm{cs}$ decreases,
and the uncertainty relation 
$(\Delta x)_\textrm{cs} (\Delta p)_\textrm{cs} \geq \frac{\hbar}{2}$ 
is kept.

In addition, the expected values of the 
superpotential 
($\langle \hat{\Phi}_\gamma \rangle_{\textrm{cs}}$
and $\langle \hat{\Phi}_\gamma^2 \rangle_{\textrm{cs}}$)
and linear pseudomomentum 
($\langle \hat{\Pi}_\gamma \rangle_{\textrm{cs}}$
and $\langle \hat{\Pi}_\gamma^2 \rangle_{\textrm{cs}}$)
for coherent states are
\begin{subequations}
\label{eq:Pi-pi^2-cs}
\begin{align}
\label{eq:cs_expected_values_Phi}
\langle \hat{\Phi}_\gamma \rangle_{\textrm{cs}} &= 
	\frac{\sigma_0}{\sqrt{2}} 
	(\alpha_{\gamma}^{\ast} + \alpha_{\gamma}), \\
\label{eq:cs_expected_values_Phi^2}
\langle \hat{\Phi}_\gamma^2 \rangle_{\textrm{cs}} &=
	\frac{\sigma_0^2}{2} 
	\left[ 1+(\alpha_{\gamma}^{\ast} + \alpha_{\gamma})^2 
	 - \frac{\gamma \sigma_0}{\sqrt{2}} 
	(\alpha_\gamma^{\ast} + \alpha_\gamma)
	- \frac{\gamma^2 \sigma_0^2}{2} \right], \\
\label{eq:cs_expected_values_Pi}
\langle \hat{\Pi}_\gamma \rangle_{\textrm{cs}} &= 
	\frac{i\hbar}{\sqrt{2}\sigma_0} 
	(\alpha_{\gamma}^{\ast} - \alpha_{\gamma}), \\
\label{eq:cs_expected_values_Pi^2}
\langle \hat{\Pi}_\gamma^2 \rangle_{\textrm{cs}} &=
	\frac{\hbar^2}{2 \sigma_0^2} 
	\left[ 1-(\alpha_{\gamma}^{\ast} - \alpha_{\gamma})^2 
	 - \frac{\gamma \sigma_0}{\sqrt{2}} 
	(\alpha_\gamma^{\ast} + \alpha_\gamma)
	- \frac{\gamma^2 \sigma_0^2}{2} \right].
\end{align}
\end{subequations}
It is easy to check that the results of a oscillator with constant mass 
are recovered:
$\lim_{\gamma \rightarrow 0} \langle \hat{x} \rangle_{\textrm{cs}} =
\lim_{\gamma \rightarrow 0} \langle \hat{\Phi}_\gamma \rangle_{\textrm{cs}} = 
\frac{\sigma_0}{\sqrt{2}} (\alpha_0^{\ast} + \alpha_0)$,
$\lim_{\gamma \rightarrow 0} \langle \hat{x}^2 \rangle_{\textrm{cs}} = 
\lim_{\gamma \rightarrow 0} \langle \hat{\Phi}_\gamma^2 \rangle_{\textrm{cs}} = 
\frac{\sigma_0^2}{2} [ 1 + (\alpha_0^{\ast} + \alpha_0)^2]$,
$\lim_{\gamma \rightarrow 0} \langle \hat{p} \rangle_{\textrm{cs}} = 
\lim_{\gamma \rightarrow 0} \langle \hat{\Pi}_\gamma \rangle_{\textrm{cs}}=
\frac{i\hbar}{\sqrt{2}\sigma_0} (\alpha_0^{\ast} - \alpha_0)$
and
$\lim_{\gamma \rightarrow 0} \langle \hat{p}^2 \rangle_{\textrm{cs}} = 
\lim_{\gamma \rightarrow 0} \langle \hat{\Pi}_\gamma^2 \rangle_{\textrm{cs}} =
\frac{\hbar^2}{2\sigma_0^2} [ 1 - (\alpha_0^{\ast} - \alpha_0)^2]$.

From Eqs.~(\ref{eq:x-p-x^2-p^2-cs})
and (\ref{eq:Pi-pi^2-cs}), 
we readily obtain that the dispersions 
$(\Delta x )_{\textrm{cs}}^2 =  
\langle \hat{x}^2 \rangle_{\textrm{cs}}
-\langle \hat{x} \rangle_{\textrm{cs}}^2$
and
$(\Delta \Pi_\gamma)_{\textrm{cs}}^2 =  
\langle \hat{\Pi}_\gamma^2 \rangle_{\textrm{cs}}
-\langle \hat{\Pi}_\gamma \rangle_{\textrm{cs}}^2$
are expressed by
\begin{subequations}
\label{eq:Dx_DPi_quad_cs}
\begin{align}
(\Delta x )_{\textrm{cs}}^2 &= \frac{\sigma_0^2}{2} 
		\left( 1 +\gamma \langle \hat{x} \rangle_{\textrm{cs}} \right)^2
		\left( \frac{1}{1+\gamma \langle \hat{x} \rangle_{\textrm{cs}}}
		-\frac{\gamma^2 \sigma_0^2}{2}
		\right)^{\! \! -1} \\
(\Delta \Pi_\gamma )_{\textrm{cs}}^2 &= \frac{\hbar^2}{2\sigma_0^2} 
		\left( 
		\frac{1}{1+\gamma \langle \hat{x} \rangle_{\textrm{cs}}}
		+\frac{\gamma^2 \sigma_0^2}{2}
		\right) 
\end{align}
\end{subequations}
In the limit $|\alpha_\gamma| / \gamma \sigma_0  \gg 1$ 
(or $\sigma_0 \rightarrow 0$), 
we have that the squared of the deviations
(\ref{eq:Dx_DPi_quad_cs}) behave like
$(\Delta x )_{\textrm{cs}}^2 \rightarrow  \frac{\sigma_0^2}{2}
( 1 +\gamma \langle \hat{x} \rangle_{\textrm{cs}})^3$
and
$(\Delta \Pi_\gamma )_{\textrm{cs}}^2 \rightarrow  \frac{\sigma_0^2}{2}
( 1 +\gamma \langle \hat{x} \rangle_{\textrm{cs}})^{-1}$,
so that 
$(\Delta x )_{\textrm{cs}} (\Delta \Pi_\gamma )_{\textrm{cs}}
= \frac{\hbar}{2} (1+\gamma \langle \hat{x} \rangle_{\textrm{cs}})$.
Therefore, the coherent states minimize the generalized uncertainty relation
between $\hat{x}$ and $\hat{\Pi}_\gamma$
(Eq.~(\ref{eq:GUP}) for $g(\hat{x}) = \hat{1} + \gamma \hat{x}$).

Because of the importance of the superpotential operator 
together with linear pseudomomentum in the factorization of $\hat{H}$, 
we also calculate, for the sake of completeness, the expected values
$\langle \hat{\Phi}_\gamma \rangle$ and $\langle \hat{\Phi}_\gamma^2 \rangle$
for the eigenstates $\psi_n(x)$, such that
\begin{subequations}
\label{eq:Phi-expected-value-psi_n}
\begin{align}
\langle \hat{\Phi}_\gamma \rangle &= n \gamma \sigma_0^2, \\
\langle \hat{\Phi}_\gamma^2 \rangle &= 
	\sigma_0^2 \left[ 
		(1-\gamma^2 \sigma_0^2 ) \left( n + \frac{1}{2} \right) 
		+ \frac{\gamma^2 \sigma_0^2}{4}				
	\right],
\end{align}
\end{subequations}
whose mean square deviation is
$
(\Delta \Phi_\gamma)^2 = 
	\sigma_0^2 \left( n + \frac{1}{2} \right) \left[ 
		1-\gamma^2 \sigma_0^2  \left( n + \frac{1}{2} \right) 
	\right].
$
From Eqs.~(\ref{eq:Pi-expected-value-psi_n}) and (\ref{eq:Phi-expected-value-psi_n}),
the product of the uncertainties of $\hat{\Phi}_\gamma$ and $\hat{\Pi}_\gamma$ 
for the eigenstates satisfies
\begin{align}
\Delta \Phi_\gamma \Delta \Pi_\gamma
&= \hbar  \left( n + \frac{1}{2} \right)  \left[ 
		1-\gamma^2 \sigma_0^2  \left( n + \frac{1}{2} \right) 
	\right] 
	\nonumber \\
&= \hbar \left( n + \frac{1}{2} \right)
  \left\langle \frac{1}{\hat{1}+\gamma \hat{x}} \right\rangle,
\end{align}
i.e.,
$\Delta \Phi_\gamma \Delta \Pi_\gamma
\geq \frac{\hbar}{2} \left\langle \frac{1}{\hat{1}+\gamma \hat{x}} \right\rangle.$
On the other hand, from Eqs.~(\ref{eq:Pi-pi^2-cs}) we arrive at
\begin{align}
(\Delta \Phi_\gamma)_{\textrm{cs}} (\Delta \Pi_\gamma)_{\textrm{cs}}
&=\frac{\hbar}{2} \left[ 1 - \frac{\gamma \sigma_0}{\sqrt{2}} 
	(\alpha_\gamma^{\ast} + \alpha_\gamma)
	- \frac{\gamma^2 \sigma_0^2}{2} \right]  \nonumber \\
&= \frac{\hbar}{2}
\left\langle \frac{1}{\hat{1}+\gamma \hat{x}} \right\rangle_{\textrm{cs}}.
\end{align}
Thus, the coherent states minimize the uncertainty relation 
between the superpotential and the linear pseudomomentum
in agreement with Ref.~\onlinecite{Ruby-Senthilvelan-2010}. 
In particular, the minimization of the uncertainty product 
$\Delta \Phi_\gamma \Delta \Pi_\gamma$
for the coherent states of the Morse oscillator 
was realized by Cooper through the algebraic method
(see Ref.~\onlinecite{Copper-1992} for more details).
Here, we verify that this property 
can also be extended to the asymmetric oscillator with PDM.

\subsection{Dynamics of quasi-classical states}
 
In order to calculate the time evolution of coherent states
$|\alpha_\gamma (t) \rangle$, 
let us start with the equation of motion 
of the operator $\hat{a}_\gamma$,
\begin{equation}
\frac{\textrm{d}}{\textrm{d}t}\hat{a}_\gamma
	= \frac{1}{i\hbar} [\hat{a}_\gamma, \hat{H}]
	= -i\omega_0 \left(\frac{1}{1+\gamma \hat{x}} \right)\hat{a}_\gamma.
\end{equation}
Considering
$\hat{a}_\gamma (t) = 
\hat{a}_\gamma (t_0) e^{-i\Theta_{\textrm{cs},\gamma} (t)}$
the temporal rate of the phase
$\Theta_{\textrm{cs},\gamma} (t)$
for coherent states becomes
\begin{align}
\label{eq:time-derivative-deformed-phase_CS}
\frac{\textrm{d}\Theta_{\textrm{cs},\gamma} (t)}{\textrm{d}t}  
	&= \omega_0 \left\langle
	\frac{1}{\hat{1} +\gamma \hat{x}}
	\right\rangle_{\textrm{cs}}
	\nonumber \\
	&= \omega_0 \left[
		1-\sqrt{2} \gamma \sigma_0 |\alpha_\gamma | \cos \Theta_{\textrm{cs},\gamma} (t)
		-\frac{\gamma^2 \sigma_0^2}{2}
	\right],
\end{align}
with
$\alpha_\gamma^{\ast}(t)+\alpha_\gamma(t)=
2|\alpha_\gamma| \cos \Theta_{\textrm{cs},\gamma} (t)$ and $\Theta_{\textrm{cs},\gamma}(t_0)=0$.
Like the classic oscillator, by integrating 
Eq.~(\ref{eq:time-derivative-deformed-phase_CS}), we arrive at
\begin{equation}
\label{eq:deformed-phase_CS}
\Theta_{\textrm{cs},\gamma} (t) = 
	2\textrm{tan}^{-1} \left\{ 
	\sqrt{\frac{1 - \sqrt{2} \gamma \sigma_0 |\alpha_\gamma| -\frac{\gamma^2 \sigma_0^2}{2}}{
		         1 + \sqrt{2} \gamma \sigma_0 |\alpha_\gamma| -\frac{\gamma^2 \sigma_0^2}{2}}}
	\textrm{tan} \left[ 
	\frac{1}{2} \Omega_{{\textrm{cs}},\gamma} (t-t_0)
	\right] \right\}
\end{equation}
and oscillation frequency
$\Omega_{\textrm{cs},\gamma} = \omega_0 
\sqrt{\left(1-\frac{\gamma^2 \sigma_0^2}{2}\right)^2 
            - 2\gamma^2 \sigma_0^2 |\alpha_\gamma|^2}.$

\begin{figure}[htb]
\centering
\includegraphics[width=0.36\linewidth]{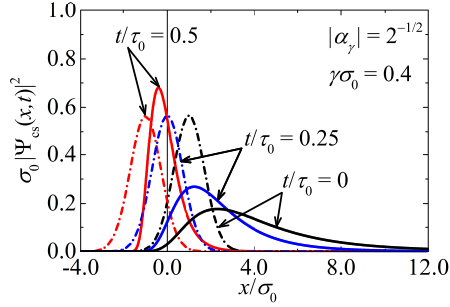}
\caption{\label{fig:9}
Snapshots of the probability density $|\Psi_{\textrm{cs}} (x,t)|^2$ 
for coherent states $| \alpha_\gamma \rangle$ 
of the asymmetric oscillator with PDM and deformation parameter 
$\gamma \sigma_0 = 0.4$ and $|\alpha_\gamma| = 1/\sqrt{2}$.
The frames are represented at the times 
$t=0$, $\tau_0/4$ and $\tau_0/2$ ($\tau_0 = 2\pi/\omega_0$).
The usual case ($\gamma = 0$, dashed-dotted line) is also shown 
for comparison.
}
\end{figure}

The temporal evolution of wavefunction $\Psi_{\textrm{cs}} (x,t)$ 
for coherent states is obtained by making the change
$\alpha_\gamma \rightarrow \alpha_\gamma (t) 
= |\alpha_\gamma| e^{-i\Theta_{\textrm{cs},\gamma} (t)}$
in Eqs.~(\ref{eq:psi_cs(x)})
with the deformed phase $\Theta_{\textrm{cs},\gamma}(t)$ given by
Eq.~(\ref{eq:deformed-phase_CS}).
In figure~\ref{fig:9} we plot three frames
of the motion of the probability density 
$|\Psi_{\textrm{cs}} (x, t)|^2$ with the deformation parameter 
$\gamma \sigma_0 = 0$ (standard oscillator) and $0.4$.

Since
$\textrm{Re} [\alpha_\gamma (t)] =  
|\alpha_\gamma| \cos \Theta_\gamma (t)$
and
$ \textrm{Im} [\alpha_\gamma (t)] =  
|\alpha_\gamma| \sin \Theta_\gamma (t),$
from Eqs.~(\ref{eq:expected-value-x-cs}),
(\ref{eq:expected-value-p-cs}) and (\ref{eq:cs_expected_values_Pi}) 
we can express the time evolution of 
the expected values of the position, the linear momentum 
and the linear pseudomomentum as
\begin{subequations}
\begin{align}
\label{eq:x_cs}
\langle \hat{x} (t) \rangle_{\textrm{cs}} & = 
		\frac{A_{\textrm{cs}} \cos \Theta_{\textrm{cs},\gamma} (t) + \gamma \sigma_0^2}{
			  1 - \gamma A_{\textrm{cs}} \cos \Theta_{\textrm{cs},\gamma} (t) - \gamma^2 \sigma_0^2},
\\
\label{eq:p_cs}
\langle \hat{p} (t) \rangle_{\textrm{cs}} 
&= - m_0 \omega_0 A_{\textrm{cs}} \sin{\Theta_{\textrm{cs},\gamma}} (t)
	\left[
		1 - \gamma A_{\textrm{cs}} \cos \Theta_{\textrm{cs},\gamma} (t) - \frac{\gamma^2 \sigma_0^2}{2}
	\right]
\\
\label{eq:Pi_gamma_cs}
\langle \hat{\Pi}_\gamma (t) \rangle_{\textrm{cs}} 
&= - m_0 \omega_0 A_{\textrm{cs}} \sin{\Theta_{\textrm{cs},\gamma}} (t)
\end{align}
\end{subequations}
with $A_{\textrm{cs}} = \sqrt{2} \sigma_0 |\alpha_\gamma|$
being an amplitude oscillation for the coherent states.
As predicted, 
$\langle \hat{x} (t) \rangle_{\textrm{cs}}$,
$\langle \hat{p} (t) \rangle_{\textrm{cs}}$ and
$\langle \hat{\Pi}_\gamma (t) \rangle_{\textrm{cs}}$
for $A_{\textrm{cs}} / \sigma_0  \gg 1$ 
(or $|\alpha_\gamma| \gg 1$)
evolve over time in the same way 
as their respective classical analog
[Eqs.~(\ref{eq:x_classic(t)}), (\ref{eq:p_classic(t)}) 
and (\ref{eq:Pi_classic(t)})].
A correspondence between the temporal evolution 
for the expected values of the coherent states 
and their trajectory in classical formalism was previously obtained 
by Kais and Levine for the Morse oscillator.\cite{Kais-Levine-1990}
Here, we realize that the correspondence 
is valid for the asymmetric oscillator with PDM.

\begin{figure}[htb]
\centering
\includegraphics[width=0.32\linewidth]{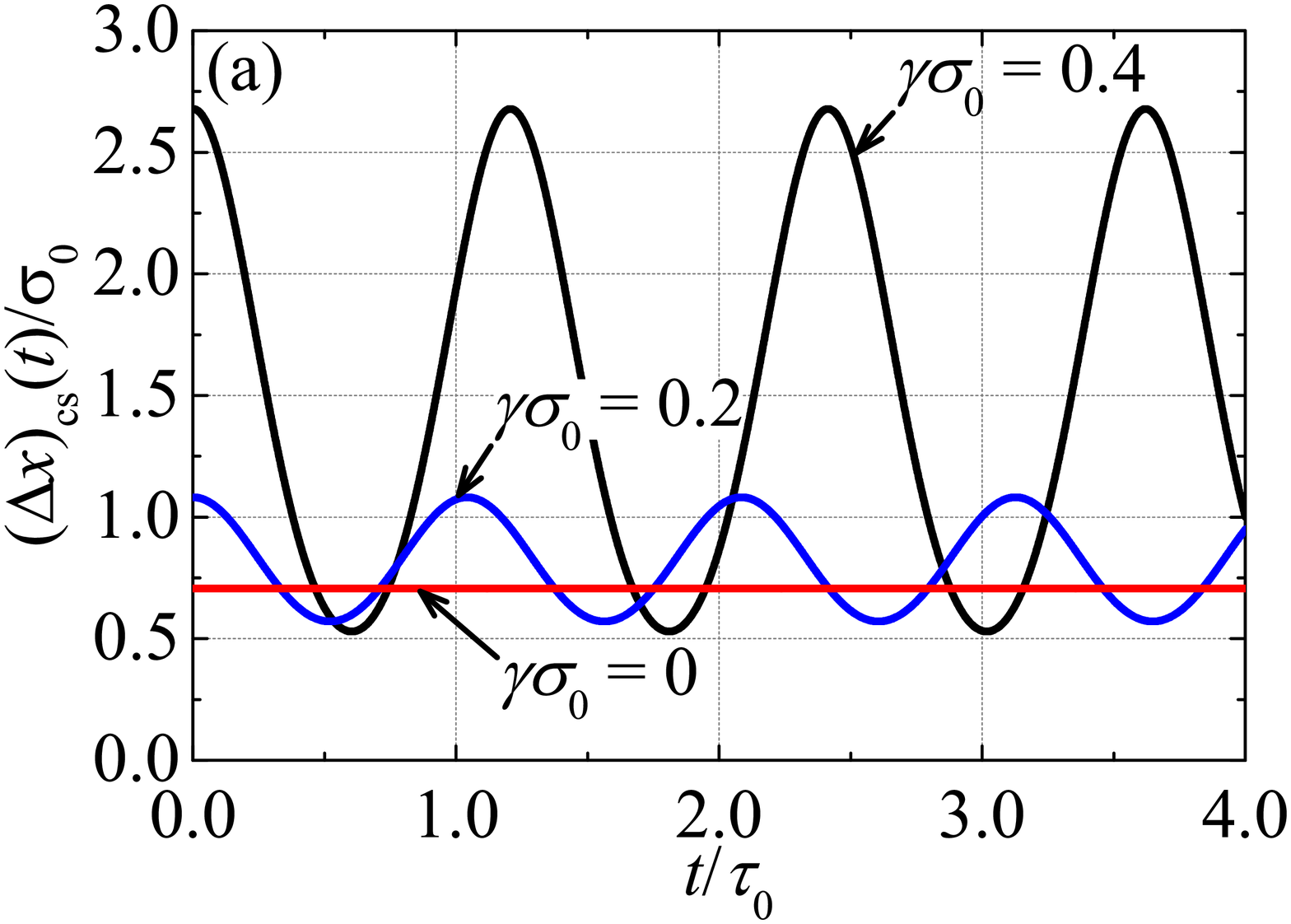}
\includegraphics[width=0.32\linewidth]{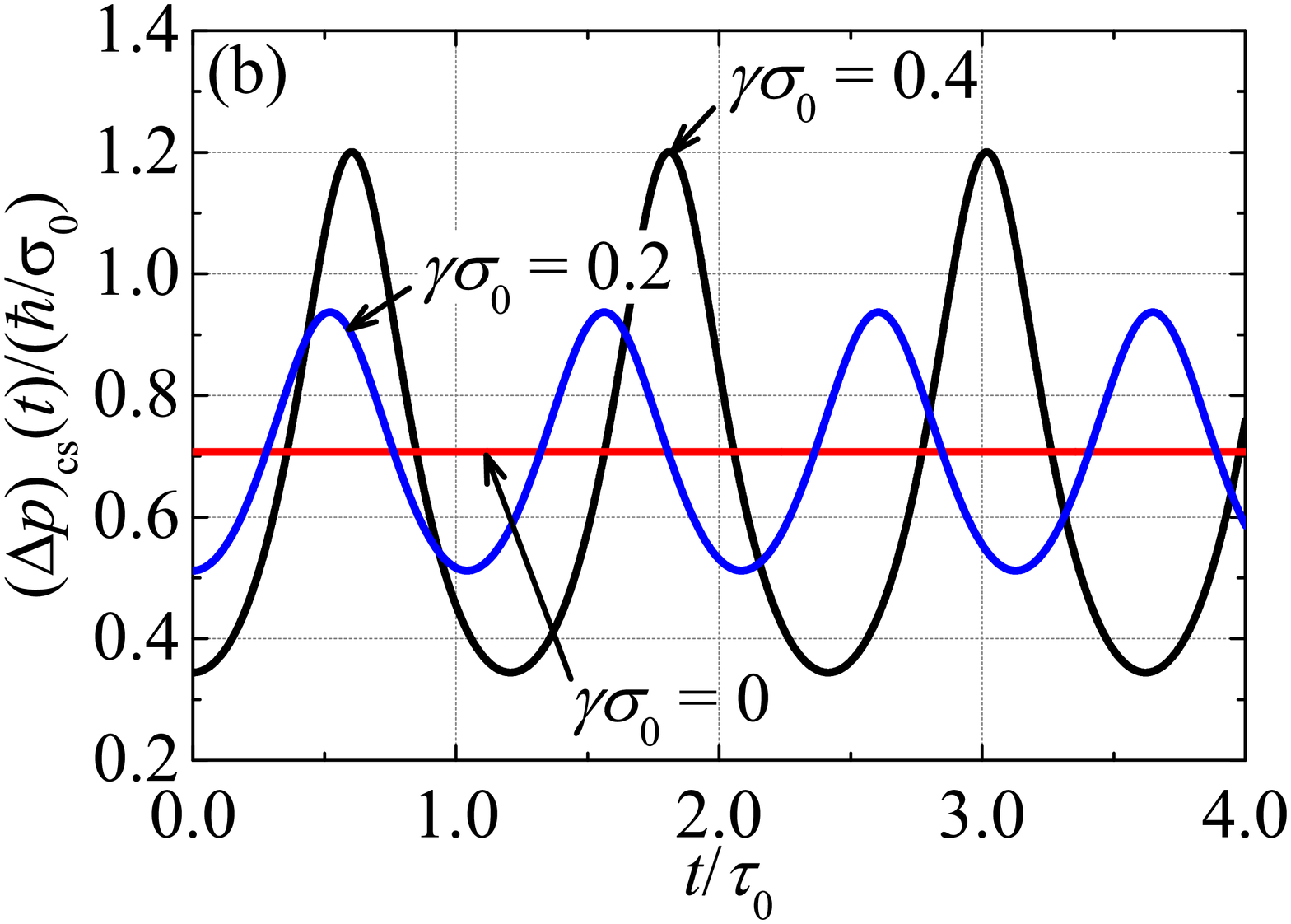}
\includegraphics[width=0.32\linewidth]{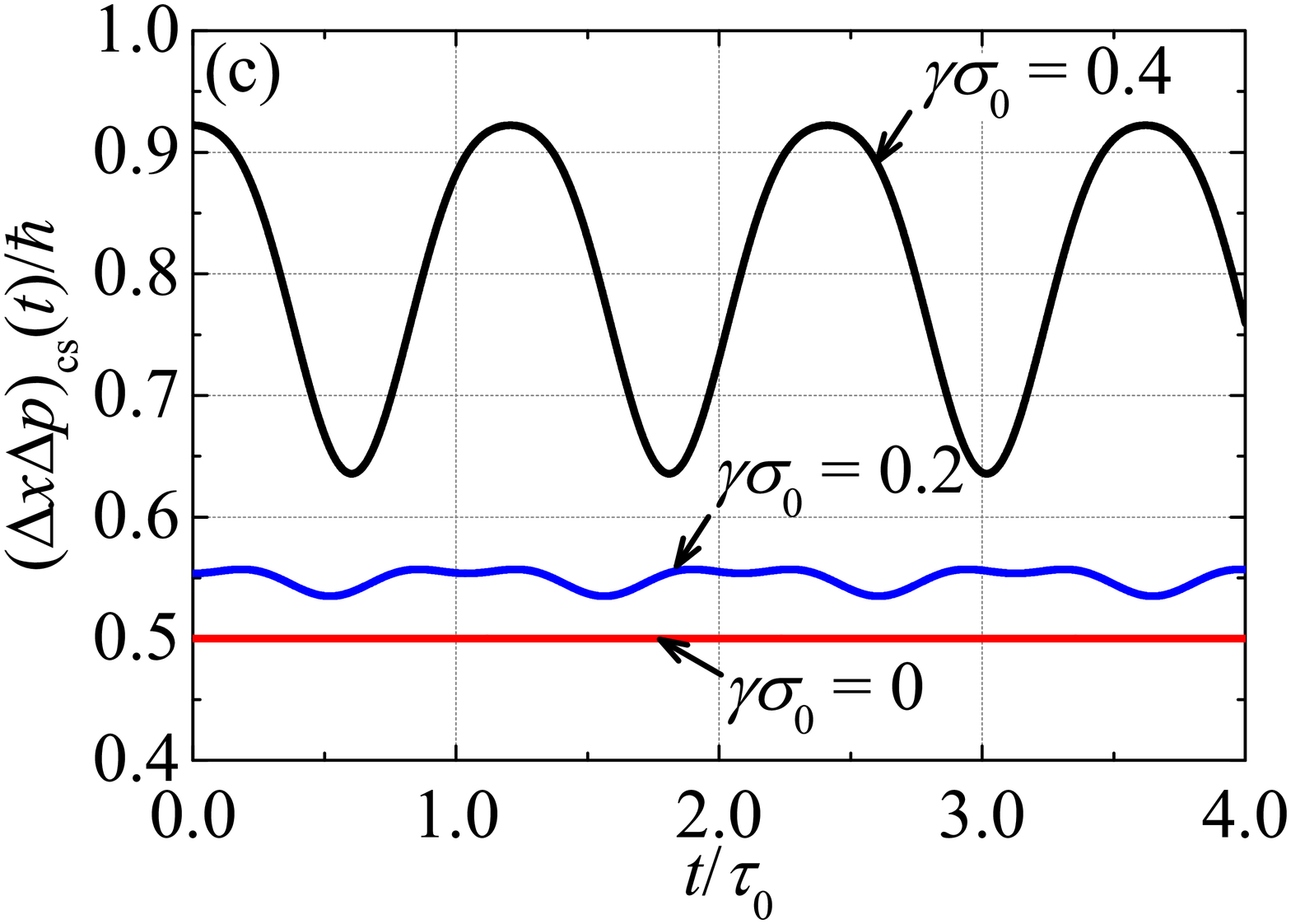}
\caption{\label{fig:10}
Time evolution of the uncertainties of 
(a) the position $(\Delta x)_{\textrm{cs}}(t)$ and 
(b) the linear momentum $(\Delta p)_{\textrm{cs}}(t)$, and 
(c) of the product $(\Delta x)_{\textrm{cs}}(t)(\Delta p)_{\textrm{cs}}(t)$
for the coherent states of the asymmetric oscillator with PDM,
with $|\alpha_\gamma| = 1/\sqrt{2}$ and deformation parameters
$\gamma \sigma_0=0$ (usual case), 
$0.2$ (dashed red) and 
$0.4$ (dashed-dotted blue).
}
\end{figure}

From Eqs.~(\ref{eq:expected-value-x-cs})--(\ref{eq:expected-value-p^2-cs})
with $\alpha_\gamma (t) = |\alpha_\gamma| e^{-i \Theta_\gamma (t)}$
and deformed phase (\ref{eq:deformed-phase_CS}),
we present in Fig.~\ref{fig:10} 
the time evolution of the uncertainties of the position 
$
(\Delta x)_{\textrm{cs}}(t) = 
\sqrt{\langle \hat{x}^2  (t) \rangle_{\textrm{cs}}
      -\langle \hat{x}  (t) \rangle_{\textrm{cs}}^2},
$
the linear momentum
$
(\Delta p)_{\textrm{cs}}(t) = 
\sqrt{\langle \hat{p}^2  (t) \rangle_{\textrm{cs}}
      -\langle \hat{p}  (t) \rangle_{\textrm{cs}}^2},
$
and the product
$(\Delta x)_{\textrm{cs}}(t) (\Delta p)_{\textrm{cs}}(t)$
for coherent states of the asymmetric oscillator
with different values of $\gamma \sigma_0$.
The standard case ($\gamma =0$) 
[$(\Delta x)_{\textrm{cs}} (t) = \frac{\sigma_0}{\sqrt{2}}$,
 $(\Delta p)_{\textrm{cs}} (t) = \frac{\hbar}{\sqrt{2} \sigma_0}$
 and
 $(\Delta x)_{\textrm{cs}} (t) (\Delta p)_{\textrm{cs}} (t) = \frac{\hbar}{2}$] 
is shown for comparison.

As a final application, let 
$\psi_{\pm \alpha_\gamma}(x) = \langle{x}|{\pm \alpha_\gamma} \rangle$
be the wavefunctions
of the corresponding states $|\pm \alpha_\gamma \rangle$ 
in position representation $\{ |\hat{x}\rangle \}$, 
and the even and odd coherent states are given by
\cite{Dodonov-Malkin-ManKo-1974,Buzek-VidiellaBarranco-Knight-1992}
\begin{subequations}
\label{eq:psi_even_odd}
\begin{align}
\psi_{\textrm{even}}(x) &= C_{\textrm{even}} 
[\psi_{\alpha_\gamma}(x) + \psi_{-\alpha_\gamma}(x)], \\
\psi_{\textrm{odd}}(x) &= C_{\textrm{odd}}
[\psi_{\alpha_\gamma}(x) - \psi_{-\alpha_\gamma}(x)].
\end{align}
\end{subequations}
with the normalization constants given by
\begin{equation}
\left\{
\begin{array}{c}
C_{\textrm{even}} \\
C_{\textrm{odd}} 
\end{array}
\right\}
= \frac{1}{2[1\pm \langle -\alpha_\gamma | \alpha_\gamma \rangle]}.
\end{equation}
For the deformed oscillator, we obtain that the inner product 
$\langle -\alpha_\gamma | \alpha_\gamma \rangle
= \int_{-\infty}^{\infty} \psi_{-\alpha_\gamma}^{\ast}(x)
\psi_{\alpha_\gamma}(x)\textrm{d}x$
is expressed as
\begin{equation}
\langle -\alpha_\gamma | \alpha_\gamma \rangle = \sqrt{
\frac{B(\widetilde{\lambda}_{\textrm{cs}} (\alpha_\gamma),\widetilde{\lambda}_{\textrm{cs}} (-\alpha_\gamma))}{
B(\lambda_{\textrm{cs}} (\alpha_\gamma),\lambda_{\textrm{cs}} (-\alpha_\gamma))}},
\end{equation}
whose arguments are
$\lambda_{\textrm{cs}} (\alpha_{\gamma}) 
	= \frac{2}{\gamma^2 \sigma_0^2} 
	  [1-\sqrt{2}\gamma \sigma_0 \textrm{Re}(\alpha_{\gamma})] -1
$
and
$\widetilde{\lambda}_{\textrm{cs}} (\alpha_{\gamma}) 
	= \frac{2}{\gamma^2 \sigma_0^2} 
	  [1-i\sqrt{2}\gamma \sigma_0 \textrm{Im}(\alpha_{\gamma})] -1.$
For $\gamma = 0$ , we recover the standard case
$\lim_{\gamma \rightarrow 0}\langle - \alpha_\gamma | \alpha_\gamma \rangle = e^{-2|\alpha_0|^2}$.

\begin{figure}[htb]
\centering
\includegraphics[width=0.32\linewidth]{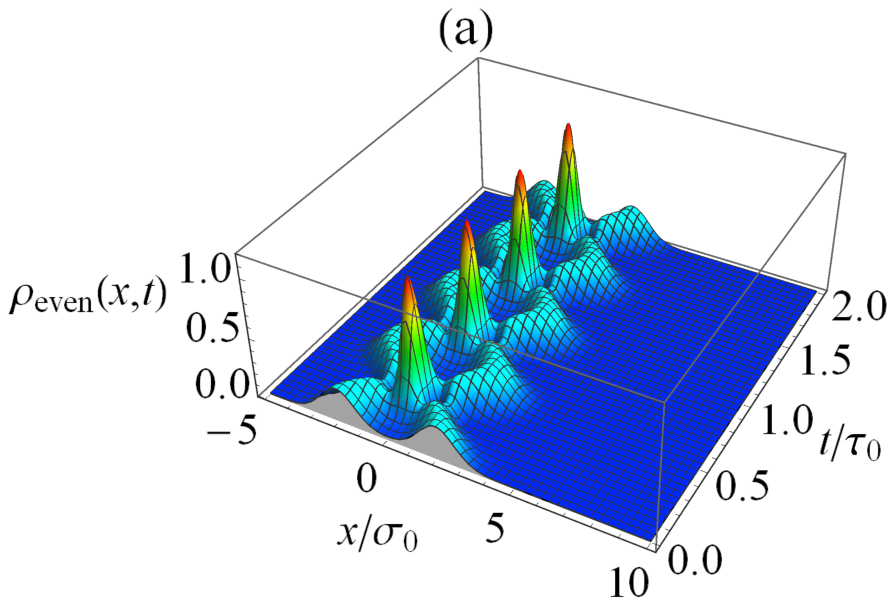}
\includegraphics[width=0.32\linewidth]{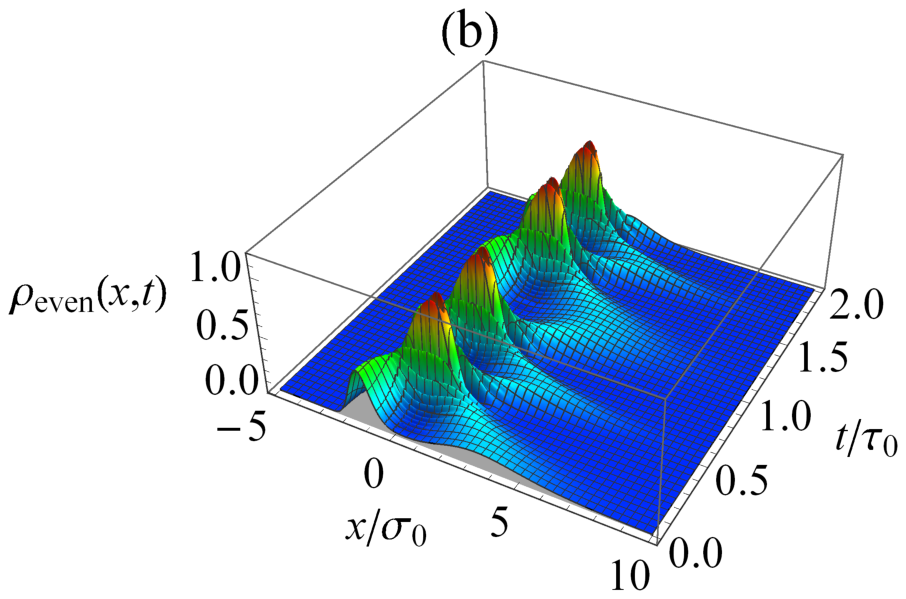}
\includegraphics[width=0.32\linewidth]{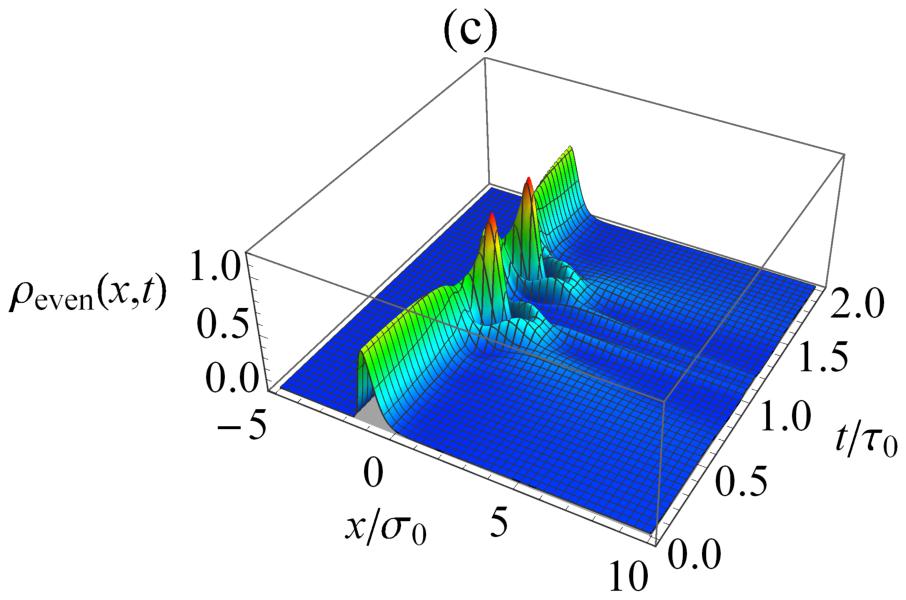}\\
\includegraphics[width=0.32\linewidth]{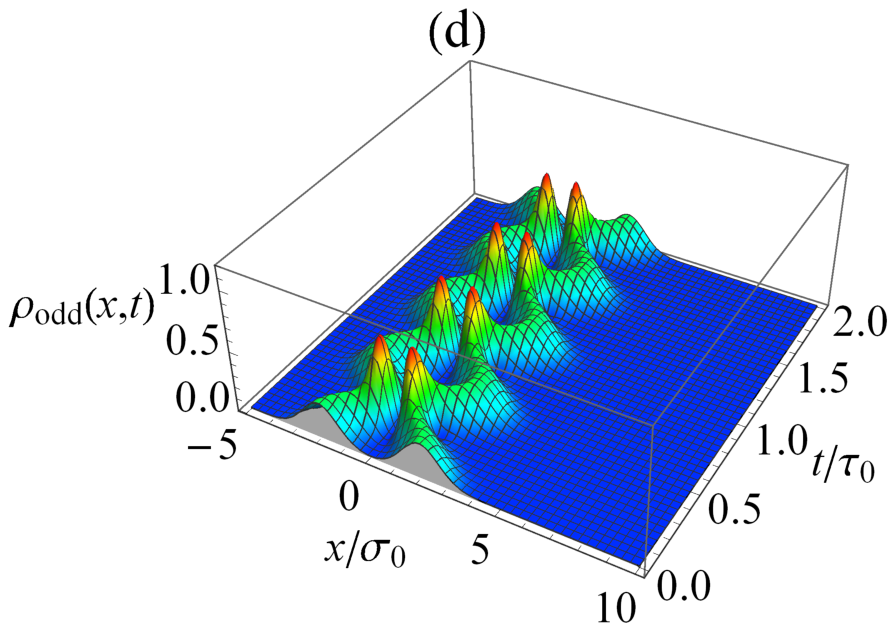} 
\includegraphics[width=0.32\linewidth]{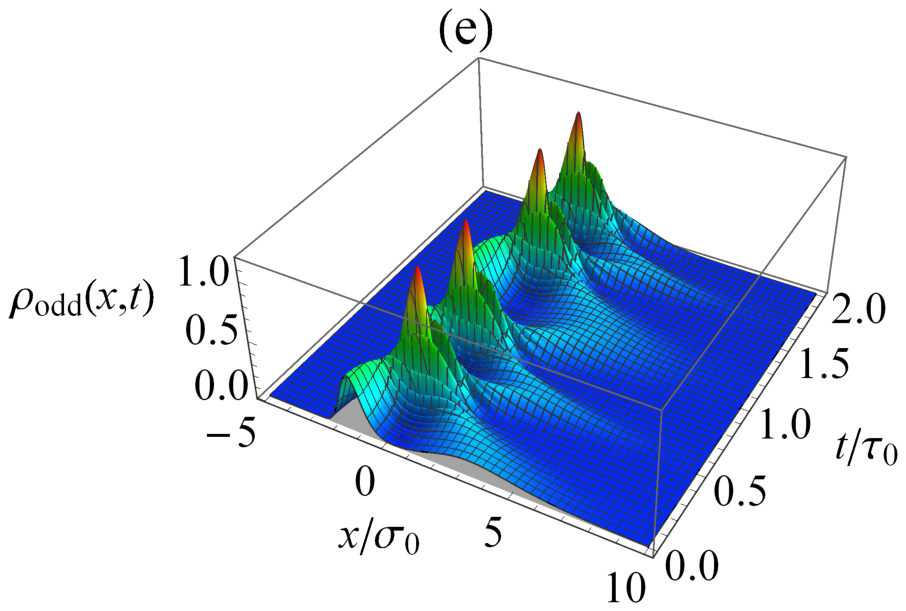}
\includegraphics[width=0.32\linewidth]{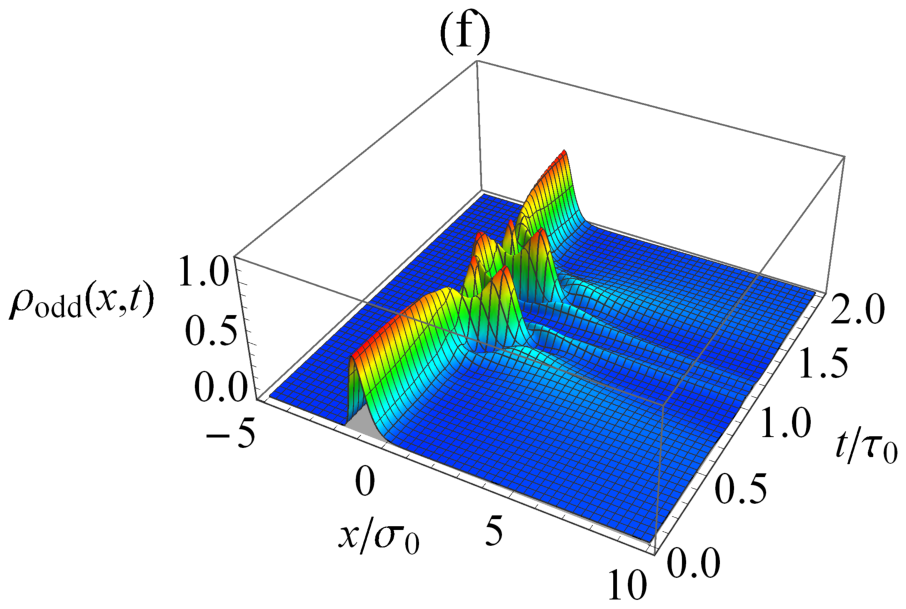}
\caption{\label{fig:11}
Probability behavior of even (upper line) and odd (bottom line) 
coherent states with $|\alpha_\gamma| = \sqrt{2}$ and
$\gamma \sigma_0 = 0$ [(a) and (d) -- standard case],
$\gamma \sigma_0 = 0.1$ [(b) and (e)], and
$\gamma \sigma_0 = 0.2$ [(c) and (f)].
}
\end{figure}

The temporal evolution $\Psi_{\textrm{even}} (x,t)$
and $\Psi_{\textrm{odd}} (x,t)$ of the coherent states are 
obtained by changing
$\alpha_\gamma \rightarrow \alpha_\gamma (t) 
= |\alpha_\gamma| e^{-i\Theta_{\textrm{cs},\gamma} (t)}$
in Eqs.~(\ref{eq:psi_even_odd}).
In figure~\ref{fig:11} we plot the density probabilities
$\rho_{\textrm{even}} (x,t) = |\Psi_{\textrm{even}} (x,t)|^2$
and $\rho_{\textrm{odd}} (x,t) = |\Psi_{\textrm{odd}} (x,t)|^2$
for different values of the parameters 
$\alpha_\gamma$ and $\gamma \sigma_0$.

\section{\label{sec:final-remarks}
         Final remarks}

In this work, we have addressed the properties of a classical 
and quantum deformed oscillator provided with 
an asymmetrical position-dependent mass $m(x)=m_0/(1+\gamma x)^2$ 
in the harmonic potential term of the Hamiltonian. 
As a continuation of our previous work
(Ref.~\onlinecite{Costa-Silva-Gomez-2021}), 
we obtain eigenfunctions, energy levels, mean values, 
coherent states, and uncertainty relationships.
We summarize our contributions as follows:
(i) The classical and quantum Hamiltonians of the deformed 
oscillator result equivalent to the Morse 
oscillator Hamiltonian expressed in the deformed space
due to the introduction of the position-dependent mass 
$m(x)=m_0/(1+\gamma x)^2$
in the potential term given by 
$V(x) =\frac{1}{2}m(x)\omega_0^2x^2$. 
(ii) The evolution of the mean values of the position 
and the momentum remains as its oscillatory character,
but the phase space deforms continuously and asymmetrically 
from a circle (null deformation)
to a closed curve with peaks (Fig. 1). 
In addition, the wavefunctions and their 
probability functions result asymmetrical around $x=0$.  
(iii) For large values of the quantum number 
$n$, the classical limit is recovered (Fig. 3). 
(iv) Uncertainty relationships have a lower bound 
that increases as the deformation increases (Fig. 5).

Overall, there exist two controlled ways 
to introduce a deformation in the context 
of the harmonic oscillator Hamiltonian: 
(1) by means of a deformation in the momentum 
$p\rightarrow p_\gamma$
and (2) through a deformation in the harmonic potential 
$V_0(x)=\frac{1}{2}m_0\omega_0^2x^2\rightarrow V(x)=\frac{1}{2}m(x)\omega_0^2x^2$. 
By employing both (1) and (2) we have shown that, 
analogously to the accomplished in Ref.~\onlinecite{Costa-Silva-Gomez-2021}, 
the deformation affects the stationary aspects, 
expressed by the asymmetry of the energies and eigenfunctions 
and the increasing lower bounds of the uncertainty relationships 
as well as the temporal aspects regarding the localization 
of the probability distributions of the eigenfunctions 
and the oscillations of the uncertainty relationships. 
It remains to be seen whether this is due to 
the fact that the same position-dependent mass was used 
in the momentum and in the potential, 
which we hope will be studied future research.

We also investigate the coherent states of the deformed oscillator.
The probability density of the coherent states 
localizes rapidly regarding the oscillation period 
$\tau_0=2\pi/\omega_0$ because of the effect 
of the asymmetrical deformation provided by the PDM (Fig 6). 
As previously predicted in the literature for coherent states 
of the Morse oscillator (see Ref.~\onlinecite{Angelova-Hussin-2008})
--- system equivalent to the PDM oscillator that we study here---
the uncertainty principle is also satisfied for the deformed oscillator.
Periodic oscillations in the time evolution 
of the uncertainty relationships arise 
as the dimensionless deformation parameter 
$\gamma \sigma_0$ increases (Fig 7).
In addition, as in the Morse oscillator,\cite{Kais-Levine-1990} 
the time evolution of the observable position 
and linear momentum for the coherent states 
reproduce their respective classical analogs.

We have limited this work in investigating
Glauber's coherent states for the deformed oscillator.
However, the investigation of generalized coherent states 
for systems with PDM can be obtained from the approaches of
Klauder and Sudarshan,\cite{Klauder-1963a,Klauder-1963b,Sudarshan-1963}
Barut and Girardello,\cite{Barut-Girardello-1971,Popov-Dong-Pop-Sajfert-Simon-2013}
Gazeau--Klauder,\cite{Roy-Roy-2002} and
Perelomov,\cite{Perelomov-1972,Perelomov-1986}
quantum anomaly in the systems with the second order supersymmetry,\cite{Plyushchay-2017}
generalized Heisenberg algebras,\cite{Belfakir-Hassouni-Curado-2020}
and conformal quantum mechanical\cite{Inzunza-2020}
in interesting additional developments.
A group theoretical approach to the deformed oscillator, 
including both bound and scattering, 
can be obtained in future developments from the formalism introduced in 
Ref.~\onlinecite{Alhassid-Iachello-Gursey-1983} and applied 
to the Morse oscillator.


\section*{References}


\begin{thebibliography}{99}


\bibitem{Schrodinger-1926}
E. Schr\"odinger,
Naturwissenschaften {\bf 14}(28), 664 (1926).

\bibitem{Glauber-1963}
R. J. Glauber,
Phys. Rev. {\bf 131}(6), 2766 (1963).

\bibitem{Klauder-1963a}
J. R. Klauder, 
J. Math. Phys. {\bf 4}(8), 1055 (1963).

\bibitem{Klauder-1963b}
J. R. Klauder, 
J. Math. Phys. {\bf 4}(8), 1058 (1963).
 
\bibitem{Sudarshan-1963}
E. C. G. Sudarshan, 
Phys. Rev. Lett. {\bf 10}, 277 (1963).

\bibitem{Barut-Girardello-1971}
A. O. Barut and L. Girardello, 
Commun. Math. Phys. {\bf 21}, 41 (1971).


\bibitem{Perelomov-1972}
A. M. Perelomov,
Commun. Math. Phys. {\bf 26}, 222 (1972).


\bibitem{Gazeau-2009}
J. P. Gazeau,  
{\it Coherent States in Quantum Physics}
(Wiley--VCH, 2009).

\bibitem{Gerry-et-al-2005}
C. Gerry and  P. L. Knight,
{\it Introductory Quantum Optics} 
(Cambridge University Press, 2005).

\bibitem{Monroe-et-al-1996}
C. Monroe, D. M. Meekhof, B. E. King, and  D. J. Wineland,  
Science, {\bf 272}(5265), 1131 (1996).

\bibitem{Laghaout-2013}
A. Laghaout, J. S. Neergaard-Nielsen, I. Rigas, C. Kragh, A. Tipsmark, and U. L. Andersen, 
Phys. Rev. A {\bf 87}(4), 043826 (2013).

\bibitem{Lutterbach-Davidovich-1997}
L. G. Lutterbach and L. Davidovich, 
Phys. Rev. Lett. {\bf 78}(13), 2547 (1997).

\bibitem{Aragao-et-al-2004}
A. Arag\~ao, A. T. Avelar, and B.  Baseia,
Phys. Lett. A {\bf 331}(6), 366-373 (2004).

\bibitem{Zavatta-et-al-2007}
A. Zavatta, V. Parigi, and M. Bellini, 
Phys. Rev. A {\bf 75}(5), 052106 (2007).

\bibitem{Walls-1983}
D. F. Walls, 
Nature {\bf 306}(5939), 141 (1983).



\bibitem{Bastard-1975}
G. Bastard, J. K. Furdyna, and J. Mycielski,
Phys. Rev. B {\bf 12}, 4356 (1975).

\bibitem{vonroos_1983}
O. von Roos,
Phys. Rev. B {\bf 27}, 7547 (1983).

\bibitem{BenDaniel-Duke-1966}
D. J. BenDaniel and C. B. Duke,
Phys. Rev. {\bf 152}, 683 (1966).

\bibitem{Gora-Williams-1969}
T. Gora and F. Williams,
Phys. Rev. {\bf 177}, 1179 (1969).

\bibitem{Zhu-Kroemer-1983}
Q.-G. Zhu and H. Kroemer,
Phys. Rev. B {\bf 27}, 3519 (1983).

\bibitem{Li-Kuhn-1993}
T. L. Li and K. J. Kuhn,
Phys. Rev. B {\bf 47}, 12760 (1993).

\bibitem{Morrow-Brownstein-1984}
R. A. Morrow and K. R. Brownstein,
Phys. Rev. B {\bf 30}, 678 (1984).

\bibitem{Mustafa-Mazharimousavi-2007}
O. Mustafa and S. H. Mazharimousavi,
Int. J. Theor. Phys. {\bf 46}, 1786 (2007).

%
\bibitem{Li-Guo-Jiang-Hu}
K. Li, K. Guo, X. Jiang, M. Hu,
Optik {\bf 132}, 375 (2017).

%
\bibitem{Saavedra_1994}
F. Arias de Saavedra, J. Boronat, A. Polls, and A. Fabrocini,
Phys. Rev. B {\bf 50}, 4248 (1994).

\bibitem{Bencheikh-2004}
K. Bencheikh, K. Berkane and S. Bouizane,
J. Phys. A: Math. Gen. {\bf 37} (45), 10719 (2004).


\bibitem{Yu-Dong-Sun-2004}
J. Yu, S.-H. Dong, and G.-H. Sun, 
Phys. Lett. A {\bf 322}, 290 (2004).

\bibitem{Christiansen-Cunha-2014}
H. R. Christiansen and M. S. Cunha,
J. Math. Phys. {\bf 55}, 092102 (2014).

\bibitem{Yanez-Navarro}
G. Ya\~nez-Navarro, G.-H. Sun,  T. Dytrych, K. D. Launey, S.-H. Dong, and J. P. Draayer, 
Ann. Phys. {\bf 348} 153 (2014).

\bibitem{Aydogdu-Arda-Sever-2012}
O. Aydo\v{g}du, A. Arda, and R. Sever,
J. Math. Phys. {\bf 53}, 042106 (2012).

\bibitem{Merad-etal_2019}
A. Merad,  M. Aouachria, M. Merad, and T. Birkandan, 
Int. J. Mod. Phys. A {\bf 34}(32), 1950218 (2019).

\bibitem{Alimohammadi-Hassanabadi-Zare-2017}
M. Alimohammadi, H. Hassanabadi, and S. Zare,
Nucl. Phys. A {\bf 960}, 78 (2017).

\bibitem{Schmidt-2018}
A. G. M. Schmidt and A. L. de Jesus,
J. Math. Phys. {\bf 59}, 102101 (2018).

\bibitem{Jesus-2019}
A. L. de Jesus and A. G. M. Schmidt,
J. Math. Phys. {\bf 60}, 122102 (2019).

\bibitem{Mathews-Lakshmanan-1974}
P. M. Mathews and M. Lakshmanan,
Q. Appl. Math. {\bf 32}, 215 (1974).

\bibitem{Mathews-Lakshmanan-1975}
P. M. Mathews and M. Lakshmanan,
Nuovo Cimento A {\bf 26}, 299 (1975).

\bibitem{Tiwari-2013}
A. K. Tiwari, S. N. Pandey, M. Senthilvelan, and M. Lakshmanan,
J. Math. Phys. {\bf 54}(5), 053506 (2013).


\bibitem{Bagchi-2015}
B. Bagchi, A. Ghose Choudhury, and  P. Guha,
J. Math. Phys. {\bf 56}, 012105 (2015).


\bibitem{Carinena-2015}
J. F. Cari\~nena,  M. F. Ra\~nada, and M. Santander,
Ann. Phys. {\bf 322}, 434 (2007).

\bibitem{Ruby-2015}
V. Chithiika Ruby, V. K. Chandrasekar, M. Senthilvelan, and  M. Lakshmanan,
J. Math. Phys. {\bf 56}, 012103 (2015).

\bibitem{Schulze-Halberg-Roy-2016}
A. Schulze-Halberg and B. Roy,
J. Math. Phys. {\bf 57}, 102103 (2016).

\bibitem{Karthiga-et-al-2017}
S. Karthiga, V. Chithiika Ruby, M. Senthilvelan, and M. Lakshmanan,
J. Math. Phys. {\bf 58}, 102110 (2017).

\bibitem{Quesne-2022}
C. Quesne, 
Eur. Phys. J. Plus {\bf 137}, 225 (2022).

\bibitem{Jafarov-2020}
E. I. Jafarov, S. M. Nagiyev, and A. M. Jafarova, 
Rep. Math. Phys. {\bf 86}, 25 (2020).

\bibitem{Jafarov-2021}
E. I. Jafarov and J. Van der Jeugt, 
Eur. Phys. J. Plus {\bf 136}, 758 (2021).

\bibitem{Jafarov-2022}
E. I. Jafarov,
Physica E {\bf 139}, 115160 (2022).

\bibitem{Plastino-etal-1999}
A. R. Plastino, A. Rigo, M. Casas, F. Garcias, and A. Plastino,
Phys. Rev. A {\bf 60}(6), 4318 (1999).

\bibitem{Bravo-PRD-2016}
R. Bravo and M. S. Plyushchay,
Phys. Rev. D {\bf 93}, 105023 (2016).

\bibitem{Amir-Iqbal-2016}
N. Amir and S. Iqbal,
J. Math. Phys. {\bf 57}, 062105 (2016).

\bibitem{Karthiga-2018}
S. Karthiga, V. Chithiika Ruby,, and  M. Senthilvelan,
Phys. Lett. A {\bf 382}(25), 1645 (2018).

\bibitem{Mustafa-2020}
O. Mustafa,
Phys. Lett. A {\bf 384}, 126265 (2020).

\bibitem{Ruby-Senthilvelan-2010}
V. Chithiika Ruby and M. Senthilvelan,
J. Math. Phys. {\bf 51}, 052106 (2010).

\bibitem{Amir-Iqbal-2015}
N. Amir and S. Iqbal,
J. Math. Phys. {\bf 56}, 062108 (2015).

\bibitem{Amir-Iqbal-2016-CS}
N. Amir and S. Iqbal,
Commun. Theor. Phys. {\bf 66}, 615 (2016).

\bibitem{Tchoffo-2019}
M. Tchoffo, F. B. Migueu, M. Vubangsi, and L. C. Fai,
Heliyon {\bf 5}(9), e02395 (2019).


\bibitem{CostaFilho-Almeida-Farias-AndradeJr-2011}
R. N. Costa Filho, M. P. Almeida, G. A. Farias, and J. S. Andrade Jr.,
Phys. Rev. A {\bf 84}, 050102(R) (2011).

\bibitem{CostaFilho-Alencar-Skagerstam-AndradeJr-2013}
R. N. Costa Filho, G. Alencar, B.-S. Skagerstam, and J. S. Andrade Jr.,
Europhys. Lett. {\bf 101}, 10009 (2013).

\bibitem{Aguiar-2020}
V. Aguiar, S. M. Cunha, D. R. da Costa,  R. N. Costa Filho,
Phys. Rev. B {\bf 102}, 235404 (2020).


\bibitem{Costa-Borges-2014}
B. G. da Costa and E. P. Borges,
J. Math. Phys. {\bf 55}, 062105 (2014).

\bibitem{Costa-Borges-2018}
B. G. da Costa and E. P. Borges,
J. Math. Phys. {\bf 59}, 042101 (2018).

\bibitem{Costa-Silva-Gomez-2021}
B. G. da Costa, G. A. C. da Silva, I. S. Gomez
J. Math. Phys. {\bf 62}, 092101 (2021).

\bibitem{Costa-Gomez-Portesi-2020}
B. G. da Costa, I. S. Gomez, and M. Portesi,
J. Math. Phys. {\bf 61}, 082105 (2020).

\bibitem{Jamshir-Lari-Hassanabadi-2021}
N. Jamshir, B. Lari, and H. Hassanabadi,
Physica A {\bf 565}, 125616 (2021).

\bibitem{Cruz-2019}
S. Cruz y Cruz and C. Santiago-Cruz 
Math. Methods Appl. Sci. {\bf 42}, 4909 (2019).

\bibitem{Plyushchay-2017}
M. S. Plyushchay,
Ann. Phys. {\bf 377}, 164 (2017).


\bibitem{Nieto-SimmonsJr-1979}
M. M. Nieto and L. M. Simmons Jr,
Phys. Rev. D {\bf 20}, 1342 (1979).


\bibitem{Perelomov-1986}
A. Perelomov, {\it Generalized Coherent States}
(Springer, Berlin, 1986).


\bibitem{Kais-Levine-1990}
S. Kais and R. D. Levine, 
Phys. Rev. A {\bf 41}, 2301 (1990).


\bibitem{Copper-1992}
I. L. Cooper, 
J. Phys. A: Math. Gen. {\bf 25} 1671 (1992).


\bibitem{Roy-Roy-2002}
B. Roy and P. Roy, 
Phys. Lett. A {\bf 296}(4-5) 187 (2002).



\bibitem{Popov-Dong-Pop-Sajfert-Simon-2013}
D. Popov, S.-H. Dong, N. Pop, V. Sajfert, and S. \c{S}imon,
Ann. Phys. {\bf 339}, 122 (2013).



\bibitem{Angelova-Hussin-2008}
M. Angelova and  V. Hussin,  
J. Phys. A: Math. Theor. {\bf 41}(30), 304016 (2008).


\bibitem{Belfakir-Hassouni-Curado-2020}
A. Belfakir, Y. Hassouni, and E. M. F. Curado 
Phys. Lett. A {\bf 384}, 126553 (2020).




\bibitem{Dodonov-Malkin-ManKo-1974}
V. V. Dodonov, I. A. Malkin, and V. I.  Man'ko, 
Physica {\bf 72}(3), 597 (1974).

\bibitem{Buzek-VidiellaBarranco-Knight-1992}
V. Bu\v{z}ek,  A. Vidiella-Barranco, and P. L. Knight, 
Phys. Rev. A {\bf 45}, 6570 (1992).


\bibitem{Inzunza-2020}
L. Inzunza, M. S. Plyushchay, A. Wipf, 
Phys. Rev. D {\bf 101}, 105019 (2020).


\bibitem{Alhassid-Iachello-Gursey-1983}
Y. Alhassid, F. Iachello, and F. G\"ursey,  
Chem. Phys. Lett. {\bf 99}(1), 27 (1983).



\end{thebibliography}
\end{document}